\documentclass[a4paper,10pt]{article}
\usepackage[margin=1in]{geometry}
\usepackage[utf8]{inputenc}
\usepackage{mathtools}
\usepackage{amsmath}
\usepackage{amssymb}
\usepackage{bm}
\usepackage[toc,page]{appendix}
\newtheorem{thm}{Theorem}[section]
\newtheorem{defn}[thm]{Definition}
\usepackage{graphicx}
\usepackage{booktabs}  
\usepackage{float}
\usepackage{caption}
\usepackage{etoolbox}
\AtBeginEnvironment{figure}{\setcounter{subfigure}{0}}
\usepackage{booktabs}

\usepackage[center]{subfigure}
\usepackage{algorithm2e}
\usepackage[utf8]{inputenc}
\usepackage[nottoc,numbib]{tocbibind}
\graphicspath{{ResultFigs/}}

\usepackage{multirow}
\newtheorem{proposition}{Proposition}

\title{Monte Carlo Calculation of Exposure Profiles and Greeks for Bermudan and Barrier Options under the Heston Hull-White Model}
\author{Q.Feng \thanks{CWI, center for mathematics and computer science, Amsterdam, the Netherlands, qian@cwi.nl}
\and C.W.Oosterlee\thanks{CWI center of mathematics and computer science, and Delft University of Technology, Delft, the Netherlands, c.w.oosterlee@cwi.nl}}

\date{\today}

\begin{document}
\maketitle

\begin{abstract}
Valuation of Credit Valuation Adjustment (CVA) has become an important field as its calculation is required 
in Basel III, issued in 2010, in the wake of the credit crisis. 
\textit{Exposure}, which is defined as the potential future loss of a default event without any recovery, is one of the key elements
for pricing CVA. This paper provides a backward dynamics framework for assessing exposure profiles of European, Bermudan and barrier options
under the Heston and Heston Hull-White asset dynamics. We discuss the potential of an efficient and adaptive Monte Carlo approach, the
 \textit{Stochastic Grid Bundling Method} (SGBM), which employs the techniques of
\textit{simulation}, \textit{regression} and \textit{bundling}. Greeks of the exposure profiles can be calculated in the same backward iteration with little extra effort.
Assuming independence between default event and exposure profiles, we give  examples of calculating exposure, CVA and Greeks for Bermudan and barrier options. 
\end{abstract}


\section{Introduction}
\textit{Counterparty Credit Risk} (CCR) refers to the risk that a counterparty of a financial contract will default prior to the 
expiration of the contract, and thus cannot make all payments required by the contract. 
The management of CCR has caught special attention since the financial crisis which started in 2007 \cite{JonGregoryCVA}. OTC (over-the-counter) derivative contracts are especially subject to CCR, since trades are made directly between two parties without supervision of a third party. 
Basel III issued in 2010 introduced an additional capital charge requirement, called \textit{Credit Valuation Adjustment} (CVA) 
to cover the risk of losses on a counterparty default event for OTC derivatives. 
The CVA equals the difference between the risk-free value and the market value 
of a contract with the possibility of counterparty default, which indicates that CVA materializes the market value of CCR \cite{JonGregoryCVA}.

One difficulty in pricing CVA arises from the uncertainty of the losses in the event of default, which is commonly defined as \textit{Exposure}.
As the market moves, exposure values of contracts evolve over time, diverting away from the initial book value \cite{JonGregoryCVA}. 

The presence of CCR may have significant impact, for example, on the exercise-strategy of a Bermudan option, as the holder can decide to 
exercise the option earlier to reduce the risk of a future default event \cite{PeterKlein}. 
In the present paper however, we assume independence between default and exposure values, and
we further assume that the presence
of counterparty default has no influence on the holder's decision of exercising options. 

Monte Carlo simulation is a widely used approach to determine an empirical distribution of the underlying asset,  
but also other pricing approaches have been adapted to the computation of exposure.  In \cite{yanbinshen}, 
the COS method is adapted for pricing the exposure of Bermudan options under L\'{e}vy process ; in \cite{keesqian}, the finite difference method is employed  
for calculation of the exposure values under the Heston model. 

One focus of this paper is to develop and understand an adaptive and efficient Monte Carlo approach, the \textit{Stochastic Grid Bundling Method} (SGBM)
 for computation of exposure profiles of European, Bermudan, barrier options with a single asset 
under stochastic volatility and stochastic interest rate dynamics (Heston and Heston Hull-White model, respectively).

We will study the convergence and performance of SGBM for valuing mainly the exposure of Bermudan options, determine accurate Greeks, and also present the corresponding results for barrier options. 

SGBM was proposed for pricing Bermudan multiple asset derivative contracts under Black-Scholes dynamics, by Jain and Oosterlee in~\cite{shashijain}.
The method is based on \emph{simulation}, \emph{regression} and \emph{bundling}. 
The idea of using simulation and regression for pricing American options has been used earlier by Carriere(1996) \cite{Carriere}, Tsitsiklis and Van Roy(2001) \cite{Tsitsiklis},
or Longstaff and Schwartz(2001) \cite{Longstaff-Schwartz}.
All these methods, though different in some aspects, approximate the continuation value at each time point based on regression to determine the 
\textit{optimal} exercise-strategy. 
There are several modifications and comparisons of the original pricing techniques, like in \cite{Stentoft}, or Broadie, Glasserman and Ha (2000)~\cite{Broadie2}.
In \cite{Glasserman}, Glasserman and Yu(2002) compare the so-called 'regression later' approach 
with a 'regression now' approach, and conclude the superiority of 'regression later'. SGBM is also based on the 'regression later approach'; in combination with bundling of grid points at each simulated 
time step, the regression step is enhanced.
An important feature of SGBM is that at each time point the option value (and thus the exposure value) is available at each grid point of the stochastic grid, not only for in-the-money points.
This is beneficial within the dynamic programming principle giving accurate direct estimators, and facilitates the accurate computation of Greeks at all time points.
In section \ref{secSGBM}, the framework to apply SGBM for computing exposure profiles for European, Bermudan and barrier options is explained;
we will give insight in the approximation of the option and continuation values on a polynomial space, and thus on the choice of basis functions for the regression step; based on convergence analysis, the need to use bundles can be explained.
Computational details and numerical results for Heston and the Heston Hull-White (HHW) models are given in sections \ref{sec:bundles}, respectively. 

The HHW model is a useful model when pricing long-term contracts in which we observe an implied volatility smile in the asset, and we require a stochastic interest rate. The model is encountered in FX markets, for example, as well as in embedded options or inflation options in the pension and insurance industry.
Our SGBM exposure pricing technique for the HHW model is based on approximations of the full-scale dynamics in \cite{Lech}.

Before we start to focus on SGBM, we will give an overview of the concepts and mathematical formulation of CVA and exposure, 
and provide a backward pricing dynamics of pricing in section \ref{pricing}.

\section{CVA, exposure and option pricing}\label{pricing}
Basically, there are three key elements in pricing CVA: (1) loss given default (LGD), which is the percentage of loss 
in a default event, (2) expected exposure (EE), which quantifies the expectation of the losses at a future time point and (3) probability 
of the counterparty default (PD). It is often assumed that LGD is a fixed ratio based on market information, and 
we also make the assumption that LGD is constant \cite{JonGregoryCVA}. 

In a general real-life situation however these three elements are correlated and not independent, as indicated by the concept {\em wrong-way (or right-way) risk}. Here, we assume independence and focus on the accurate computation of exposure and the corresponding Greeks under stochastic volatility/stochastic interest rate asset dynamics.
The wrong-way risk concept is left for our future research.

Credit \textit{exposure}  represents the economic loss for the derivatives contract holder in a default event without any recovery. 
The contract holder suffers a loss only if the contract has a positive mark-to-market (MtM) value when the other party defaults, i.e. the exposure is defined as the maximum of 
the contract's market value and zero.
We are concerned with unilateral CVA in this paper, although CCR is bilateral between two parties in the contract.

The future exposure value is uncertain as the market moves unpredictably. We can generate an empirical exposure density via simulation by generating a large number of 
asset paths. A general framework is presented for calculating exposure distributions  on OTC derivative products at each future point in \cite{MichaelPykhtin}. 
There are three basic steps in the valuation of exposure profiles: (1) Monte Carlo path generation for a series of simulation dates under some underlying dynamics; 
(2) valuation of mark-to-market values of 
the contract for each realization at each simulation date, applying some numerical method; (3) calculation of exposure for each simulation at each simulation date.

In order to quantify exposure and CVA, we need mathematical expressions.  
We use an $n$-dimensional market state variable,
$\textbf{X}_t:=[x_t^{(1)},x_t^{(2)},\dots,x_t^{(n)}]^T$, to 
present the market information, i.e. the log-asset price ($x_t$), volatility($v_t$) and/or the interest rate($r_t$) ($n=2$ or $n=3$). 
Suppose the random variable $\textbf{X}_t$ follows stochastic dynamics given by
\begin{equation}\label{eq:dynamic}
d\textbf{X}_t=\mu\left(\textbf{X}_t\right)dt+\sigma(\textbf{X}_t)d\widetilde{\textbf{W}}_t,
\end{equation}
where $\widetilde{\textbf{W}}_t$ is an $\mathcal{F}_t$-vector of independent Brownian motions 
in $\mathbb{R}^n$, $\mu\left(\textbf{X}_t\right)\rightarrow\mathbb{R}^n$, $\sigma(\textbf{X}_t)\rightarrow\mathbb{R}^{n\times n}$. 

By definition, the option exposure is the positive part of the market value of the contract, i.e., the maximum of the contract value and zero,
\begin{equation}
\label{eq:exposure}
\text{E}(t,\textbf{X}_t):=\max(V(t,\textbf{X}_t),0),
\end{equation}
where $ V(t, \textbf{X}_t)$ is the  option value at time $t$ depending on market variable $\textbf{X}_t$.
It is assumed that the exposure value becomes zero immediately when the option is not active
(either exercised in the case of Bermudan options or knocked out in the case of barrier options).

Some well-known quantities of the exposure distribution are used for measurement. 
\textit{Expected Exposure} (EE),  the expectation of the exposure, provides an estimate of the expected value of loss; while \textit{Potential Future Exposure} (PFE),
the quantile of exposure at a certain fixed level, is used to measure the 'worst' loss for risk management purposes. 
Both EE and PFE are deterministic functions with respect to time \cite{JonGregoryCVA}.
The mathematical formulas for the functions EE and PFE, conditioned on $\textbf{X}_{0}$, are given by: 
\begin{align}
\text{EE}(t)&:=\mathbb{E}^{\mathbb{Q}}\left[\text{E}(t,\textbf{X}_t)\big|\textbf{X}_0\right], \\
\text{PFE}^{\alpha}(t)&:=\text{inf}\left\{x\Big|\mathbb{Q}\left\{\text{E}(t,\textbf{X}_t)<x|\textbf{X}_{0} \right\}>\alpha \right\},
\end{align}
where $\mathbb{Q}$ is the risk-neutral measure, 
and $\alpha$ is the confidence level. For calculating PFE, the confidence level $\alpha=97.5\%$ is commonly used to measure the 'worst' losses. 

There is a discussion about using the real world measure $\mathbb{P}$ or the risk-neutral measure $\mathbb{Q}$ in \cite{Brigo1}. 
The measure $\mathbb{Q}$ is preferred  for pricing, but
when simulating under measure $\mathbb{P}$, one can include the ``correlation'' between the default probability and the underlying asset, that
allows modeling the \textit{wrong way risk}.  
Default probabilities under $\mathbb{Q}$, inferred from market prices of CDS or
corporate bonds, are typically larger than those under the measure $\mathbb{P}$~\cite{Brigo1}. 
As we will not consider wrong-way risk, and we choose measure $\mathbb{Q}$ here for the pricing purposes.

To calculate CVA, we define another quantity to represent the discounted EE at time $t$:
\begin{equation}
\text{EE}^*(t):=\mathbb{E}^{\mathbb{Q}}\left[D(0,t)\text{E}(t,\textbf{X}_t)\big|\textbf{X}_0\right],
\end{equation}
where the discount factor is defined by
\begin{equation}
D(s,t):=\exp\left(-\int_s^t r_udu\right), \quad s<t,
\end{equation}
with the interest rate $r_u$ at time $u$.

The difference between $\text{EE}^*(t)$ and $\text{EE}(t)$ is that $\text{EE}(t)$ denotes the future exposure, while $\text{EE}^*(t)$ indicates today's value of the future exposure.
When interest rate is deterministic, we can write $\text{EE}^*(t)=D(0,t)\text{EE}(t)$. Here, however, we will deal with stochastic interest rates under the Hull-White model when focusing on the HHW dynamics. In that case $\text{EE}^*(t)$ in a natural quantity, as we will see.

When exposure values and default events are independent, the expression for CVA is given by\cite{JonGregoryCVA}
\begin{equation}\label{eq:CVAconti}
\text{CVA}=(1-\delta)\int_0^T \text{EE}^*(t)d \text{PD}(t),
\end{equation}
where $\delta$ is the recovery rate, generally considered as a constant; $\text{PD}(s)$ is the default probability function.
The default probability function is modeled as follows here:
\begin{align}\label{eq:default}
 \text{PD}(t)=1-\exp\left(-\int_0^t h(t)dt\right),
\end{align}
where $h(t)$ is the hazard rate(intensity), which we will set constant, $h=0.03$, in the numerical results section.

The market value of the contract is observed at a set of discrete times, 
  $\mathcal{T}=\left\{t_m \big |m=0,1,\dots,M\right\}$,
with $t_0=0, t_M=T$, $T$ being the tenor of the contract. The notation of the variables at time $t_m$ is simplified by $\textbf{X}_m:=\textbf{X}_{t_m}$.
A discrete version of the CVA formula in (\ref{eq:CVAconti}) can be written as 	
\begin{equation}
\label{eq:CVAdiscrete}
\text{CVA} \approx (1-\delta)\sum_{m=0}^{M-1}\text{EE}^*(t_m)\left(\text{PD}(t_{m+1})-\text{PD}(t_{m})\right).
\end{equation}
In (\ref{eq:CVAdiscrete}), the key elements to calculate CVA are the values of the $\text{EE}^*(t)$ function and the default probability at the discrete observation dates. 
We are also interested in values of the PFE function and the corresponding Greeks. 
Hence it becomes crucial to validate exposure of each realized scenarios at all simulated dates. 
\subsection{Backward pricing dynamics of  options}\label{sec:exposurefunction}
In this section, we present the backward pricing dynamics framework for financial derivatives. 
When an option is exercised, the holder receives the payoff value, i.e.
\begin{align}
 &g(S_m):=\max\left(\omega(S_m-K),0\right), \quad \text{with}
 \begin{cases}
 \omega=1, & \text{for a call};\\
 \omega=-1, & \text{for a put},
 \end{cases}
\end{align}
where $S_m=\exp(x_m)$ is the underlying asset variable at time $t_m$.

When the option contract is still alive, the discounted option value, the {\em continuation value} w.r.t. state vector $\textbf{X}_m$, reads
\begin{equation}\label{eq:thecontinuation}
c_m( \textbf{X}_m):=\mathbb E^{\mathbb{Q}} \left[V_{m+1}(\textbf{X}_{m+1})\cdot D(t_m,t_{m+1}) \bigg |\textbf{X}_m\right], 
\end{equation}
where $V_{m+1}(\cdot)$ represents the option value at time $t_{m+1}$. At time $t_M=T$,  $c_M:=0$.
A filtration satisfies $\mathcal{F}_0 \subset \cdots \mathcal{F}_m \subset \cdots \mathcal{F}_M$, 
which describes the information flow. 

We establish the backward pricing dynamics framework based on the \emph{Snell Envelope} concept \cite{D.Lamberton}. 
The owner of the American option makes the exercise decision for the option 
based on the information at time $t_m$. 

For European options, the option value equals the continuation value before  maturity and the holder receives the payoff value only at time $t_M$, i.e. 
\begin{equation}\label{eq:euro}
V^{\text{Euro}}_m(\textbf{X}_m)=
\begin{cases}
g(S_{M}), & \text{for } t_M,\\
c_m(\textbf{X}_m), & \text{for } t_m\in \mathcal{T}-t_M.\\
\end{cases}
\end{equation}

For Bermudan options, the problem becomes more involved as the option holder has the right to exercise the option at a series of times before maturity. 
We  make the assumption here that the credit information of the counterparty does not influence the exercise decision of the option holder:
the holder makes the exercise decision based on the maximum profit he/she can gain. 
So, at each exercise date
the holder compares the payoff value with the continuation value of the option, conditioned on the current market information. 
Once the payoff value is higher, the option will be exercised; otherwise the holder 
will hold the option. 

Suppose that the option holder is allowed to exercise the option at the exercise dates, $\mathcal{T}_e=\left\{t_e \big|t_e \in \mathcal{T},e>0\right\}$.
When the option is still active at time $t_m$, the Bermudan option can be computed via
\begin{equation}\label{eq:bermdan}
V^{\text{Berm}}_m(\textbf{X}_m)=
\begin{cases}
\max\left\{c_m(\textbf{X}_m),g(S_{m})\right\}, & \text{for } t_m\in \mathcal{T}_e,\\
c_m(\textbf{X}_m), & \text{for } t_m\in \mathcal{T}-\mathcal{T}_e.\\
\end{cases}
\end{equation}

We construct the pricing dynamics for barrier options in a similar way. 
Barrier options become active/knocked out when the underling reaches a predetermined level (the barrier). There are four main types of barrier options: up-and-out, down-and-out,
up-and-in, down-and-in options. Here we focus on down-and-out put barrier options. 

A \textit{down-and-out} put barrier option is active initially and gets value zero when the underlying reaches the barrier; if the option is not knocked out during its lifetime, the holder will receive the payoff value at the end of the contract. At time $t_m$, when the option is still active, the pricing dynamics are given by,
\begin{equation}\label{eq:barrier}
V^{\text{barr}}_m(\textbf{X}_m)=
\begin{cases}
 g(S_m)\cdot\emph{1}_{S_m\leq L}, & \text{for } t_M,\\
 c_m(\textbf{X}_m)\cdot\emph{1}_{S_m\leq L}, & \text{for } t_m\in \mathcal{T}-t_M,\\
\end{cases}
\end{equation}
where $\emph{1}\left(\cdot\right)$ is the indicator function. 

The value of the exposure is defined as the value of the option when the option is still active. When the option is exercised/knocked out, the exposure value 
becomes $0$. The pricing dynamics of the exposure can be put in the following formulas:
\begin{equation}\label{eq:exp}
\text{E}_m( \textbf{X}_m)=
\begin{cases}
0, & \text{when the option is knocked out};\\
V_m(\textbf{X}_m), & \text{when the option is alive};\\ 
\end{cases}
\end{equation}
where $\text{E}_m(\cdot)$ represents the exposure at time $t_m$, $m=1,2,\cdots, M-1$. We define $\text{E}_M=0$.

Once options have been exercised/knocked out at time $t_m$, the exposure later than time $t_m$ becomes zero. 

\section{Stochastic Grid Bundling Method}\label{secSGBM}
We present the Stochastic Grid Bundling Method (SGBM). SGBM is based on \textit{simulation}, \textit{bundling} and \textit{regression}.
A stochastic grid is defined via generation (simulation) of a large number of stochastic paths, where we can determine the empirical distribution of the state variable at each future time point. 
Denote the values of the state variables of the $i$-th path at time $t_m$ as $\hat{\textbf{x}}_m(i)$, $i=1,\dots, N$, and once we computed the 
exposure values of these scenarios at times $t_m$, $m=0, \dots, M-1$ the values of the EE function can be approximated by 
\begin{equation}\label{eq:EE1}
 \text{EE}(t_m)\approx \frac{1}{N}\sum_{i=1}^N \text{E}_m( \hat{\textbf{x}}_m(i)),
\end{equation}
where $N$ represents the number of paths. The values of the PFE function can also be approximated by the corresponding quantiles based on the realized exposure values
of the generated scenarios. 

The exposure profiles presented in formula (\ref{eq:EE1}) are values without discount factor. However, for pricing CVA, it is the discounted exposure value that is needed. 
When the stochastic interest rate is deterministic, the discounted exposure is the product of the discount factor and EE. When the interest rate 
is stochastic, we need to discount the realized values of the generated scenarios as follows:
\begin{align}
 \text{EE}^*(t_m)\approx \frac{1}{N} \sum_{i=1}^N \left( \exp\left(\sum _{k=0}^{m-1} r_k(i) \left(t_{k+1}-t_k\right)\right)\cdot E_m(\hat{\textbf{x}}_m(i))\right),
\end{align}
where $r_k(i)$ is the realized interest value at time $t_k$ in the $i$-th path.

The option values at the paths at each time point are needed for the calculation of the exposure distribution,
and thus the continuation values, defined in (\ref{eq:thecontinuation}), of all paths at each time point need to be calculated. 
In the following sections, we discuss how to approximate the continuation values in SGBM. 	

\subsection{Least-squares approximation}\label{polynomialmeasure}
Option function $V_{m+1}(\cdot)$ at time $t_{m+1}$  is $L_2$-measurable on a bounded domain $\textbf{I}_{m+1}$ 
, and we approximate it by a simpler function $\hat V_{m+1}(\cdot)$, i.e.
\begin{equation}
  \hat V_{m+1}(\textbf{X}_{m+1})\approx   V_{m+1}(\textbf{X}_{m+1}),\quad \textbf{X}_{m+1} \in \textbf{I}_{m+1}.
\end{equation}

The approximation is made by projecting the option onto a polynomial space, where the values are linear combinations of $H$ basis functions, defined by 
\begin{equation}
 \mathcal{P}_p(\textbf{I}_{m+1})=\{f|f(x)=\sum_{k=1}^H \beta(k) \psi_k(x), x \in \textbf{I}_{m+1}, \forall k, \beta(k) \in\mathbb{R} \},
\end{equation}
where $p$ is the order of the polynomial subspace, and $H$ represents the number of basis functions. Thus, the approximated option value can be written as
\begin{equation}
  \hat V_{m+1}(\textbf{X}_{m+1})= \sum_{k=0}^{H-1} \beta(k) \psi_k(\textbf{X}_{m+1}), \quad \textbf{X}_{m+1} \in \textbf{I}_{m+1},
\end{equation}
where the coefficient set $\{\beta(k)\}_{k=0}^{H-1}$ can be determined in least-squares sense by regression when option values at time $t_{m+1}$ are available, 
\begin{equation}\label{eq:leastsqureapproximation}
\min\limits_{\forall k, \beta(k)\in \mathbb{R}} \sum_{i=1}^N
\left(V_{m+1}(\hat{\textbf{x}}_{m+1}(i))-\sum_{k=0}^{H-1} \beta(k) \psi_k(\hat{\textbf{x}}_{m+1}(i))\right)^2. 
\end{equation}

Essentially, formula (\ref{eq:leastsqureapproximation}) gives the \textit{best} approximation on polynomial space $\mathcal{P}_p(\textbf{I})$ of the option function in the $L_2$ norm, which is called the {\em projection} on polynomial space $\mathcal{P}_p(\textbf{I})$. 
Notice that  approximating a function by interpolation and approximating a function in least-squares sense are two different concepts.
By interpolation, the values of the approximation function are exact at discrete nodes,  while when approximating in least-squares sense, 
we study the approximation in average.
In the latter case, we find the projection of the function $f$ onto some subspace $\Omega$, denoted by $P_{\Omega}$, such that the difference
$f-P_{\Omega}f$ is orthogonal to all functions in subspace $\Omega$ \cite{Matlarson1}. 
The $L_2$ projection does not need to be continuous, or have well-defined nodal values, which
is convenient in our case, as the grid nodes have been generated via simulation. 

When we approximate the option function by a linear combination of basis functions, the continuation function at the earlier time point can be approximated
by a linear combination of conditional expectations of the discounted basis functions, i.e.
\begin{align}\label{eq:thecon}
c_m\left(\textbf{X}_m\right)\approx\hat c_m\left(\textbf{X}_m\right)&=\mathbb{E}^{\mathbb{Q}}\left[\hat V_{m+1}(\textbf{X}_{m+1})\cdot D(t_m,t_{m+1})\bigg|\textbf{X}_m\right]\notag \\
&= \mathbb{E}^{\mathbb{Q}}\left[\sum_{k=0}^{H-1} \beta(k) \psi_k(\textbf{X}_{m+1})\cdot D(t_m,t_{m+1})\bigg|\textbf{X}_m\right] \notag \\
&= \sum_{k=0}^{H-1} \beta(k)\mathbb{E}^{\mathbb{Q}}\left[ \psi_k(\textbf{X}_{m+1})\cdot D(t_m,t_{m+1})\bigg|\textbf{X}_m\right] \notag \\
&= \sum_{k=0}^{H-1} \beta(k) \phi_k(\textbf{X}_m),
\end{align}
where we denote the conditional expectations of the discounted basis functions  by
\begin{equation}\label{eq:discountedbasis}
\phi_k(\textbf{X}_m):=\mathbb{E}^{\mathbb{Q}}\left[ \psi_k(\textbf{X}_{m+1})\cdot D(t_m,t_{m+1})\bigg|\textbf{X}_m\right] ,\quad k=0,\dots,H.
\end{equation}

In fact, it is easy to see that the span of the series $\{\phi_k\}_{k=0}^k$ forms a closed subspace of functions in $L_2$ space, denoted as:
\begin{equation}
\mathcal{E}\mathcal{P}(\textbf{I}_m)=\{Ef| Ef(x)=\sum_{k=1}^H \beta(k) \phi_k(x), x \in \textbf{I}_m, \beta(k) \in\mathbb{R},\forall k\}.
\end{equation}

In other words, we approximate the continuation function on space $\mathcal{E}\mathcal{P}(\textbf{I}_m)$, and  the coefficients 
are obtained via approximation of the option function at a later time point $t_{m+1}$. 

When the basis functions are chosen such that analytic formulas of the corresponding conditional expectations of the discounted basis functions
are available, the continuation value can be calculated immediately, and the option value at the same point can be calculated  applying formulas (\ref{eq:euro}), (\ref{eq:bermdan}) and  (\ref{eq:barrier})
for European, Bermudan or barrier options, respectively.

Though there are many possibilities for choosing the set of basis functions for a polynomial space of order $p$, we choose \textit{monomials} of degree  equal or lower than 
order $p$ as the basis functions. 
A \textit{monomial} is a polynomial with only one term, which can be defined as a product of powers of variables with non-negative integer exponents. 
The degree of a monomial is defined as the sum of all exponents of the variables. Considering an $n$-dimensional problem with a polynomial space of order $p$,
it is easy to see that the set of monomials with degree less or equal to
order $p$ is a span of polynomials\footnote{The number of monomials of $n$ variable with order $d$ is $\frac{(d+n-1)!}{(n-1)!d!}$.}, and the total number of these basis functions
is equal to $\frac{1}{(n-1)!}\sum_{d=0}^p\frac{(d+n-1)!}{d!}$.

It is convenient to construct the polynomial subspace with monomials. Moreover, we then have analytic formulas for the expectations of the discounted monomials: when we use monomials as the basis functions, the conditional expectations of the discounted basis functions
are \textit{discounted moments}. For 2-d dynamics these moment are given in definition \ref{defn:moments}.

\begin{defn}[Discounted moments]\label{defn:moments}
In a time period $[t,T]$, $t<T$, conditioned on the information at time $t$, the discounted moments of a state variable $\textbf{Y}_t=[y_t,z_t]^T$  of order $p+q$ 
is defined as $\mathbb{E}^{\mathbb{Q}}\left[\left(y_T\right)^{p} \cdot \left(z_T\right)^{q} \cdot D(t,T)\big| \textbf{Y}_t\right]$. 
\end{defn}

There is a useful link between the discounted moments and the discounted characteristic function (dChF)  of the dynamics.
By derivatives w.r.t $y_{T}$ and $z_{T}$, respectively, we find
\begin{align}
 \mathbb{E}^{\mathbb{Q}}\left[\left(y_{t}\right)^{p} \cdot \left(z_{t}\right)^{q} \cdot D(t,T)\big| \textbf{Y}_t\right] 
 =\frac{1}{(i)^{p+q}}\cdot \frac{\partial ^{p}\Phi}{\partial u_1^{p}} \cdot\frac{\partial ^{q}\Phi}{\partial u_2^{q}}(u_1,u_2;\textbf{Y}_t, t,T) \bigg |_{u_1=0,u_2=0},
\end{align}
where $\Phi(;)$ is the dChF. The higher-dimensional case can be defined in a similar way. 

For the constant basis function (of degree $0$), the discounted first moment equals the zero coupon bond in interval $[t,T]$:
\begin{equation}
 \phi_0(\textbf{X}_t)= \phi_0:=\mathbb{E}^{\mathbb{Q}}\left[D(t,T)|\textbf{X}_t\right]= \mathbb{E}^{\mathbb{Q}}\left[e^{-\int_{t}^{T}r_udu}\big |\textbf{X}_t\right]=: P(t,T).
\end{equation}

Once analytic formulas of the discounted moments are derived, multiplication of the coefficient sets determined at time $t_{m+1}$ gives us
the formula for approximating the continuation value at time $t_m$. 

\subsection{Greeks}
The state variable $\textbf{X}_m$ must at least contain the underlying asset information, and we always put the log-asset variable as the first element in the vector that 
$\textbf{X}_m=[x_m,\dots]^T$, where $x_m:=\log(S_m)$. Here we present the sensitivity w.r.t the initial asset value $S_0$,
which can be applied for all models discussed in this paper. 

It is direct to approximate the sensitivities of the exposure profile in SGBM. At time $t_m$, the sensitivity \textit{Delta} ($\Delta $) of the EE
w.r.t the change in the underlying asset price $S_0$ can be derived by 
\begin{align}
\Delta_{\text{EE}}(t_m)&=\frac{\partial \text{EE}(t_m)}{\partial S_0}
                     \approx \frac{1}{N}\sum_{i=1}^N\frac{\partial \text{E}_m }{\partial S_0}\left(\hat{\textbf{x}}_m(i)\right)  \notag \\
                     &=\frac{1}{N}\sum_{i=1}^N\frac{\partial \text{E}_m }{\partial x_m}\cdot \frac{\partial x_m}{\partial S_m}\cdot \frac{\partial S_m}
                     {\partial S_0} \left(\hat{\textbf{x}}_m(i)\right), \quad {m=0,\dots,M-1},
\end{align}
where $\hat{\textbf{x}}_m(i)=[\log(S_m(i)),\dots]^T$ is the $i$-th realization at time $t_m$ of the state variable;
applying the chain rule and calculating the partial derivative, we get\footnote{
Since 
\begin{align}
 S_m=S_0\exp{\left(\left(r_m-\frac{1}{2}\sigma_m^2\right)t_m+\sigma_m W^x_{t_m}\right)},
\end{align}
it is easy to derive that:
\begin{align}
\frac{\partial S_m}{\partial S_0}=\exp{\left(\left(r_m-\frac{1}{2}\sigma_m^2\right)t_m+\sigma_m W^x_{t_m}\right)}=\frac{S_m}{S_0}, 
\end{align}
where $\sigma_m=\sqrt{v_m}$ and $r_m$, are the volatility and interest rate at time $t_m$ respectively. 
Here the variance variable and interest rate can be either a constant or a stochastic value. 
}
\begin{equation}
\frac{\partial x_m}{\partial S_m}=\frac{1}{S_m}, \quad \frac{\partial S_m}{\partial S_0}=\frac{S_m}{S_0}, \quad {m=0,\dots,M-1}. 
\end{equation}

The first-derivative of the exposure profile is defined as:
\begin{align}\label{eq:derivativeE}
\frac{\partial \text{E}_m }{\partial x_m}:=
\begin{cases}
0 & \text{when the option is exercised},\\
{\displaystyle \frac{\partial c_m}{\partial x_m}} & \text{when the option is active},\\ 
\end{cases}\quad {m=0,\dots,M-1},
\end{align}
with formula (\ref{eq:thecon}), the first derivative of the continuation function w.r.t $X_m$ is approximated by
\begin{equation}
\frac{\partial c_m}{\partial X_m} \approx \frac{\partial \hat c_m}{\partial X_m}=\sum_{k=0}^{H} \beta(k) 
\frac{\partial \phi_k}{\partial X_m},  \quad {m=0,\dots,M-1},
\end{equation}
where the coefficient set is the same as in (\ref{eq:thecon}). Hence
\begin{align}
\Delta_{\text{EE}}(t_m)
\approx \frac{1}{N}\sum_{i=1}^N\frac{\partial \text{E}_m }{\partial x_m}\left(\hat{\textbf{x}}_m(i)\right) \cdot \frac{1}{S_0}, \quad {m=0,\dots,M-1},
\end{align}
Further, we can get an easy formula for Gamma ($\Gamma$), as
\begin{align}
\Gamma_{\text{EE}}(t_m)
\approx \frac{1}{N}\sum_{i=1}^N\left(\frac{\partial^2 \text{E}_m }{\partial X_m^2}\left(\hat{\textbf{x}}_m(i)\right)-\frac{\partial \text{E}_m }{\partial x_m}
\left(\hat{\textbf{x}}_m(i)\right)\right) \cdot \frac{1}{S_0^2}, \quad {m=0,\dots,M-1},
\end{align}
\textbf{Comment}: In a similar way, we are able to calculate Rho($\rho$), which is defined as the sensitivity to the interest rate $r_0$, and Vega($\nu$), which 
is the sensitive to the volatility $v_0$. However, it is a somewhat involved to calculate Vega($\nu$) along the whole time horizon as it is nontrivial to calculate
$\frac{\partial v_m}{\partial v_0}$. 

\subsection{Convergence and choice of bundles}
There are interesting studies on the convergence and bias of Monte Carlo based regression methods.  In \cite{Celement}, Clement, Lamberton and Protter prove
the almost surely convergence of the Longstaff-Schwartz algorithm and determine the rate of the convergence. 
Other available studies show that there is an upward-bias in the direct estimates, and the result based on the obtained \textit{optimal} exercise-strategy 
is a lower bound estimate of the option price \cite{Longstaff-Schwartz, Tsitsiklis, Stentoft}. In \cite{Broadie}, Broadie and Cao propose to apply a local simulation to improve
the lower and upper bound algorithms  and reduce the variance. 

When approximating the option value, there are two types of errors appearing \cite{Celement}:
\begin{itemize}
 \item Type 1 : the error source is the difference between the true and the approximate option values, as 
 the value is  approximated by its projection onto a polynomial space.  This error is related to the properties of the polynomial space;
 \item Type 2:  the error source is the noise in estimating the coefficients via regression based on sampled data, which 
 related to the accuracy of the Monte Carlo simulation and thus to the number of paths.
\end{itemize}
We need to reduce the errors of Type 1 and Type 2. 
According to Central Limit Theorem, the error of Type 2 can be diminished by increasing the number of paths to infinity with probability $1$ a.s. \cite{Celement}.
Some available results show that the error of Type 1 goes to zero when the number of basis functions goes to infinity as well \cite{shashijain}. 
In this paper, we will present an accurate upper bound for the error of Type 1 related to the properties of the polynomial space, and give a reasoning for the use of bundles.

In \cite{Celement}, the discussion is on the error of Type 2. Although it is focused on the Longstaff-Schwartz method, the analysis can be applied to our method as well.
As in \cite{Celement}, we add the error of Type 2 as a noise term: 
\begin{align}
 \hat V (\textbf{X}_{m+1})&=\widetilde V (\textbf{X}_{m+1})+\epsilon_{m+1}, \\
 \widetilde V (\textbf{X}_{m+1})&= \sum_{k=0}^{H} \tilde \beta(k) \psi_k(\textbf{X}_{m+1}),
\end{align}
where when the number of paths $N$ is sufficiently large, $\epsilon_{m+1} \sim \mathcal{N}(0, \frac{\sigma_{m+1} ^2}{N})$ i.i.d, and independent of variable $\textbf{X}_{m+1}$ ;
$\hat V (\cdot)$ is the 'true' value of the option on the polynomial space, whereas
$\widetilde V (\cdot)$ is the approximation of the option with the coefficients set $\{\tilde \beta(k)\}_{k=1}^H$. 

We futher define the approximation of the continuation function as:
\begin{align}
 \tilde c_m(\textbf{X}_{m}):=\mathbb{E}^{\mathbb{Q}}\left[\widetilde V_{m+1}(\textbf{X}_{m+1})\cdot D(t_m,t_{m+1})\bigg|\textbf{X}_m\right],
\end{align}
hence
\begin{align}
 c_m(\textbf{X}_{m})\approx \hat c_m(\textbf{X}_{m})= \tilde c_m(\textbf{X}_{m})+ \epsilon_{m+1}, 
\end{align}

We first give a proposition regarding the boundedness of the error of the approximating continuation value by the error of the approximating option value in the $L_2$ norm. 
\begin{proposition}\label{prop:estimation2}
Assuming that $\forall \textbf{X}_m \in \textbf{I}_m$, the conditional density function satisfies 
\begin{equation}
 \int_{\textbf{I}_{m+1}} f(\textbf{X}_{m+1};\textbf{X}_m) d\textbf{X}_{m+1}=1-\varepsilon, \quad \textbf{X}_{m+1} \in \textbf{I}_{m+1},
\end{equation}
where $f(\cdot;\textbf{X}_m)$ is the density function conditioned on $\textbf{X}_m$, $\epsilon$ is a small error generated by the truncation of the integration range.
The error of approximating the continuation function at time $t_m$ can be bounded by the error of approximating the option function at time $t_{m+1}$ as:
 \begin{equation}
\|c_{m}-\tilde c_{m}\|_{L_2(\textbf{I}_m)} \leq \|V_{m+1}-\widetilde V_{m+1}\|_{L_2(\textbf{I}_{m+1})} \cdot \sqrt{(1-\varepsilon)h(\textbf{I}_m)},
 \end{equation}
 where $h(\textbf{I}_m)$ is the size of the bounded domain $\textbf{I}_m$.
\end{proposition}

Proof is given in Appendix~\ref{app:proof1}. 

Proposition \ref{prop:estimation2} ensures that, when the approximation of the option function is accurate, the approximation of 
the continuation function at an earlier stage will be accurate as well. The boundedness in Proposition \ref{prop:estimation2} includes 
errors of both Type 1 and Type 2. 

For Error of Type 1, we give two well-known insights, from \cite{Matlarson1} and \cite{Matlarson2} , which gives a bound of the error by projection onto a linear polynomial space
in a 1-d and 2-d domain, respectively. 

\begin{thm}\label{theorem:estimation1}
In a one-dimensional domain $I$, suppose that the function $f(x)$ is twice differentiable, and $\mathcal{P}_1(I)$ is the linear polynomial space. 
The projection onto the space $\mathcal{P}_1(I)$ in interval $I$ in the $L_2$ norm satisfies the best approximation result,
 \begin{equation}
 \min_{\hat f \in \mathcal{P}_1(\textbf{I})}\|f-\hat f\|_{L_2(\textbf{I})}\leq C h^{2}\|f^{(2)}\|_{L_2(\textbf{I})},
 \end{equation}
 where $h$ is the length of the interval $I$ and $f^{(2)}$ is the second derivative function.
\end{thm}

\begin{thm}\label{theorem:estimation2}
In a two-dimensional domain $\textbf{I}$, suppose that the function $f(x_1,x_2)$ is twice differentiable, and $\mathcal{P}_1(\textbf{I})$ is the linear polynomial space. 
The projection onto the space $\mathcal{P}_1(\textbf{I})$ in interval $\textbf{I}$ in the $L_2$ norm satisfies the best approximation result
 \begin{equation}
 \min_{\hat f \in \mathcal{P}_1(\textbf{I})}\|f-\hat f\|_{L_2(\textbf{I})}\leq C h^{2}\|D^2f\|_{L_2(\textbf{I})},
 \end{equation}
 where $h$ is a representative size of the mesh and  $D^2$ is a 2-d differential operator with mixed derivatives.

\end{thm}
The details of the  construction of the mesh can be found in \cite {Matlarson2}.

Theorems \ref{theorem:estimation1} and \ref{theorem:estimation2} provide upper bounds of the error by approximating 
a function by its projection on a linear polynomial space. Though we do not know much about the properties of the function, 
the error can be reduced by approximating the function on some
discrete subdomains $\textbf{I}^1$, $\dots$, $\textbf{I}^J$, such that $ \textbf{I}=\cup_{i=1}^J\textbf{I}^i$, the projection of 
the function in each subdomain $\textbf{I}^i$ is written as 
\begin{align}
 f(x) \approx \hat f_i(x), \quad x \in  \textbf{I}^i, \quad, i=1,\dots,J. 
\end{align}
In 1-d notation, the error can be bounded by,
\begin{equation}
\sum_{i=1}^J  \min_{\hat f_i\in \mathcal{P}_1\left(\textbf{I}^i\right)} 
\|f-\hat f_i\|_{L_2(\textbf{I}^i)}
\leq C \sum_{i=1}^J h_i^2\|f_i^{(2)}\|_{L_2(\textbf{I}^i)}
\leq C \left(\max_{i=1}^J h_i\right)^2\|f^{(2)}\|_{L_2(\textbf{I})},
 \end{equation}
where $\textbf{I}^i$ is the size of $\textbf{I}^i$. 
 
It tells us that the error can be reduced efficiently by approximating the function in subdomains.
Theorems \ref{theorem:estimation1} and \ref{theorem:estimation2} give upper bounds for projection on a linear polynomial space ($p=1$). 
We generalize the results to  a polynomial space of order $p$, in Proposition \ref{prop:estimation}. 
\begin{proposition}\label{prop:estimation}
In a 1-d interval $\textbf{I}$,
suppose that the function $f(x)$ is $(p+1)$ times differentiable, 
and $\mathcal{P}_p(\textbf{I})$ is a polynomial of order $p$ on $\textbf{I}$. 
The projection on the space $\mathcal{P}_p(\textbf{I})$ on interval $\textbf{I}$ in the $L_2$ norm satisfies the best approximation result as follows
 \begin{equation}
 \min_{\hat f \in \mathcal{P}_p(\textbf{I})}\|f-\hat f\|_{L_2(\textbf{I})}\leq C h^{p+1}\|f^{(p+1)}\|_{L_2(\textbf{I})},
 \end{equation}
 where $h$ is the length of interval $I$.
\end{proposition}
Proof is given in Appendix~\ref{app:proof2}. 

Proposition \ref{prop:estimation} provides a rough estimate of the accuracy of the approximation function  
in the polynomial subspace $\mathcal{P}_p(\textbf{I})$ for a 1-d problem. 
The accuracy depends on the length of the interval,
the polynomial order and the value of the $(p+1)$-th derivative. 

In practice, the determination of subdomains is done in SGBM by {\em making bundles} from the simulated paths.  We define the subdomains at time $t_{m+1}$  by clustering paths  into bundles at time $t_m$. 
 One aim  is to satisfy the assumptions in Proposition  \ref{prop:estimation2}. The paths at time $t_m$ are clustered based on criteria, so that the values of paths in the same bundle fall into the same subdomain at time $t_{m+1}$.  
When all subdomains are of the same size, the 1-d accuracy can be approximated by 

\begin{equation}
 \sum_{i=1}^J  \min_{\hat V_{m+1}^{j} \in \mathcal{P}_p\left(\textbf{I}^i_{m+1}\right)} 
\|V_{m+1}-\hat V_{m+1}^{j}\|_{L_2(\textbf{I}^i_{m+1})} \sim \frac{C_2}{J^{(p+1)}}\|f^{(p+1)}\|_{L_2(\textbf{I}_{m+1})},
\end{equation}
where $J$ is the number of bundles. 

$\hat V_{m+1}^{j} $ is the exact projection of the function on the polynomial space within the $j$-th bundle. 
In practice, we estimate it by regression, denoted by $\widetilde V_{m+1}^{j} $, based on cross-sectional data, where
the error of Type 2 plays its role. We need to ensure that there are enough paths within each bundle, so that the error of Type 2 is sufficiently small, 
otherwise it is meaningless to analyze the error of Type 1. 

We propose a \textit{principle for defining the bundles}: \textit{cohesive} in value and \textit{equal} in number. In short, within each bundle,
the state variable path values should be as close as possible, and the paths should be distributed into each bundle in a way such 
that there are 'sufficient' paths within each bundle. In section \ref{sec:bundles}, methods of making bundles  are presented.

\subsection{Backward iteration for exposure values}\label{subsec:calofcontinuation}

We present the procedure for calculating the exposure values in a backward iteration, starting at final time $T$. Suppose the path generation has finished and the realized values of the state variable of $N$ paths at all observation dates $\{t_m\}_{m=0}^M$ are available,
denoted by $\{\hat{\textbf{x}}_m(i), m=0,\dots,M, i=1,\dots,N\}$. At time $t_M$,  the exposure values are set to $0$ as the contract ended. The option value of the $i$-th path is calculated as $ V_{M}(\hat{\textbf{x}}_{M}(i))$, for European, Bermudan or barrier options, applying 
formulas (\ref{eq:euro}), (\ref{eq:bermdan}) and (\ref{eq:barrier}), respectively. 

At time $t_{M-1}$, the paths are clustered into $J$ bundles applying some bundling method, and we denote the $j$-th bundle at time $t_{M-1}$ by $\mathcal{B}_{M-1,j}$. 
On a polynomial space of order $p$, where there are $H$ basis functions in the span, we can find the 'best coefficients set' via OLS regression within the same bundle by minimizing the following expression,

\begin{equation}
\min\limits_{\forall \beta(k,\mathcal{B}_{M-1,j}) \in \mathbb{R}} \sum_{\text{paths in bundle } \mathcal{B}_{M-1,j}} 
\left(V_{M}(\hat{\textbf{x}}_{M}(i))-\sum_{k=0}^{H} \beta(k,\mathcal{B}_{M-1,j}) \psi_k(\hat{\textbf{x}}_{M}(i))\right)^2, 
\end{equation}
where  the $k$-th coefficient at time $t_{M-1}$ in the $j$-th bundle is denoted by $\beta(k,\mathcal{B}_{M-1,j})$, and thus at time $t_M$, within the $j$-th bundle,
the option function can be approximated  by 
\begin{equation}
  V_{M}(\hat{\textbf{X}}_{M})\approx \sum_{k=0}^{H-1} \beta(k,\mathcal{B}_{M-1,j}) \psi_k(\hat{\textbf{X}}_{M}), \quad \text{for values of paths in bundle } 
  \mathcal{B}_{M-1,j},
\end{equation}
With (\ref{eq:thecon}), the analytic formula of the approximation of the continuation function at time $t_{M-1}$, 
\begin{equation}
  c_{M-1}(\hat{\textbf{X}}_{M-1})\approx \sum_{k=0}^{H-1} \beta(k,\mathcal{B}_{M-1,j}) \phi_k(\hat{\textbf{X}}_{M-1}), \quad \text{for values of paths in bundle } 
  \mathcal{B}_{M-1,j}\text{ at } t_{M-1},
\end{equation}
where $\phi_k(\cdot)$ defined in (\ref{eq:discountedbasis}) represents the corresponding discounted moments. 

Hence for all paths, the continuation values  $c_{M-1}(\hat{\textbf{x}}_{M-1}(i))$, $i=1,\dots,N$, can be calculated directly;
the option values at the $i$-th path at time $M-1$,  $V_{M-1}(\hat{\textbf{x}}_{M}(i))$, and the exposure values, $E_{M-1}(\hat{\textbf{x}}_{M}(i))$, can be determined as well.

Then the iteration goes back with one time step to time $t_{M-2}$, repeating the procedures of bundling and regression for the calculation of the continuation, option and exposure values.
We progress in backward direction until we reach initial time $t_0$ where we have the option/exposure values of all paths along the whole time horizon. 
The calculation of the sensitivity values can be done at the same time in the iteration as it requires the same set of coefficients. 

The backward calculation procedure is basically the same for European, Bermudan or barrier options, except that the pricing formulas for option values differ:
we apply formulas (\ref{eq:euro}), (\ref{eq:bermdan}) and (\ref{eq:barrier}), respectively, for these options. 

When computing exposure profiles for Bermudan options, we also need to determine the \textit{optimal} early-exercise strategy.  
At time $t_m$, 
if the option has not been exercised, both option and exposure values at each path are set to the corresponding continuation value;
for each path, the payoff value is calculated, and compared with  the continuation value to determine
whether the option should be exercised, conditioned on the available information.  For a specified path, if the option should be exercised, the exposure at this path from time $t_{m}$ on
will be equal to zero, and the option value at time $t_{m}$ is equal to the payoff value. 

We store the "optimal`` exercise strategy for all paths and the corresponding cash flow, and estimate the corresponding EE value at a time point by taking the mean
of the discounted cash-flow as:
\begin{equation}
\label{eq:EE2dd}
 \text{EE}(t_m)\approx \frac{1}{N}\sum_{\tau_i>t_m}\left(D(t_m,{\tau_i})\cdot \text{cash-flow}(i)\right),
\end{equation}
where $\tau_i$ is the \textit{exercise time} (or \textit{stopping time}) of the $i$-th path, and the cash flow is the payoff value at time $\tau_i$, known as

\begin{align}
 \text{cash-flow}(i):=&g(S_{\tau_i}(i))=\max\left(\omega(S_{\tau_i}(i)-K),0\right), \quad \text{with}
 \begin{cases}
 \omega=1, & \text{for a call};\\
 \omega=-1, & \text{for a put},
 \end{cases}
\end{align}
with $S_{\tau_i}(i)$ the value of the stock of the $i$-th path at time $\tau_i$.  If the option is not exercised for some paths and expiration dates,
then the cash flow value is just zero. Notice that the discount factor needs to be incorporated in the realized values along each simulated path 
when dealing with stochastic interest rates. 

In practice, we call the EE values calculated via formula (\ref{eq:EE1}) the \textit{direct estimators}. 
and EE values calculated via formula (\ref{eq:EE2dd}) based another set of simulated scenarios the \textit{path estimators}. The detailed procedure 
calculating the path estimator is as follows: 
\begin{itemize}
 \item In the regression procedure of in the backward iteration, at each time point, the coefficient sets of each bundle are saved;
 \item We simulate another set of paths of twice the size;
 \item We calculate backward from time $t_{M-1}$ till time $t_0$:
 \begin{itemize}
\item  at time  $t_m$, make bundles, and use the same coefficient set of each bundle obtained in the first simulation 
 to determine the continuation values of the paths; 
 \item calculate the option value and determine the exercise strategy at each path at time $t_m$; save the exercise strategy of each path and the corresponding cash flow values;
 \end{itemize}
\item Applying formula (\ref{eq:EE2dd}) to calculate the EE values based on the obtained "optimal'' exercise strategy. 
\end{itemize}

\subsection{Comments}

The path estimator is commonly the lower bound  while the direct estimator represents the upper bound \cite{Longstaff-Schwartz,shashijain}, 
because of the convexity property of the ``max`` function. This is a useful result and we can test the convergence of SGBM
by comparing the difference between these upper and lower bounds. 

For Bermudan options, the estimation of the exposure heavily depends on the accuracy of the "optimal'' early-exercise strategy.
There is a discontinuity in the exposure function at the early-exercise point, thus
at the exercise date,
the exposure value jumps immediately to zero for a specific path when the option is exercised; if the calculated ``optimal`` early-exercise strategy
is inaccurate, the difference between the 'true' and the calculated exposure values would be as high as the option value at this time point. 
Pricing an option at each time point is much less sensitive to the accuracy of the ''optimal`` early-exercise strategy as the option function is smooth. 

\section{Basis functions and bundles}\label{sec:bundles}
As we see,
the number of basis functions is equal to $\frac{(n+p)!}{n!p!}$, where $n$ is the dimension and $p$
is the order of the polynomial space. We will apply bundling to enhance the accuracy. 
\subsection{Heston model}
We consider the 2-d state variable $\textbf{X}_t=[x_t,v_t]^T$, with $x_t=\log(S_t)$ the log-asset variable and $v_t$ represents the variance variable. 
The dynamics are given by the Heston stochastic volatility model
\begin{align}\label{eq:heston}
dv_t&=\kappa(\bar v-v_t)dt +\gamma \sqrt{v_t}dW^v_t, \notag \\
dx_t&=\left(r-\frac{1}{2}v_t\right)dt +\sqrt{v_t}dW^x_t,
\end{align}
where $r$ is the constant interest rate; $W^x_t$ and $W^v_t$ are Brownian motions, and $dW^x_t\cdot dW^r_t=\rho_{x,r}dt$; 
the constant parameters $\kappa$, $\bar v$ represent the speed and 
the mean level of the reverting in the volatility process, and $\gamma$  is called the vol-of-vol parameter.

The option function is a function of the log-asset price and variance. For orders $p=\{0,1,2,3\}$, the set of basis functions is presented in table \ref{tab:basisfunction}.
\begin{table}[ht!]
\centering
\begin{tabular}{ c|l }
\toprule
order $p$ of the polynomial space & the corresponding basis functions\\
\midrule
0 & $\{1\}$ \\
1& $\{1, x,v\}$ \\
2& $\{1, x, v, x^2, x\cdot v, v^2\}$\\
3& $\{1, x, v, x^2, x\cdot v, v^2, x^3,x^2\cdot v, x\cdot v^2, v^3\}$\\
\bottomrule
\end{tabular}
\caption{The basis functions and order $p$.}
\label{tab:basisfunction}
\end{table}

The Heston model belongs to the so-called class of affine asset dynamics, which means that analytic formulas for the discounted moments can be obtained, for example, via the dChF. The discounted moments
are given in Appendix~\ref{app:discountedmheston}.

In \cite{shashijain}, the \textit{recursive bifurcation method} was proposed for bundling asset paths when pricing  multiple assets under the Black-Scholes model.
A reduced space-recursive bifurcation method based on the values of the payoff was introduced to reduce the complexity. 
An important different for bundling under the Heston model is that here
the log-asset and variance processes are highly correlated when there is a strong correlation between the Brownian motions of these two processes.

With a large correlation parameter, applying the 2-d recursive bifurcation method directly results in the problematic fact that we may not have accurate results,
as the number of paths in some bundles is not large enough for accurate regression, not even with a large total number of paths.  

Hence we adapt the recursive-bifurcation method by adding a rotation step, and propose  another method for bundling as well. 
\subsubsection{Recursive-bifurcation-with-rotation}
The bundles are made 
based on the cross-sectional data at each time point, and we denote the two realized data sets at time $t_m$ as: $d_1=\{x_m(i)\}_{i=1}^N$, and $d_2=\{v_m(i)\}_{i=1}^N$.
The basic idea is that we project the correlated data sets $d_1$ and $d_2$ onto two almost independent data sets $q_1$ and $q_2$ by a rotation with an angle $\alpha_1$.

Step 1: The  values of trigonometric functions of the rotation angle $\alpha_1$ are defined as:
\begin{align}
 \cos \alpha_1= \sqrt{\frac{1}{1+k_1^2}}\text{sign}(k_1), \quad
 \sin \alpha_1=\sqrt{\frac{k_1^2}{1+k_1^2}},
\end{align}
where $k_1$ is the slope defined as\footnote{
Notice that the analytic formulas for 
\begin{align}
&\mathbb{E}\left[\left(x_m-\mathbb{E}[x_m]\right)\left(v_m-\mathbb{E}[v_m]\right)\right]=\mathbb{E}\left[x_mv_m\right]-\mathbb{E}[x_m]\mathbb{E}[v_m] ,\notag \\
&\mathbb{E}\left[\left(x_m-\mathbb{E}\left[x_m\right]\right)^2\right]=\mathbb{E}\left[x_m^2\right]-\left(\mathbb{E}[x_m]\right)^2 \notag 
\end{align}
can be obtained via moments, and thus can be derived with with ChF. We calculate $k_1$ with these analytic formulas. 
}
\begin{align}
k_1:=\frac{\mathbb{E}\left[\left(x_m-\mathbb{E}[x_m]\right)\left(v_m-\mathbb{E}[v_m]\right)\right]}{\mathbb{E}\left[\left(x_m-\mathbb{E}\left[x_m\right]\right)^2\right]}
\approx \frac{\text{covariance}(d_1,d_2)}{\text{variance}(d_1)}
\end{align}

The new two sets of data are calculated by 
\begin{equation}
 \begin{pmatrix}
  q_1 \\
  q_2
 \end{pmatrix}
 =\begin{pmatrix}
   \cos \alpha_1 & \sin \alpha_1 \\
   -\sin \alpha_1&\cos \alpha_1 
  \end{pmatrix} 
   \begin{pmatrix}
  d_1 \\
  d_2
 \end{pmatrix}
\end{equation}

Step 2: We apply the standard recursive bifurcation method on the rotated data $q_1$ and $q_2$ to make bundles. Along each dimension, 
the mean of the given set of data is computed; 
the paths are bundled separately along each dimension by dividing the data into $4$ sets with the computed values of mean,
and thus we have $4$ non-overlapping bundles after one iteration. We repeat the procedures with $j$ iterations and the total number of paths is $4^j$.

\begin{figure}[ht!]
 \begin{center}
\subfigure[Step 2: applying recursive bifurcation]{\includegraphics[width=6cm]{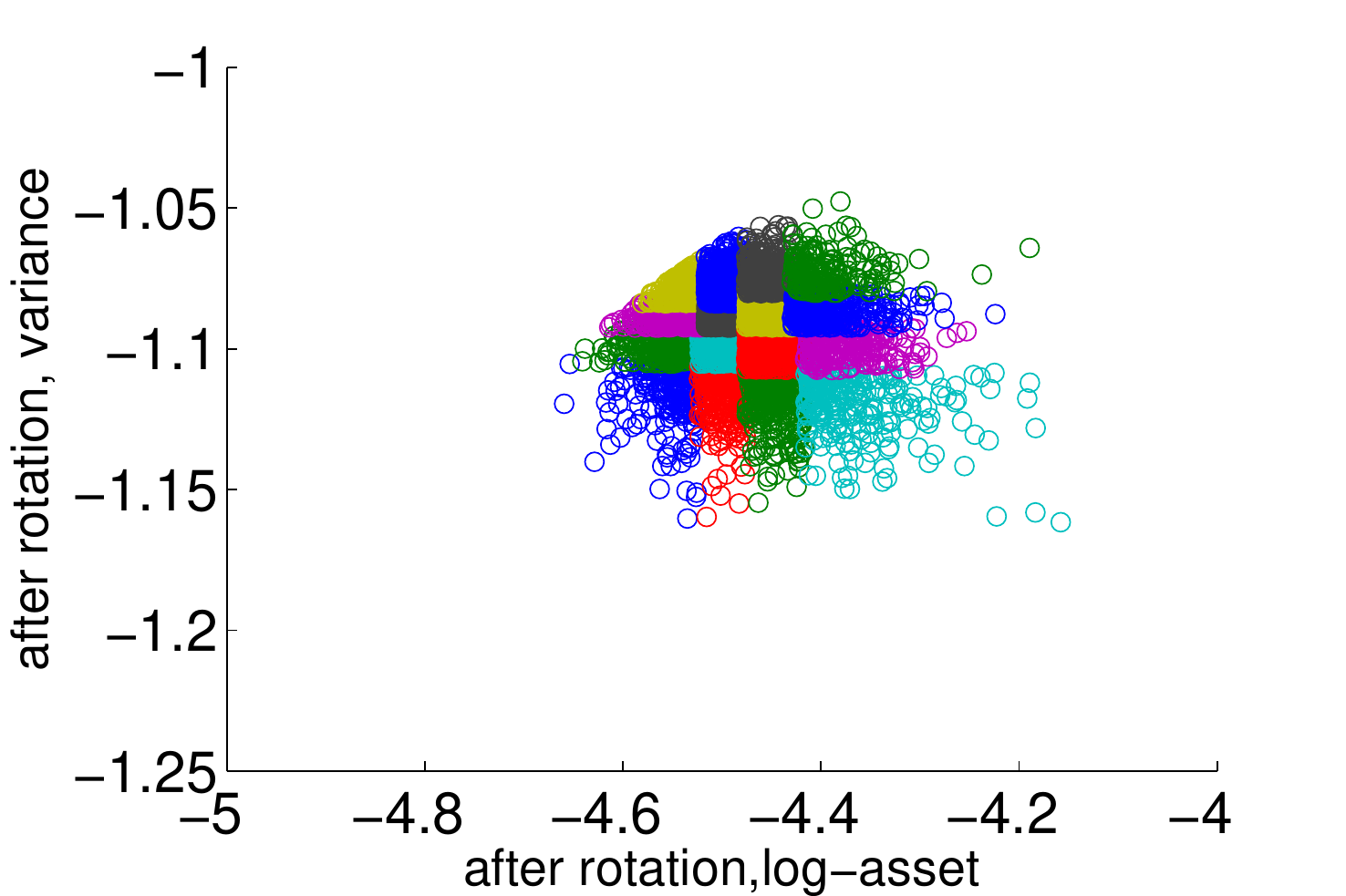}}
\hfill
\subfigure[Corresponding bundling on the original domain]{\includegraphics[width=6cm]{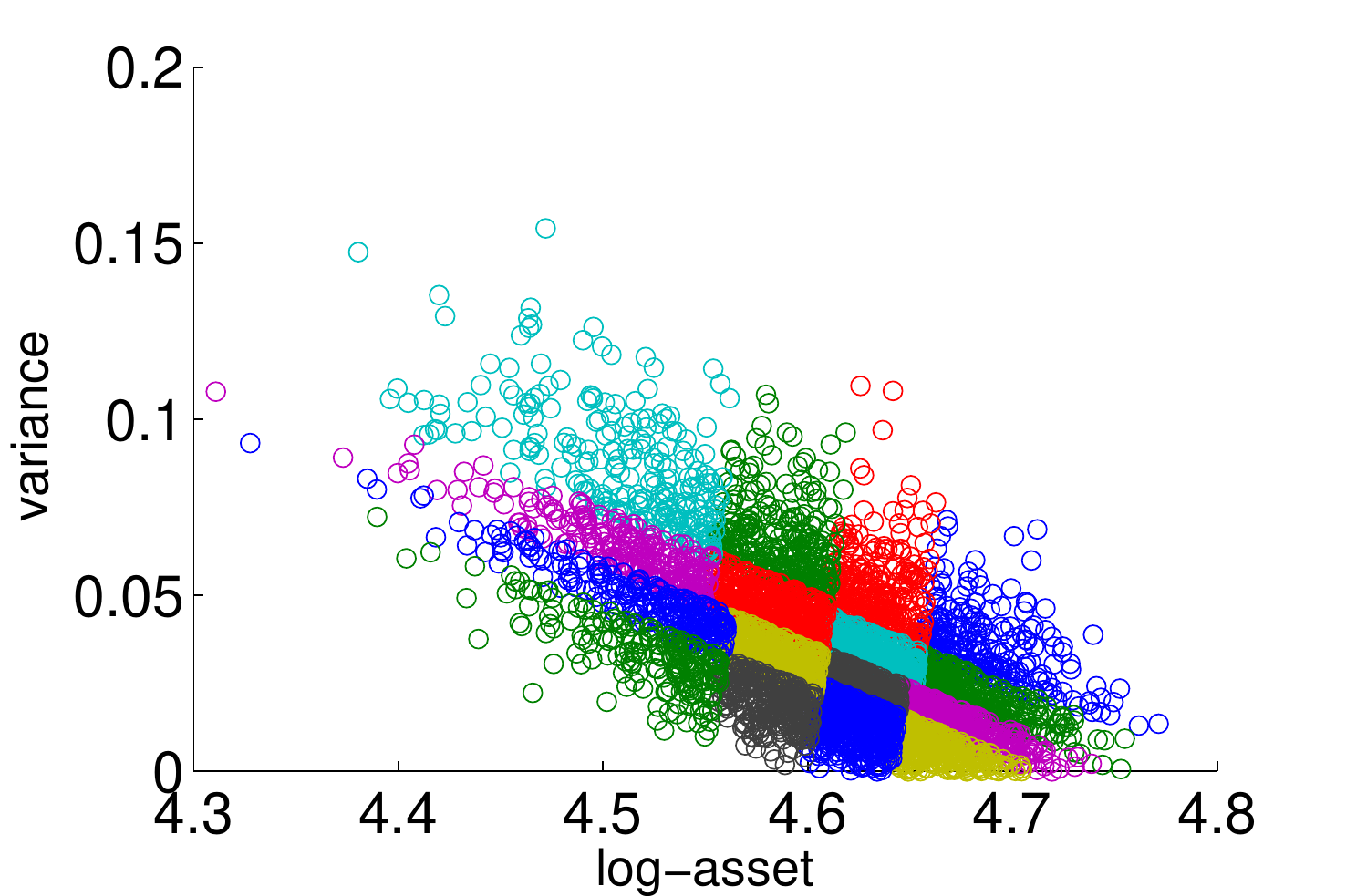}}
\end{center}
\caption{Result of recursive bifurcation with rotation.}
\label{fig:rbwr}
\end{figure}

Figure \ref{fig:rbwr} shows the results of bundling after applying the recursive bifurcation method on the rotated data.
In the left plot, the paths with values in a block of the same color belong to the same bundle at this time point; the indices of these paths are saved, and 
we also make scatter plots of the same bundles with the original (not rotated) values on the right side plot. 

The number of bundles in the recursive bifurcation method, after $j$ iterations, would be $(2^n)^j$ for an n-dimension problem. 
For the Heston model, the number of bundles is equal to $4^j$ after the $j$-th iteration. The method proposed here becomes less attractive for high-dimensional problems, as 
the rotation procedure may then become involved.

\subsubsection{Equal-number bundling}
\begin{figure}[ht!]
 \begin{center}
 \subfigure[Step 1: bundles on the log-asset domain]{\includegraphics[width=6cm]{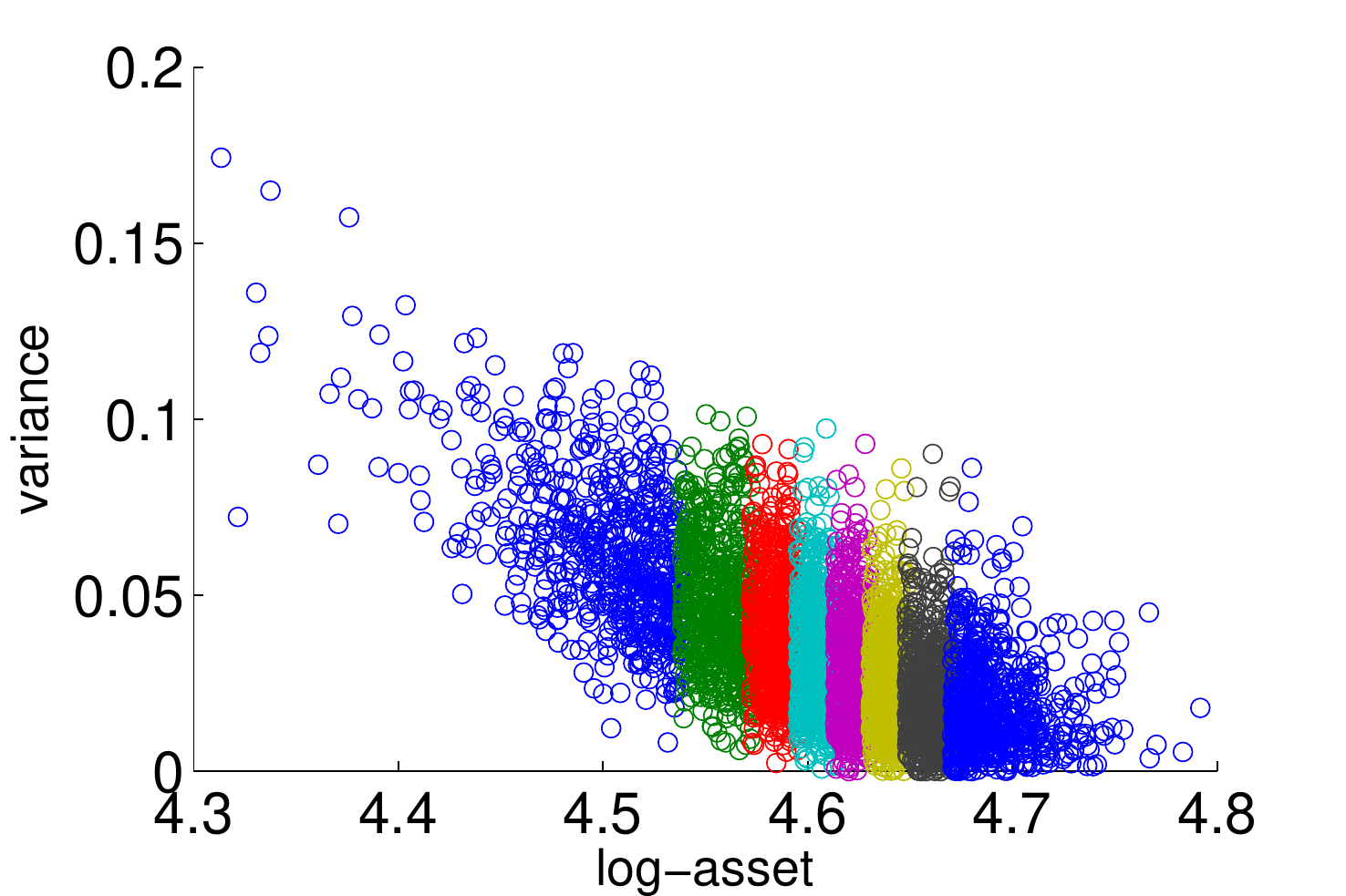}}
\hfill
\subfigure[Step 2: bundles on the variance domain]{\includegraphics[width=6cm]{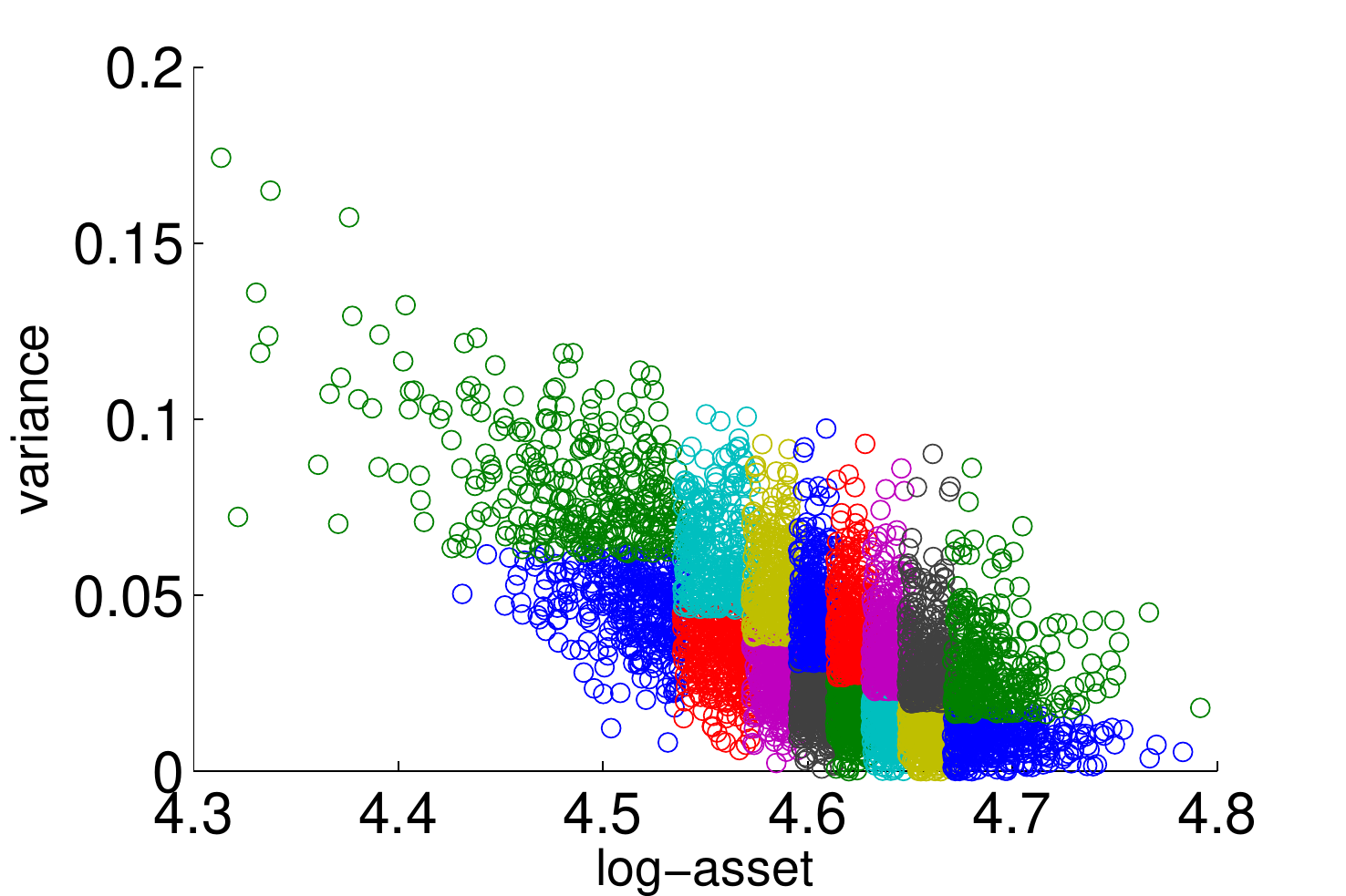}}
\end{center}
\caption{Result of equal-number bundling.}
\label{fig2:esins}
\end{figure}

Another way of making bundles is to distribute an equal number of paths to the bundles.  
In a first iteration, we make an ordering of the paths w.r.t the their log-asset variable values.
The paths with ranking between $\frac{(j-1)\cdot N}{J_1}+1$ and $\frac{j\cdot N}{J_1}$ belong to the $j$-th bundle, thus there
are $J_1$ bundles after the first iteration; then, within each bundle, we make a second ordering of the paths w.r.t their variance values,
and divide each bundle into $J_2$ smaller bundles in the same way. After these two iterations, there are $J_1\cdot J_2$ bundles with the same number 
$\frac{N}{J_1\cdot J_2}$ of paths. Figure \ref{fig2:esins} shows resulting bundles from these two iterations, where blocks of the same color 
represent log-asset and variance values of the same bundle (there is no rotation step here).

The first iteration is made based on the value of the log-asset  variable, as it is easy to see that the asset variable is dominating 
when pricing options. 
A potential problem with equal-number bundling is that when the Feller condition is not satisfied, there are zero variance values which
may be confusing when ordering these paths. 

An advantage of equal-number bundling is that one can freely choose the number of subdivisions in each dimension. We can choose a larger number 
in log-asset domain, as the log-asset value will have a more significant impact on option values. 
Equal-number bundling is also more efficient compared to the recursive bifurcation method.

\subsection{Heston-Hull-White model}
When we add interest rate as a stochastic variable, the state variable becomes $\textbf{X}_t:=[x_t,v_t,r_t]$, for which the corresponding dynamics are given, under the Heston Hull-White model, by
\begin{align}\label{eq:HHW}
dr_t&=\lambda(\theta-r_t)dt +\eta dW^r_t,\notag \\
dv_t&=\kappa(\bar v-v_t)dt +\gamma \sqrt{v_t}dW^v_t, \notag \\
dx_t&=\left(r_t-\frac{1}{2}v_t\right)dt +\sqrt{v_t}dW^x_t,
\end{align}
with $W^r_t$ Brownian motion, $dW^v_t\cdot dW^r_t=0$ and $dW^x_t\cdot dW^r_t=\rho_{x,r}$; constant parameters $\lambda$, $\theta$ and $\eta$ represent the speed of mean reversion, mean level of the interest rate and the volatility of the interest rate process, respectively;  the other parameters are the same as  in (\ref{eq:heston}). 

Basis functions of the polynomial space of order $p$, up to order 2, are presented in Table \ref{tab:basisfunction2}.
\begin{table}[ht!]
\centering
\begin{tabular}{ l|l }
\toprule
order $p$ & the basis functions\\
\midrule
0 & $\{1\}$ \\
1& $\{1, x,v,r\}$ \\
2& $\{1, x, x^2, v,v^2, r, r^2, x\cdot v,x\cdot r,r\cdot v\}$\\
\bottomrule
\end{tabular}
\caption{The basis functions and order $p$.}
\label{tab:basisfunction2}
\end{table}

A difficulty when applying SGBM for the HHW dynamics, is that analytic formulas of the discounted moments are not available as the HHW model does not belong to the affine class 
when $\rho_{x,r}\neq0$. We can see this from the covariance matrix 

\begin{equation}
\label{eq:oldco}
\sigma\left(\textbf{X}_t\right)\sigma\left(\textbf{X}_t\right)^T=
\begin{pmatrix}
v_t& \rho_{x,v}v_t &\sqrt{v_t}\eta\rho_{x,r}\\
*   &\gamma^2v_t    &0\\
*   &*               &\eta^2
\end{pmatrix}.
\end{equation}

Therefore, we will use an affine H1HW model, as proposed in \cite{Lech}, to approximate the HHW model and to find analytic formulas of the discounted moments for an {\em approximate} HHW model.

\subsubsection{Basis functions and H1HW model}\label{sec:h1hw}
To obtain an alternative affine formulation of the HHW dynamics, we approximate the stochastic term $\sqrt{v(t)}$ in (\ref{eq:oldco}) by a deterministic function, $\mathbb{E}\left[\sqrt{v_t}\big|v_s\right]$ , which is the conditional expectation of $\sqrt{v(t)}$ in time interval $[s,t]$, and 
hence the covariance matrix of the approximate model can be written as:
\begin{equation}
\label{eq:newco}
\sigma\left(\hat{\textbf{X}}_t\right)\sigma\left(\hat{\textbf{X}}_t\right)^T =
\begin{pmatrix}
v_t& \rho_{x,v}v_t &\mathbb{E}\left[\sqrt{v_t}\big|v_s\right]\eta\rho_{x,r} \\
*   &\gamma^2v_t    &0\\
*   &*               &\eta^2
\end{pmatrix}.
\end{equation}
The model connected to the covariance matrix in (\ref{eq:newco}) is called the H1HW model \cite{Lech}. 
Error analysis, comparing the full-scale HHW model with the performance of the approximate H1HW model is given in~\cite{Lech}, 
where it is shown that the errors in option values is typically very small.

In Appendix \ref{app:discountedchfhhw}, the discounted ChF of the H1HW model is presented, based on which we can determine the discounted moments. 

The expression for the conditional expectation of the volatility is given by: 
\begin{align}\label{eq:varianceexpec}
 \mathbb{E}\left[\sqrt{v_t}\big|v_s\right]=
 \sqrt{2c(\tau)}e^{-\frac{\lambda(\tau)}{2}}\sum_{k=0}^\infty \frac{1}{k!}\left(\frac{\lambda(\tau,v_s)}{2}\right)^k\frac{\Gamma\left(\frac{1+d}{2}+k\right)}{\Gamma\left(\frac{d}{2}+k\right)},
\end{align}
with $\tau:=t-s$,where 
\begin{equation}
c(\tau)=\frac{1}{4\kappa}\gamma^2(1-e^{-\kappa \tau}),\quad d=\frac{4\kappa \bar v}{\gamma^2},\quad\lambda(\tau,v_s)=\frac{4\kappa v_s e^{-\kappa \tau}}{\gamma^2(1-e^{-\kappa \tau})},
\end{equation}
which is truncated when computing this sum.
In numerical calculations, however, it turns out that the formula (\ref{eq:varianceexpec}) is not robust, particularly not  when $\tau$ is small. Hence for small values of $\tau$,
the conditional expectation of the square root of the variance is approximated as 
$$
\mathbb{E}\left[\sqrt{v_t}\big|v_s\right]\approx \sqrt{v_s},
$$
which is because 
$$
\mathbb{E}\left[\sqrt{v_t}\big|v_s\right]\rightarrow \sqrt{v_s}, \text{  when } (t-s)\rightarrow 0.
$$
When $d>\frac{1}{2}$, we can simplify the expression for the conditional expectation of the variance using the following the formula :
\begin{equation}\label{eq:determi}
\mathbb{E}\left[\sqrt{v_t}\big|v_s\right]\approx \sqrt{c(\tau)\left(\lambda(\tau,v_s)-1+d+\frac{d}{2(d+\lambda(\tau,v_s))}\right)},
\end{equation}
where $c(\tau)$, $d$ and $\lambda(\tau,v_s)$ are as defined in (\ref{eq:varianceexpec}), see~\cite{Lech}.
When $d<\frac{1}{2}$, it is suggested to use the accurate formula in (\ref{eq:varianceexpec}).

\subsubsection{Bundles}
We develop the two bundling methods in this 3-d problem. First, we present the recursive-bifurcation-with-rotation method for HHW model.

Step 1: project the  two sets of data $(d_1,d_2,d_3)$ into two independent sets of data $(q_1,q_2,q_3)$ by the following way. 
We define the rotation angles $\alpha_1, \alpha_2$ as 
\begin{align}
 &\cos \alpha_1= \sqrt{\frac{1}{1+\beta_1^2}}\text{sign}(\beta_1), 
 \sin \alpha_1=\sqrt{\frac{\beta_1^2}{1+\beta_1^2}}, \\
 & \cos \alpha_2= \sqrt{\frac{1}{1+\beta_2^2}}\text{sign}(\beta_2), 
 \sin \alpha_2=\sqrt{\frac{\beta_2^2}{1+\beta_2^2}},
\end{align}
where\footnote{These needed moments can be derived from the (non-discounted) ChF, and we calculate the slopes with the analytic formulas.} 
\begin{align}
\beta_1&:=\frac{\mathbb{E}\left[\left(x_m-\mathbb{E}(x_m)\right)\left(v_m-\mathbb{E}(v_m)\right)\right]}
{\mathbb{E}\left[\left(x_m-\mathbb{E}[x_m]\right)^2\right]}
\approx \frac{\text{covariance}(d_1,d_2)}{\text{variance}(d_1)}, \\
\beta_2&:=\frac{\mathbb{E}\left[\left(x_m-\mathbb{E}(x_m)\right)\left(r_m-\mathbb{E}(r_m)\right)\right]}
{\mathbb{E}\left[\left(x_m-\mathbb{E}[x_m]\right)^2\right]}
\approx \frac{\text{covariance}(d_1,d_3)}{\text{variance}(d_1)}, 
\end{align}

We rotate data $d_1$, $d_2$ and $d_3$ by angles $\alpha_1, \alpha_2$ using the following matrix,

\begin{equation}
 \begin{pmatrix}
  q_1 \\
  q_2 \\
  q_3
 \end{pmatrix}
 =\begin{pmatrix}
   \cos \alpha_1\sin \alpha_2 & \sin \alpha_1 &-\cos \alpha_1\cos \alpha_2 \\
   -\sin \alpha_1 \sin \alpha_2&\cos \alpha_1 & \sin \alpha_1 \cos \alpha_2 \\
   \cos \alpha_2 & 0 &\sin \alpha_2
  \end{pmatrix} 
   \begin{pmatrix}
  d_1 \\
  d_2 \\
  d_3
 \end{pmatrix}
\end{equation}

The idea of making bundles in the 3-d case with equal-numbering is also similar as for the Heston model. First, we make a ranking of the paths by their stock values, and determine $J_1$ bundles;
after this iteration, within each bundle, a ranking of the paths is made by the interest rate values to determine $J_2$ bundles in this direction followed by the ranking of the variance values to have $J_3$ sub-bundles within each bundle. After these three iterations, we have $(J_1\cdot J_2\cdot J_3)$ bundles. 

Notice that when the Feller condition is not satisfied, there are many "zeros`` in the variance values of the paths. The second bundling step is therefore made with interest rate values. The first step, however, should be based on the stock values, as the option values are dominated by stock values. 
\subsection{Bundles for barrier options}
\begin{figure}[ht!]
 \begin{center}
 \subfigure[recursive-bifurcation-with-rotation, barrier]{\includegraphics[width=6cm]{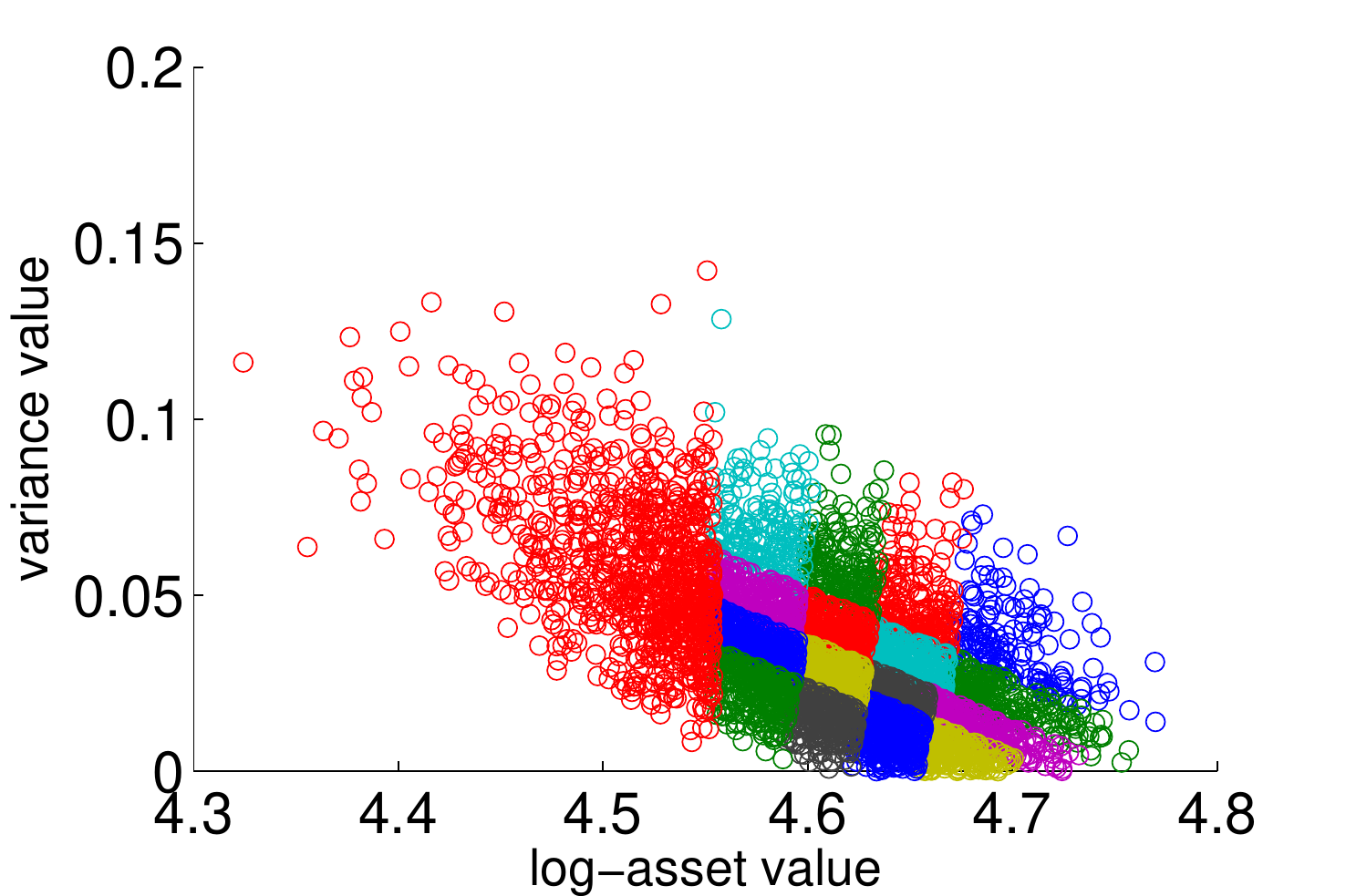}}
\hfill
\subfigure[Equal-number-bundling, barrier]{\includegraphics[width=6cm]{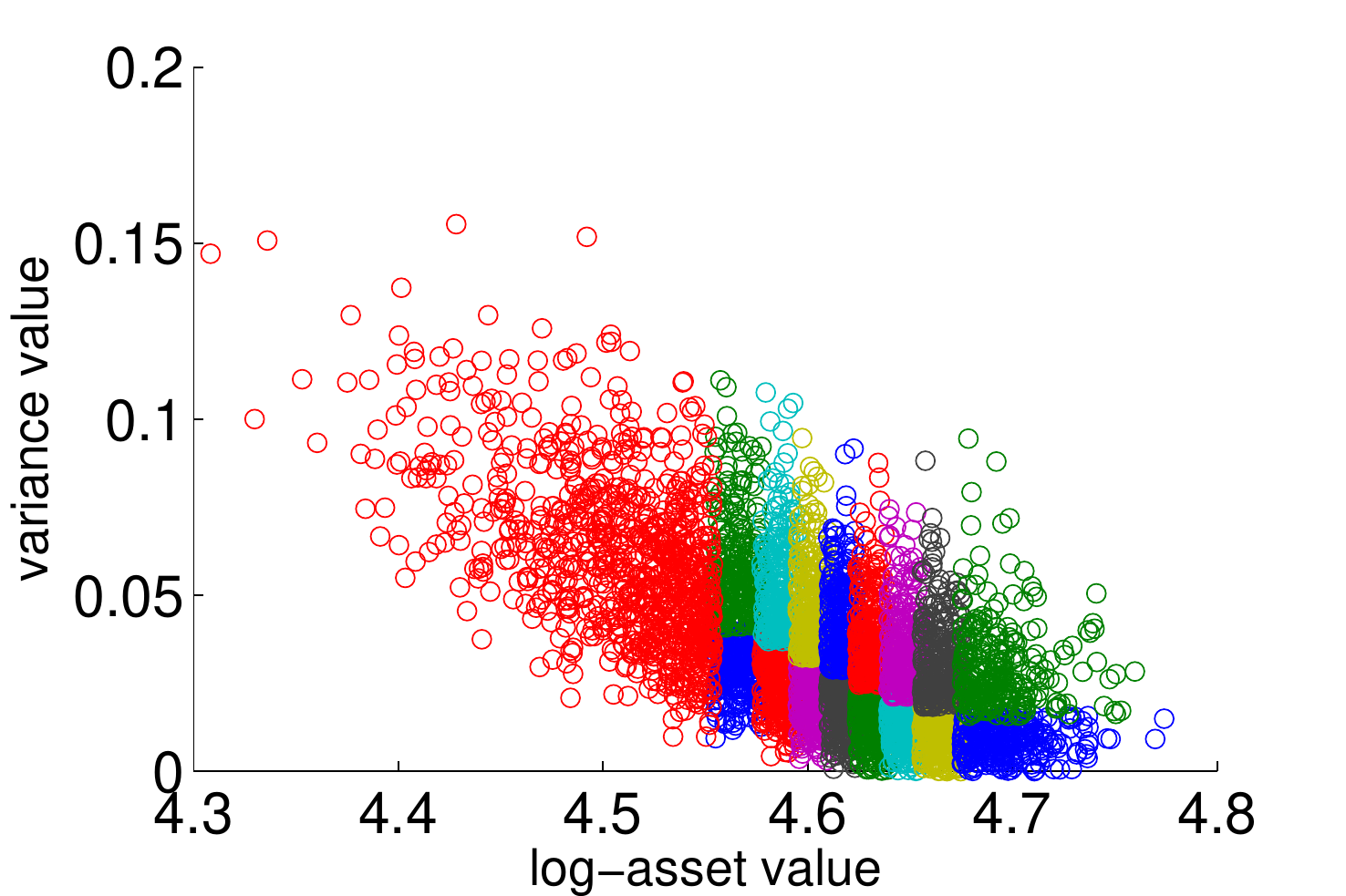}}
\end{center}
\caption{Bundling result for barrier options.}
\label{fig2:barrierbundling}
\end{figure}

When we make bundles for pricing barrier options, we only consider the 'active paths'. Hence, we find the paths that did not reach the barrier, and 
apply a regular bundling method. This is visualized in Figure \ref{fig2:barrierbundling} as a 2-D example.

\section{Numerical results}\label{sec:HHW}

In this section, several  numerical results obtained by the SGBM method are presented. We first discuss the impact of stochastic volatility and stochastic interest rate on the exposure quantities. Following this, we analyze the convergence and accuracy of SGBM, by comparing the path and direct estimators and by comparing with reference value (obtained via the COS method or the discounted cash flow of the simulated path). 

For all tests presented here, we will employ the Quadratic Exponential (QE) scheme~ \cite{Andersen} for generation of the forward MC paths, for robustness reasons. We have compared the QE scheme valuation results with the results obtained with an SDE Euler scheme and concluded that particularly in the case in which the Feller condition is not satisfied the QE scheme is superior.

We have tested several parameter sets for which the Feller condition is satisfied and for which the Feller condition is not satisfied.
We found that the Feller condition had very little impact on the performance of SGBM, due to the method components chosen (QE scheme, type of bundling and choice of basis functions).  We therefore only show results for parameter sets for which the Feller condition is not satisfied, as it is supposed to be more difficult for valuation. 
Generally the cases in which the Feller condition was satisfied were somewhat easier regarding the choice of time step.

We apply formula (\ref{eq:CVAdiscrete}) to calculate CVA with recovery rate $\delta  =0$.
The default probability function has been defined in equation (\ref{eq:default}) with a constant intensity $h=0.03$.

\subsection{Impact of the stochastic volatility and the stochastic interest rate}
We show the impact of stochastic volatility and stochastic interest rate on expected exposure(EE) and potential future exposure(PFE). Next to the already discussed Heston and HHW models, 
we also consider the Black-Scholes and the Black-Scholes Hull-White (BSHW) models in this section. The characteristic functions for these models (forming the basis for the analytic moments in SGBM) are presented in Appendix~\ref{app:discountedchfbshw}.

The parameter set chosen for the HHW model is given by $$\kappa=0.3, \gamma=0.6, v_0=\bar v=0.05, \lambda=0.01, \eta=0.01, r_0=\theta=0.02,
$$ and $S_0=100$; the correlations are set as $\rho_{x,v}=-0.3$ and $\rho_{x,r}=0.2$; $T=\{1,5\}$.

\begin{figure}[ht!]
\subfigure[T=1, EE]{\includegraphics[width=6cm]{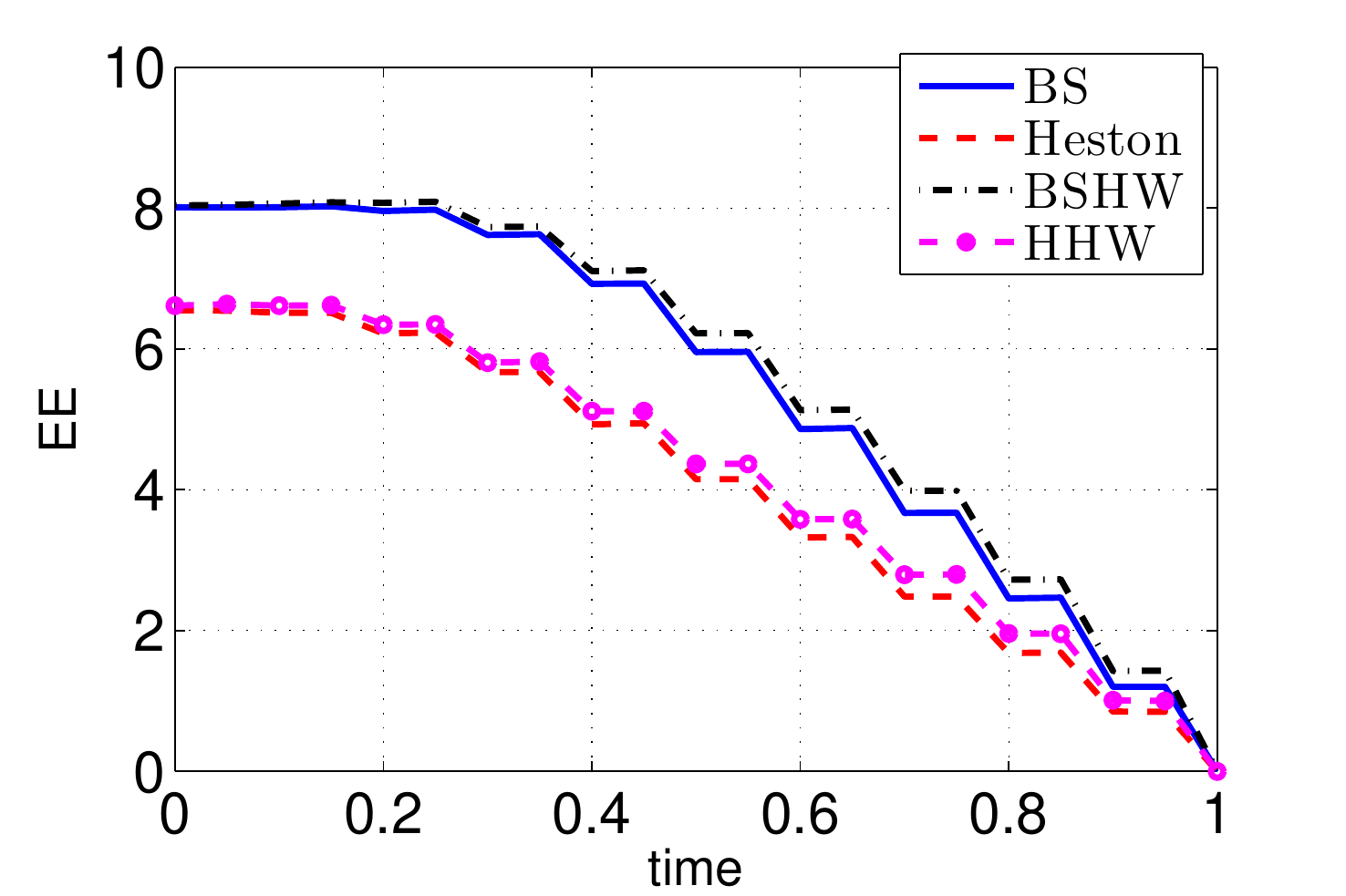}}
\hfill
\subfigure[T=5, EE]{\includegraphics[width=6cm]{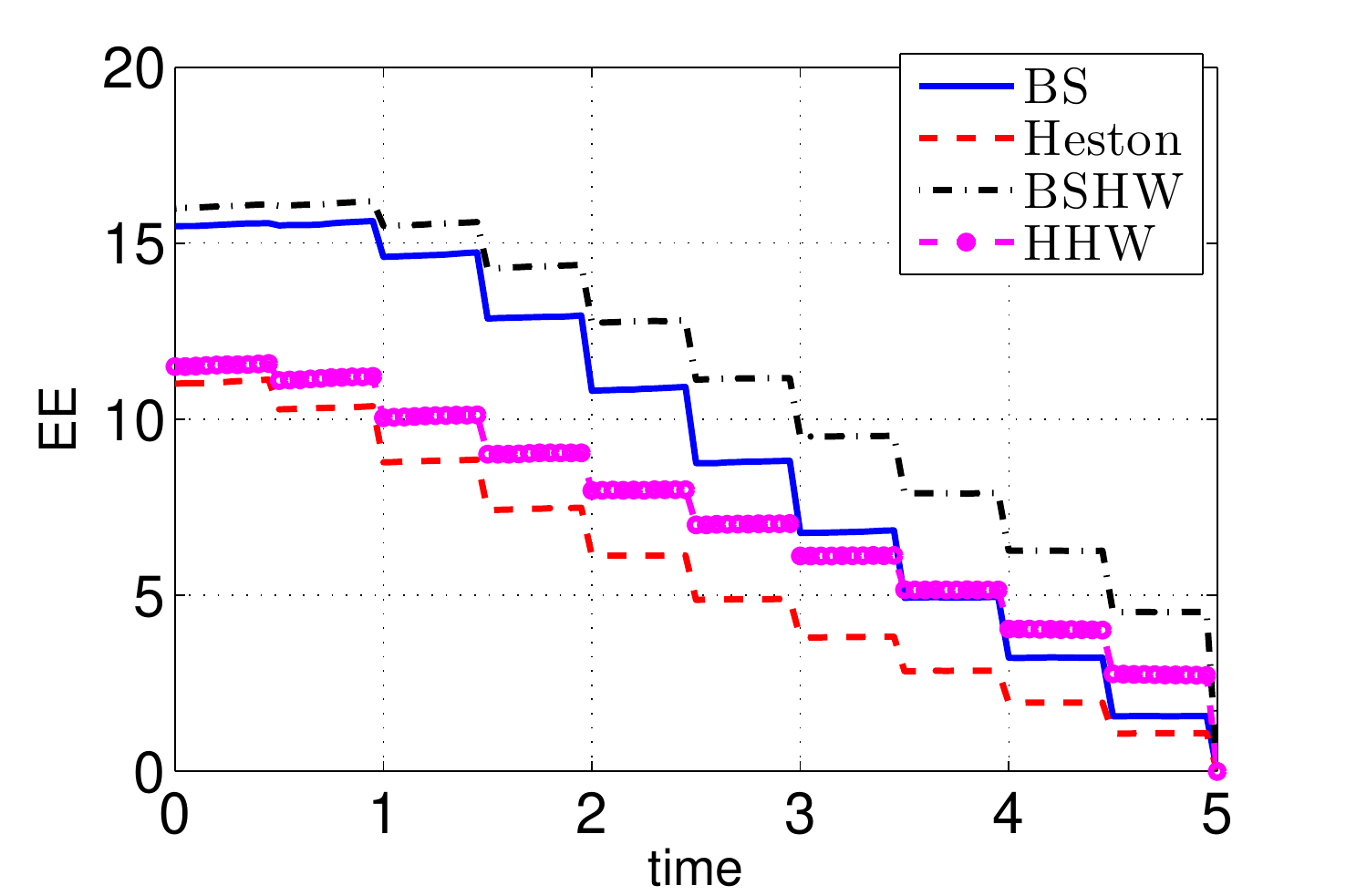}}
\hfill
\subfigure[T=1, PFE]{\includegraphics[width=6cm]{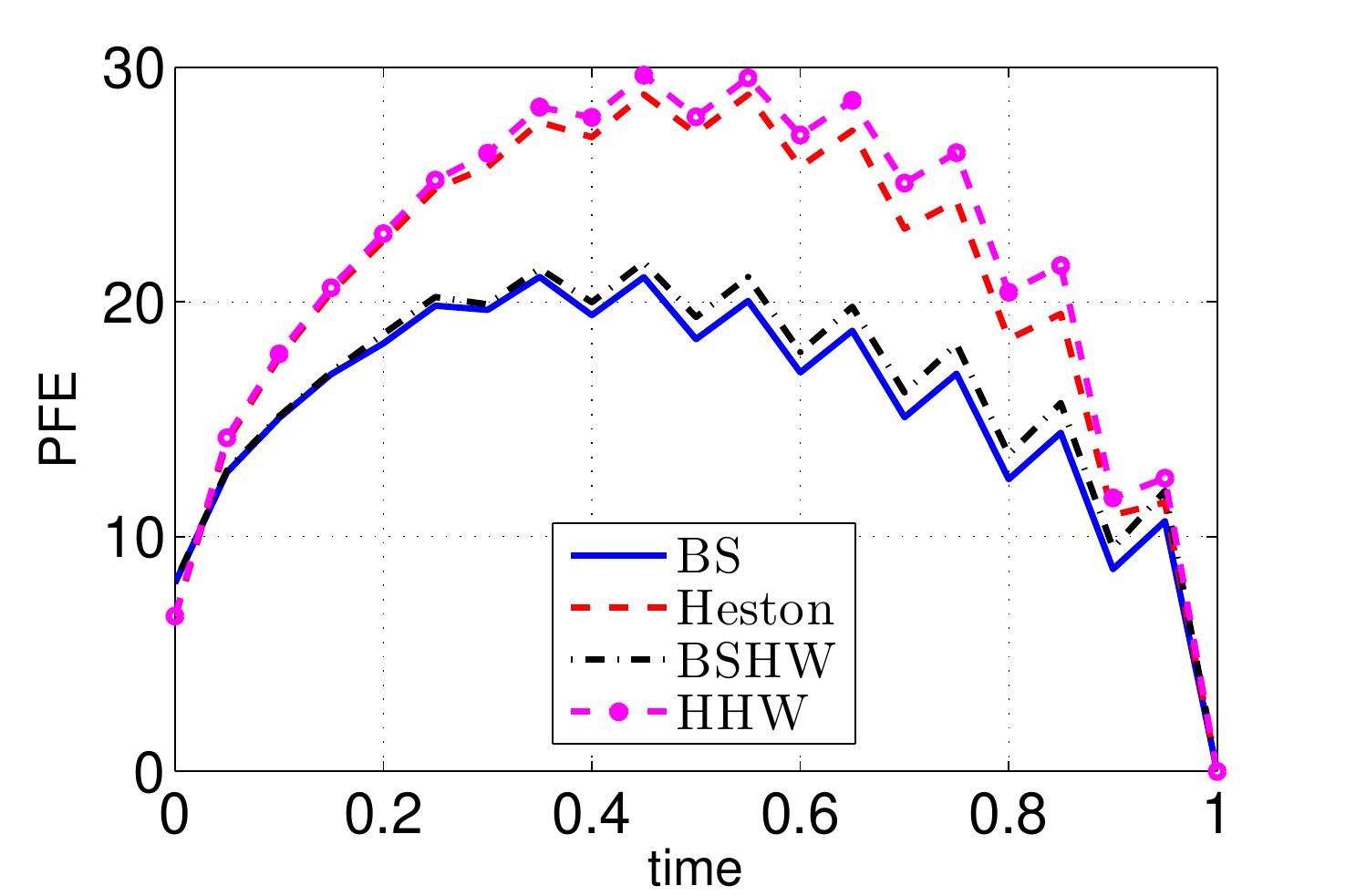}}
\hfill
\subfigure[T=5, PFE]{\includegraphics[width=6cm]{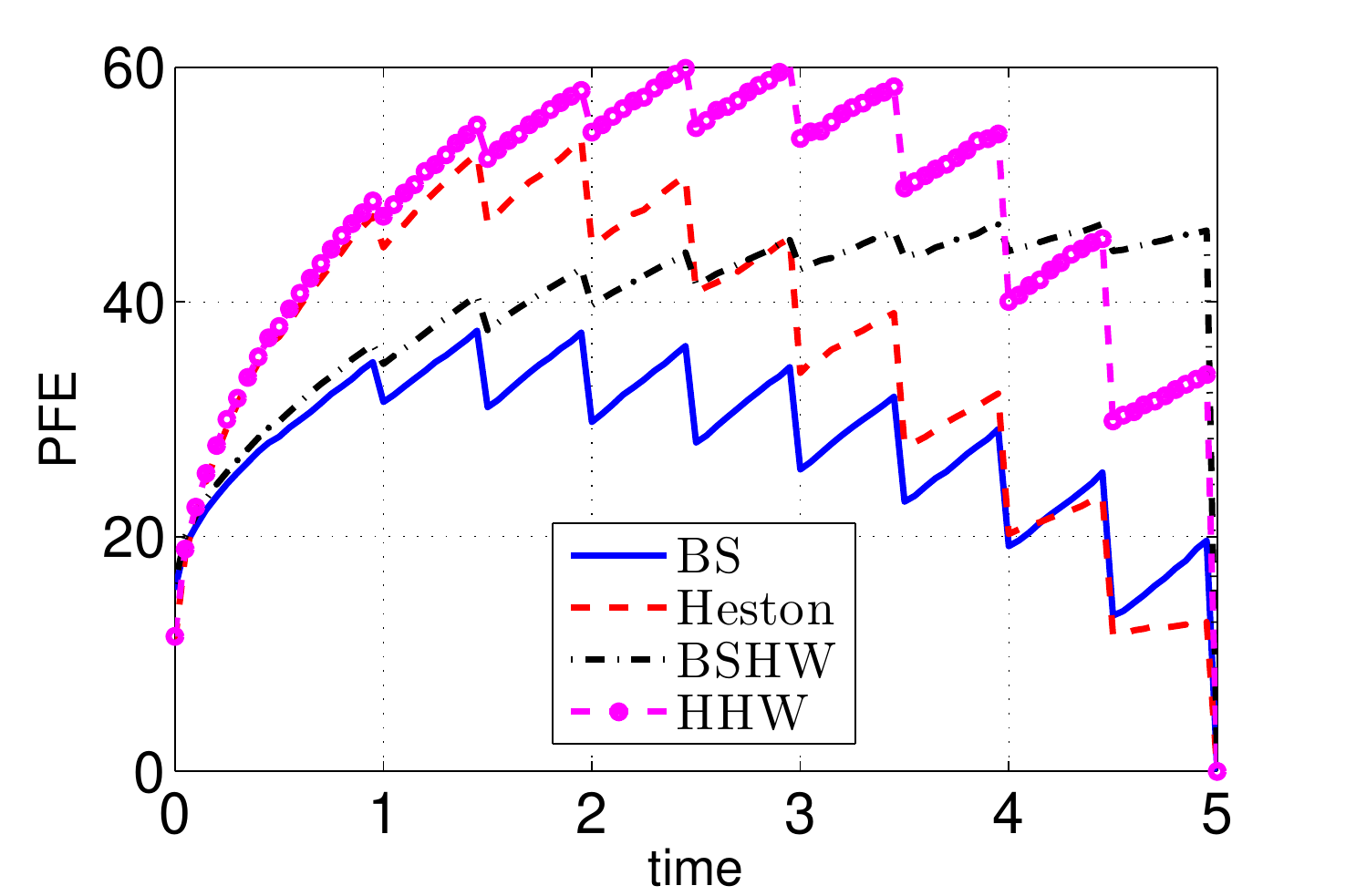}}
\caption{ Impact of the stochastic volatility and interest rate on EE and PFE with different tenors and different asset dynamics. }
\label{fig:imp2}
\end{figure}

For the comparison, we use the constant volatility, $\sigma=\sqrt{v_0}=0.2236$, for the Black-Scholes and the BSHW model, and a constant interest rate, $r=0.02$, 
for the Black-Scholes as well as for the Heston model. All other parameters in these models are set the same as in the HHW model. 
We consider a Bermudan put option with 10 exercise dates with strike $K=100$, and compare EE and PFE values for $T=1$ and $T=5$ in Figure \ref{fig:imp2}. The number of paths equals $2\cdot 10^5$, and 
the time step size of the SDE discretization, $\Delta t(\text{QE})=0.05$.
In the BS, BSHW and Heston models, the number of bundles employed is $4^3=64$, while for the HHW model, the number of bundles used is $8^3=512$.

From Figure \ref{fig:imp2} we can see the following:

\begin{itemize}
 \item When $T=1$ (as seen in (a) and (c)), the exposure values for the HHW model are relatively close to those of the Heston model, and the exposure values of the BSHW model are similar to those of the BS model. With a short time to maturity, under our model assumptions and parameters, the stochastic interest rate does not have a significant impact on the exposure profiles, whereas stochastic volatility in the asset dynamics increases the PFE values significantly. 
 \item When $T=5$ (as seen in (b) and (d)), we notice only small differences in the EE values for these models. Adding the stochastic interest rate and the stochastic volatility components, the PFE profiles have changed however without any clear pattern to be observed. 
\end{itemize}

\begin{figure}[ht!]
\subfigure[$\gamma=0.001$, PFE]{\includegraphics[width=6cm]{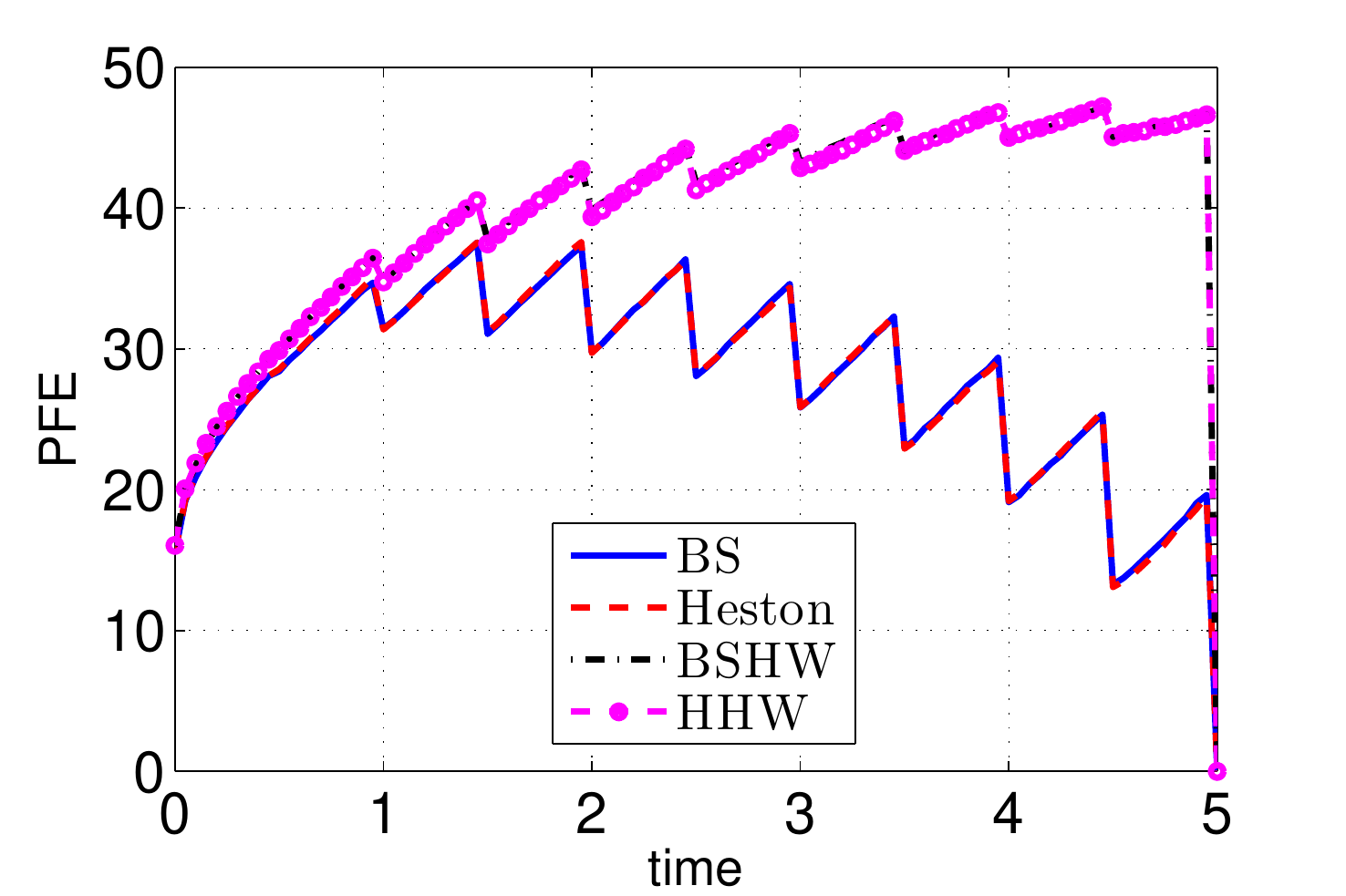}}
\hfill
\subfigure[$\eta=0.001$, PFE]{\includegraphics[width=6cm]{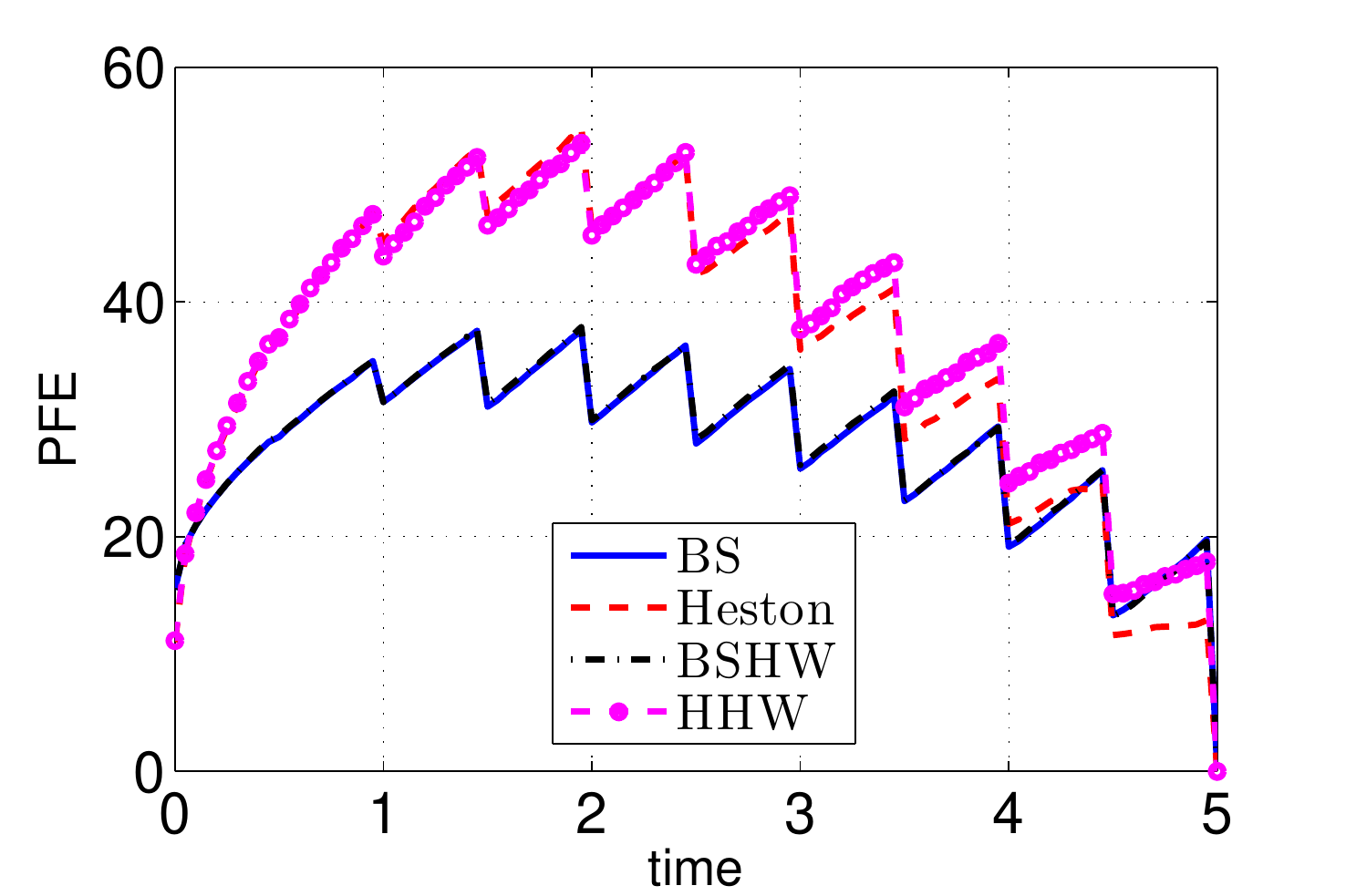}}

\caption{ Impact of the stochastic volatility and interest rate, choosing small model parameters. }
\label{fig:imp3}
\end{figure}

In order to gain additional insight in the results for the longer time to maturity, we perform some more tests. Setting either the vol-of-vol parameter $\gamma=0.001$ or the vol-of-interest-rate parameter $\eta=0.001$, respectively, but keeping the other parameters the same, we obtain the PFE values in Figure \ref{fig:imp3}. Here, in Figure~\ref{fig:imp3}(a) the PFE profile from the Heston model is very similar to that of the BS model, and the PFE profile of the HHW model is similar to that of the BSHW model.  This is clear  as the stochastic asset volatility does not play a major role in the dynamics due to the small vol-of-vol value. Thus, we can observe that the stochastic interest rate gives rise to an increasing PFE, although there is a drop in value at each exercise date. In Figure \ref{fig:imp3} (b), the PFE profile of BSHW model is close to that of the BS model, while the PFE results obtained by the HHW model are very similar to the Heston results. The PFE values increase at the early stages of the 
contract, and then drop rapidly 
towards the BS model PFE, at a later stage in the five years contract. 

The stochastic interest rate plays a significant role in the case of a longer maturity, and results in 
increasing PFE profiles; stochastic asset volatility seems to have an effect on PFE values at an early stage of a contract regardless the length of the contract. 
For asset models with stochastic interest rate and volatility, the form of the PFE profiles is not easily predictable: under the parameters chosen here, at an early stage of the 
contract (say $t<1$), the PFE profiles under the HHW model are very similar to those under the Heston model, but at later times the PFE profiles under the HHW model increase.

\subsection{SGBM convergence for exposure profiles of a Bermudan option under Heston model}\label{sec:pricingbermheston}
We study the convergence of SGBM by considering 
a Bermudan put option with 10 exercise dates. 
The number of paths equals $5\cdot 10^5$, and 
the time step is $\Delta t(\text{QE})=0.05$.
We choose the following parameter set. 
\\
\textbf{Test A}: $S_0=100$, $K=100$, $r=0.04$, $T=1$. Parameters $\kappa=1.15$, $\gamma=0.39$, $\bar v=0.0348$, $v_0=0.0348$, $\rho_{x,v}=-0.64$. (The Feller condition is not satisfied.) 

\subsubsection{Comparison of direct and path estimator }
We compare the option values, and EE values, of the direct and path estimators with respect to the number of bundles, when the order of the basis functions is chosen as $p=1$ and $p=2$. The difference of EE values is measured by the relative $L_2$-norm\footnote{The relative $L_2$-norm is defined 
as \begin{equation}
    \left(\frac{\sum_{m=0}^M \left(\text{EE}_d(t_m)-\text{EE}_p(t_m)\right)^2}{\sum_{m=0}^M \left(\text{EE}_d(t_m)\right)^2}\right)^{\frac{1}{2}},
   \end{equation}
where $\text{EE}_d(t_m)$ is the EE value obtained by the direct estimator at time $t_m$, while $\text{EE}_p(t_m)$ is the corresponding path estimator value.}, 
as EE is a time-dependent function. 

\begin{figure}[ht!]
\subfigure[Option value]{\includegraphics[width=6cm]{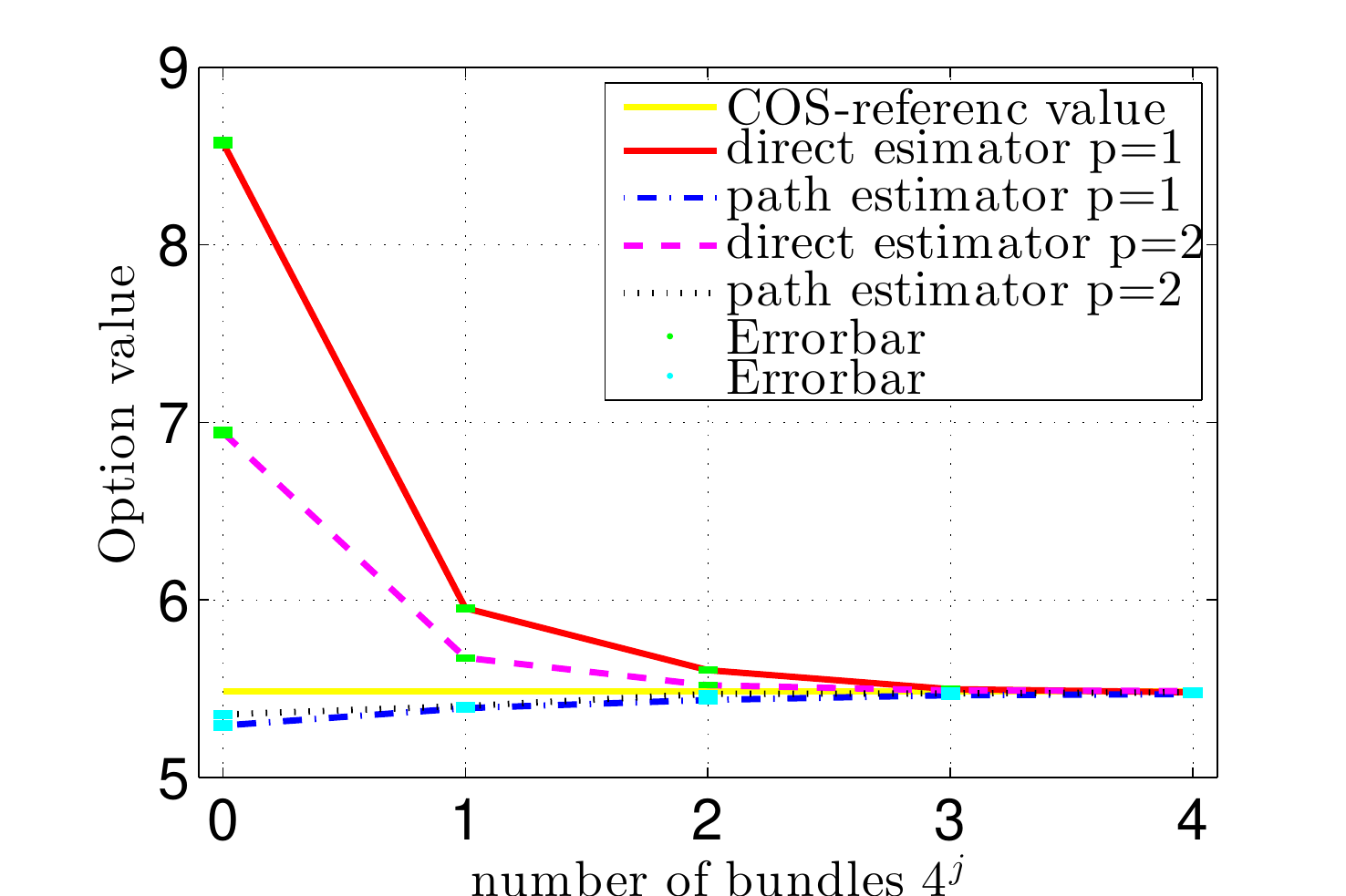}}
\hfill
\subfigure[EE difference between direct and path estimator]{\includegraphics[width=6cm]{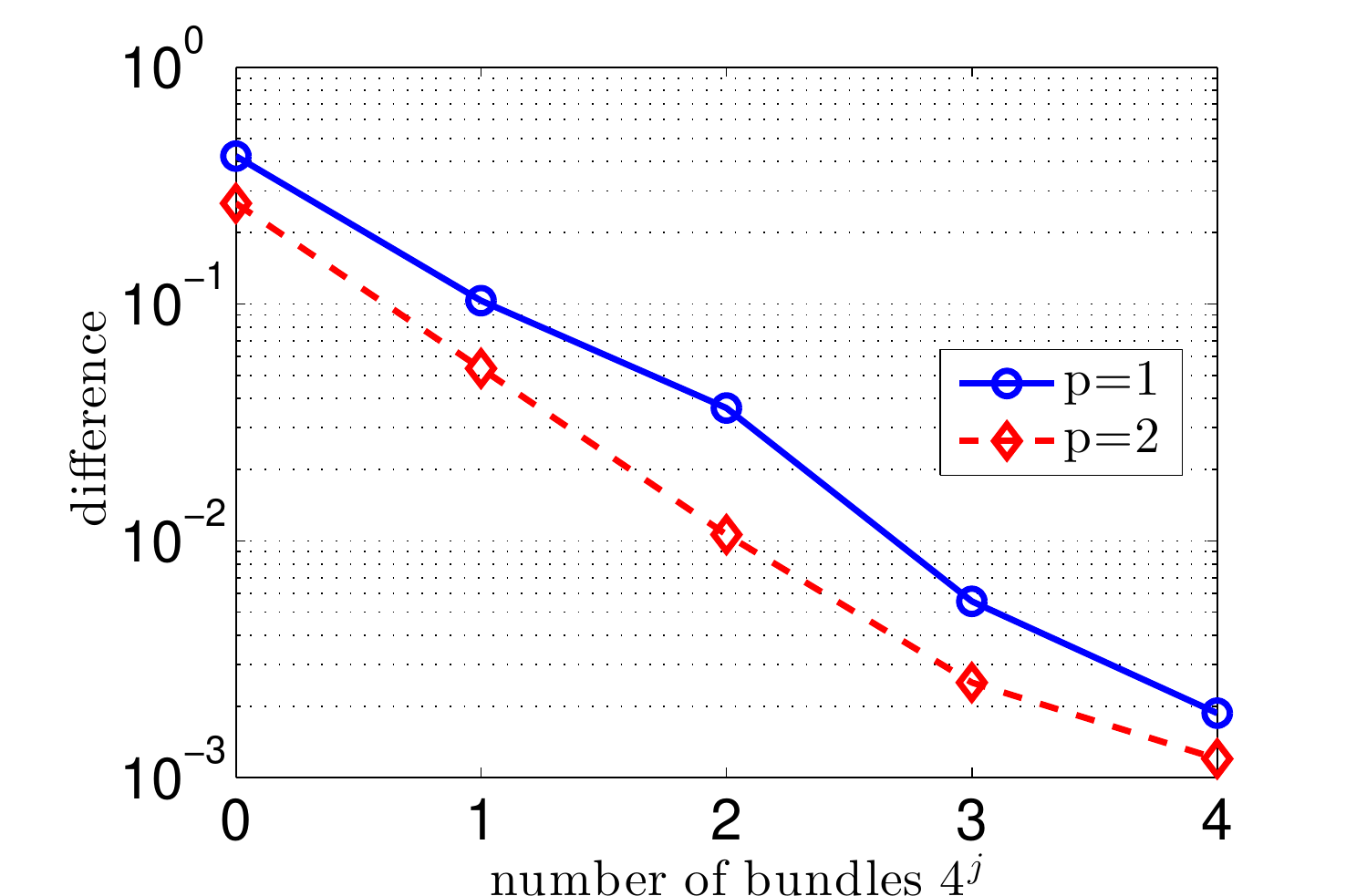}}
\caption{Option values and  relative difference of the EE values of the direct and the path estimator vs. the number of bundles.}
\label{fig:confergenceoption}
\end{figure}

Figure \ref{fig:confergenceoption} (a) shows that the path and direct estimators converge to the 'true' option values when increasing the number of bundles. When the number of bundles $J=4^3=64$, the path and the direct estimators have converged to the reference value level (calculated by the COS method). SGBM basis functions of order $p=2$ enhance the speed of convergence compared to $p=1$. 

Figure \ref{fig:confergenceoption} (b) confirms for the EE value that the path and direct estimator converge, when the number of bundles increases.  
We present the option value and Greeks of the Bermudan put option in Table \ref{tab:valueat0berm}, with the standard deviations. 
We also present the CVA value computed by Equation~(\ref{eq:CVAdiscrete}).
\begin{table}[ht!]
\centering
\begin{tabular}{ r|r|r|r}
  \toprule
  &COS & SGBM direct (std.) & SGBM path (std.) \\
 \midrule
$V(0)$                 &  5.483&5.486 (2.4e-04)  &  5.476 (4.0e-03)\\
$\Delta_{\text{EE}}(0)$& -0.327 &  -0.328 (7.9e-05) &-\\
$\Gamma_{\text{EE}}(0)$& 0.0247   &0.0247 (2.3e-05)  &-\\
 CVA                   &0.0924    & 0.0926 (8.9e-05)  &0.0949 (7.9e-05)\\
 \bottomrule
\end{tabular}
\caption{Values of Bermudan option, Greeks and CVA; SGBM based on 5 simulations.}
\label{tab:valueat0berm}
\end{table}
\subsubsection{Study of the order of basis functions}\label{sec:comparetoCOS}

The COS method is an efficient and accurate method for pricing Bermudan options based on Fourier cosine expansion \cite{fangfang}. We adapted the COS 
method in \cite{fangfang} so that it can also be applied for computing exposure profiles, see also~\cite{keesqian}. With the COS method as our reference, we study the convergence of EE and PFE and the EE Greeks. 

\begin{figure}[ht!]
\subfigure[EE]{\includegraphics[width=6cm]{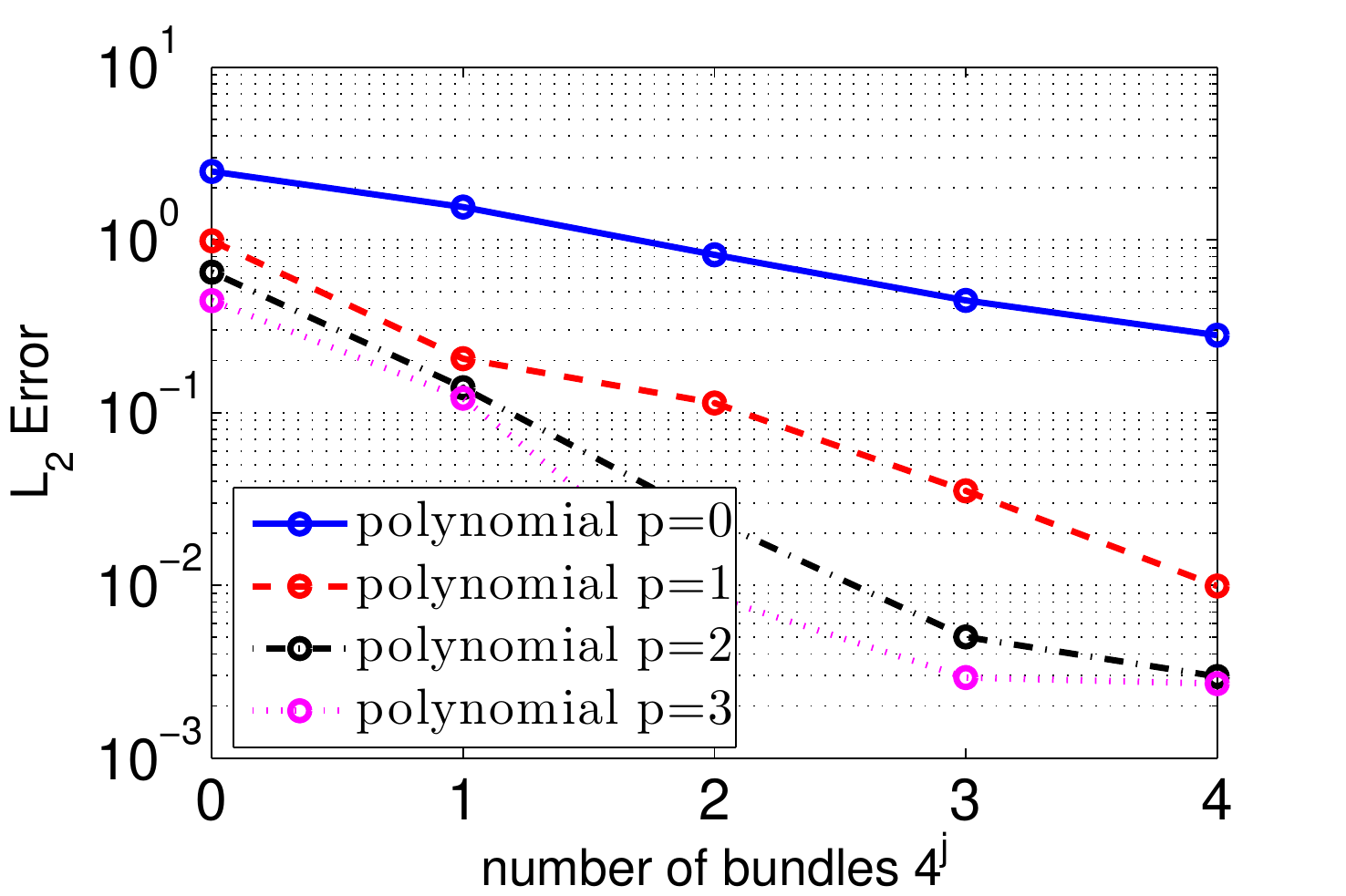}}
\hfill
\subfigure[PFE]{\includegraphics[width=6cm]{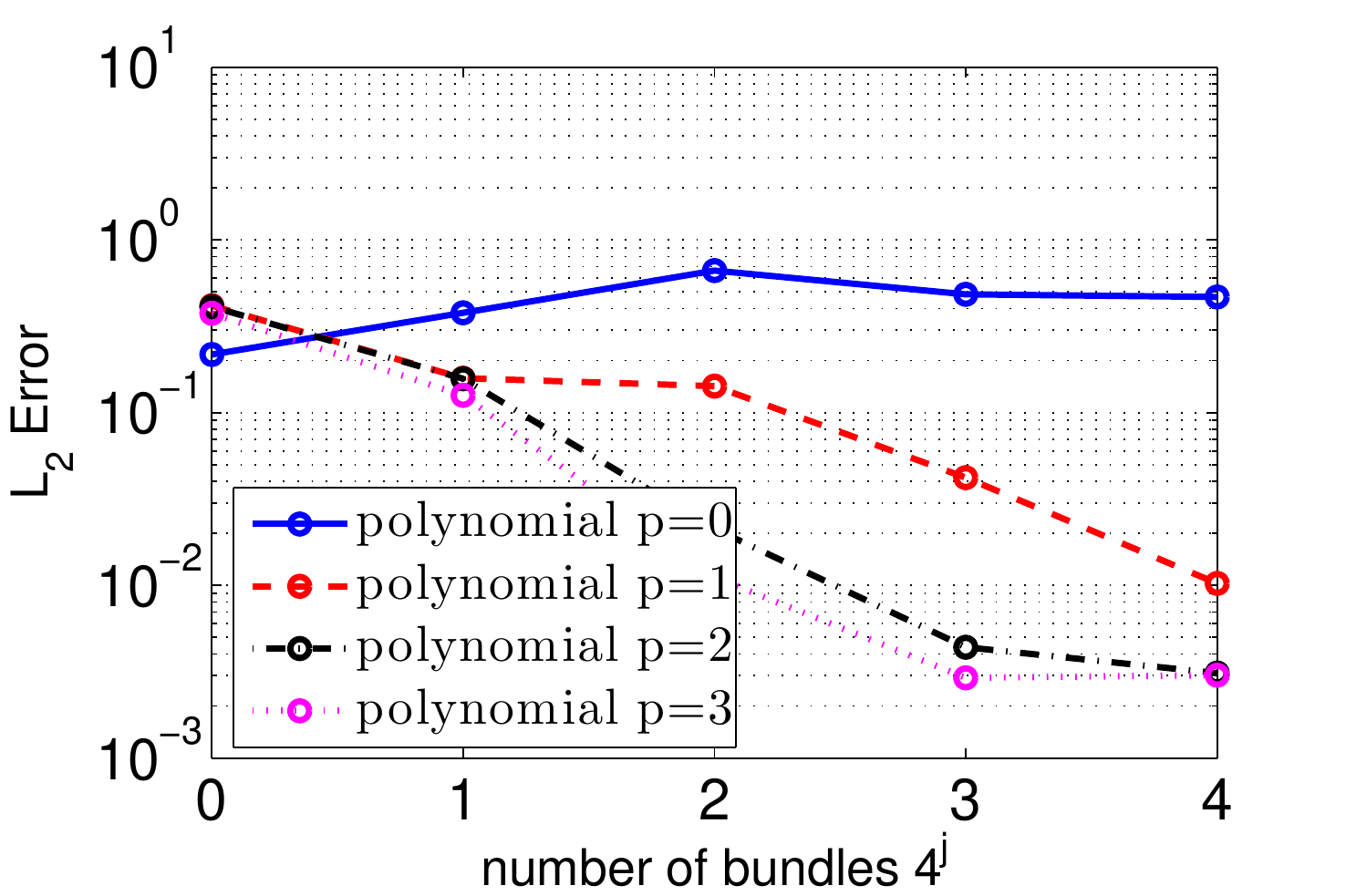}}
\hfill
\subfigure[$\Delta_{\text{EE}}$]{\includegraphics[width=6cm]{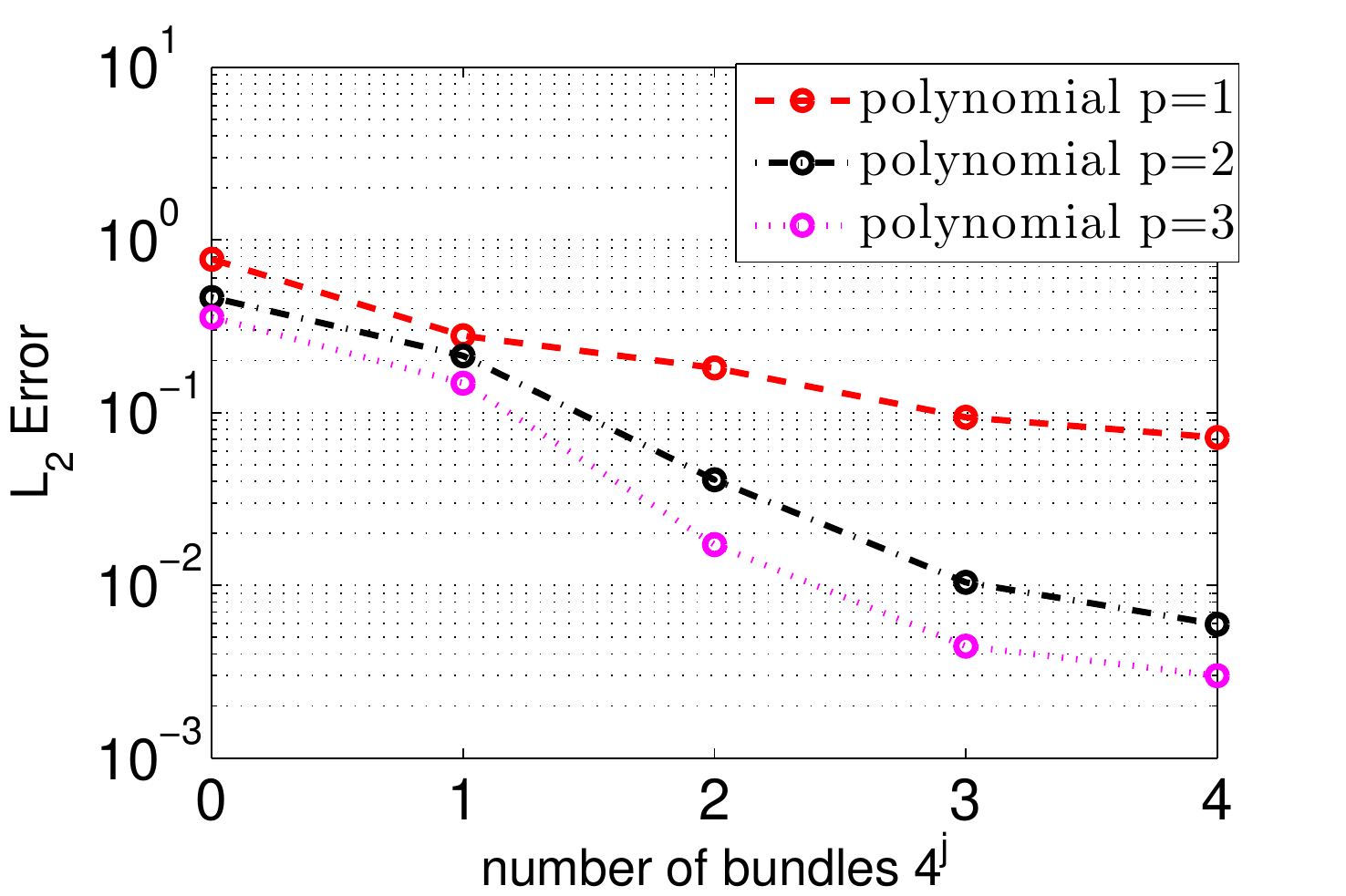}}
\hfill
\subfigure[$\Gamma_{\text{EE}}$]{\includegraphics[width=6cm]{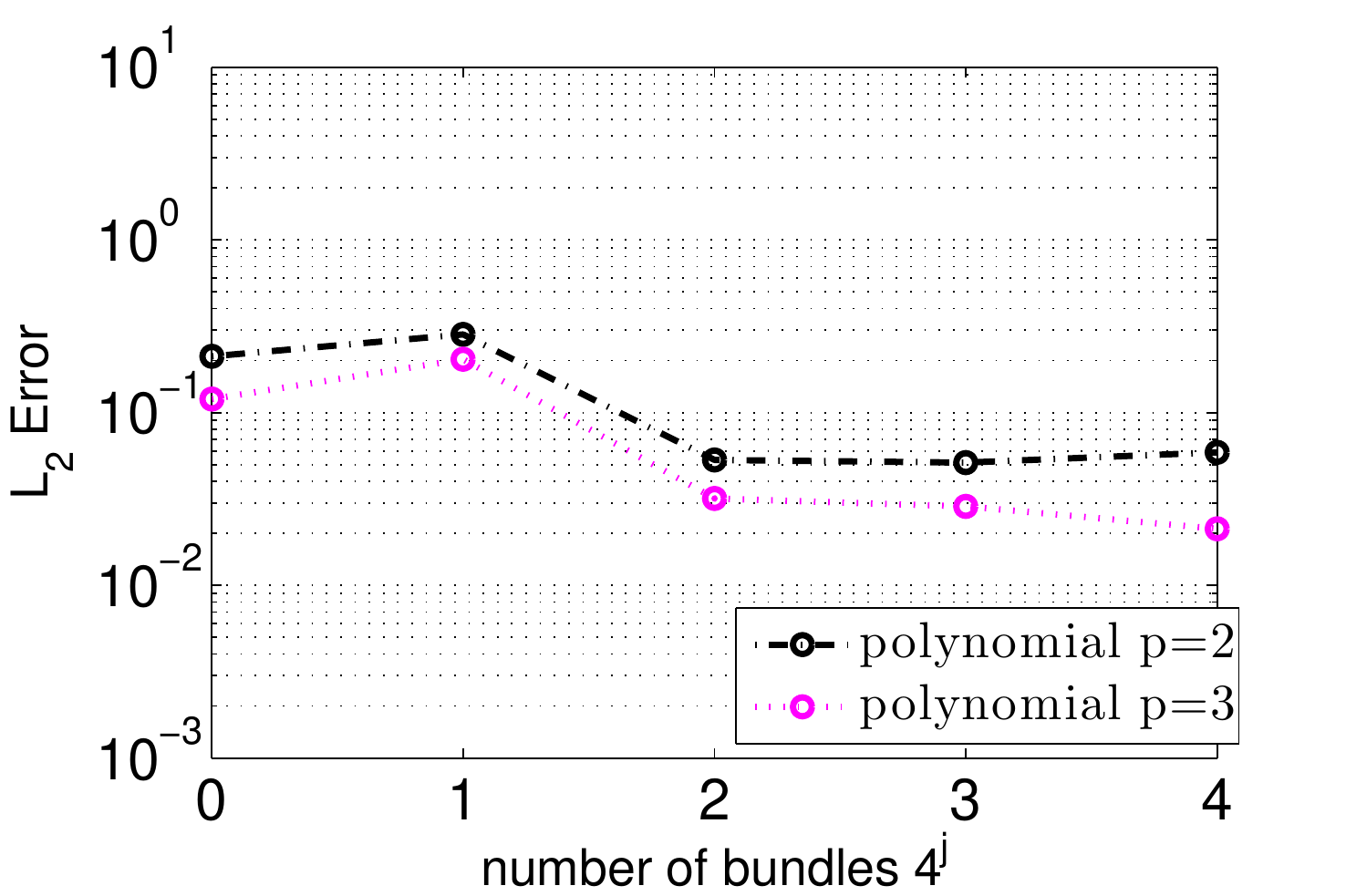}}
\caption{Relative $L_2$-error of exposure profiles and Greeks w.r.t number of bundles and the order of basis functions. {\em Bundling method: recursive-bifurcation-with-rotation}; Reference values by COS method.}
\label{fig:EEofSGBMbunglingmethod}
\end{figure}

\begin{figure}[ht!]
\subfigure[EE]{\includegraphics[width=6cm]{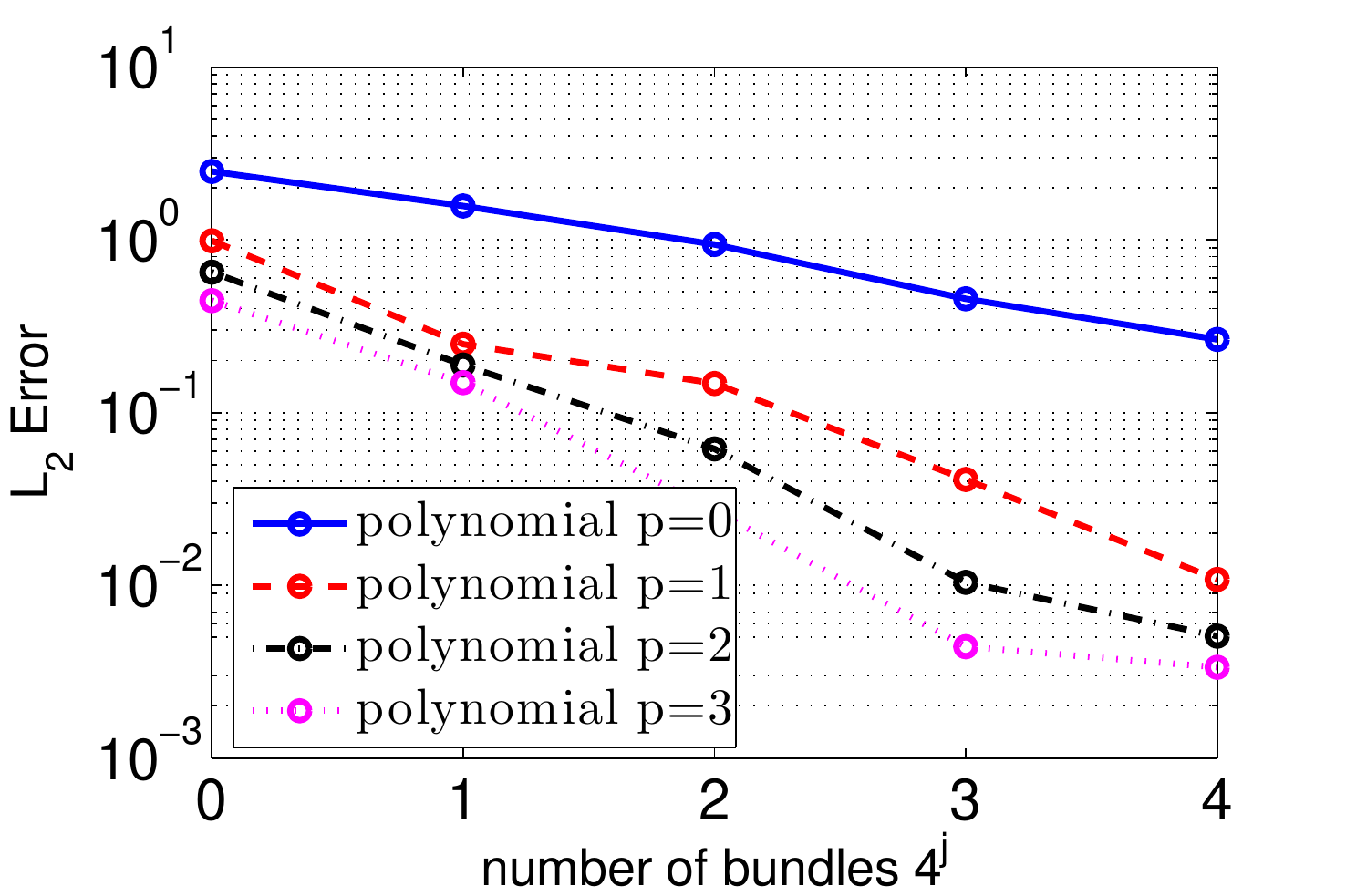}}
\hfill
\subfigure[PFE]{\includegraphics[width=6cm]{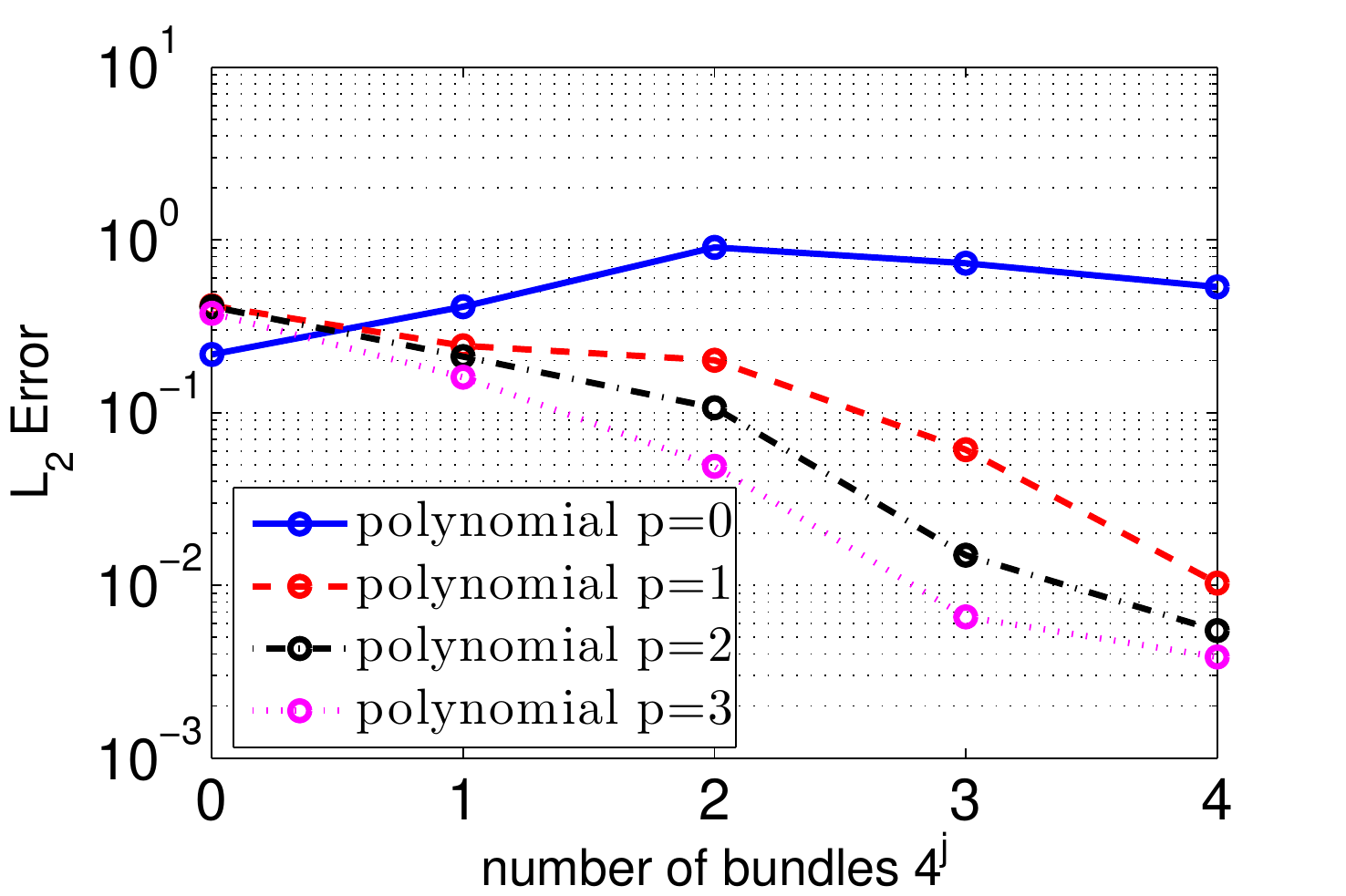}}
\hfill
\subfigure[$\Delta_{\text{EE}}$]{\includegraphics[width=6cm]{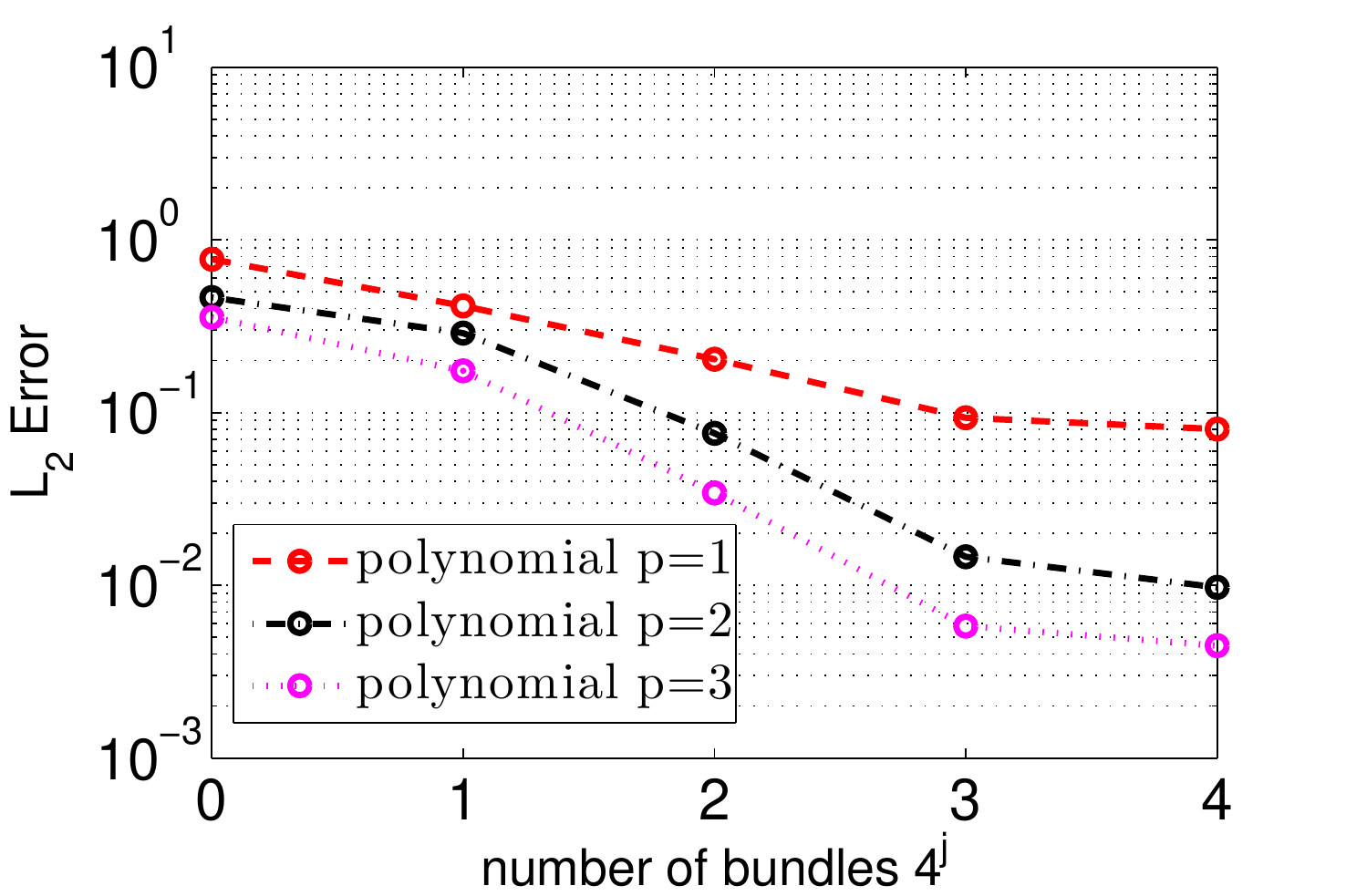}}
\hfill
\subfigure[$\Gamma_{\text{EE}}$]{\includegraphics[width=6cm]{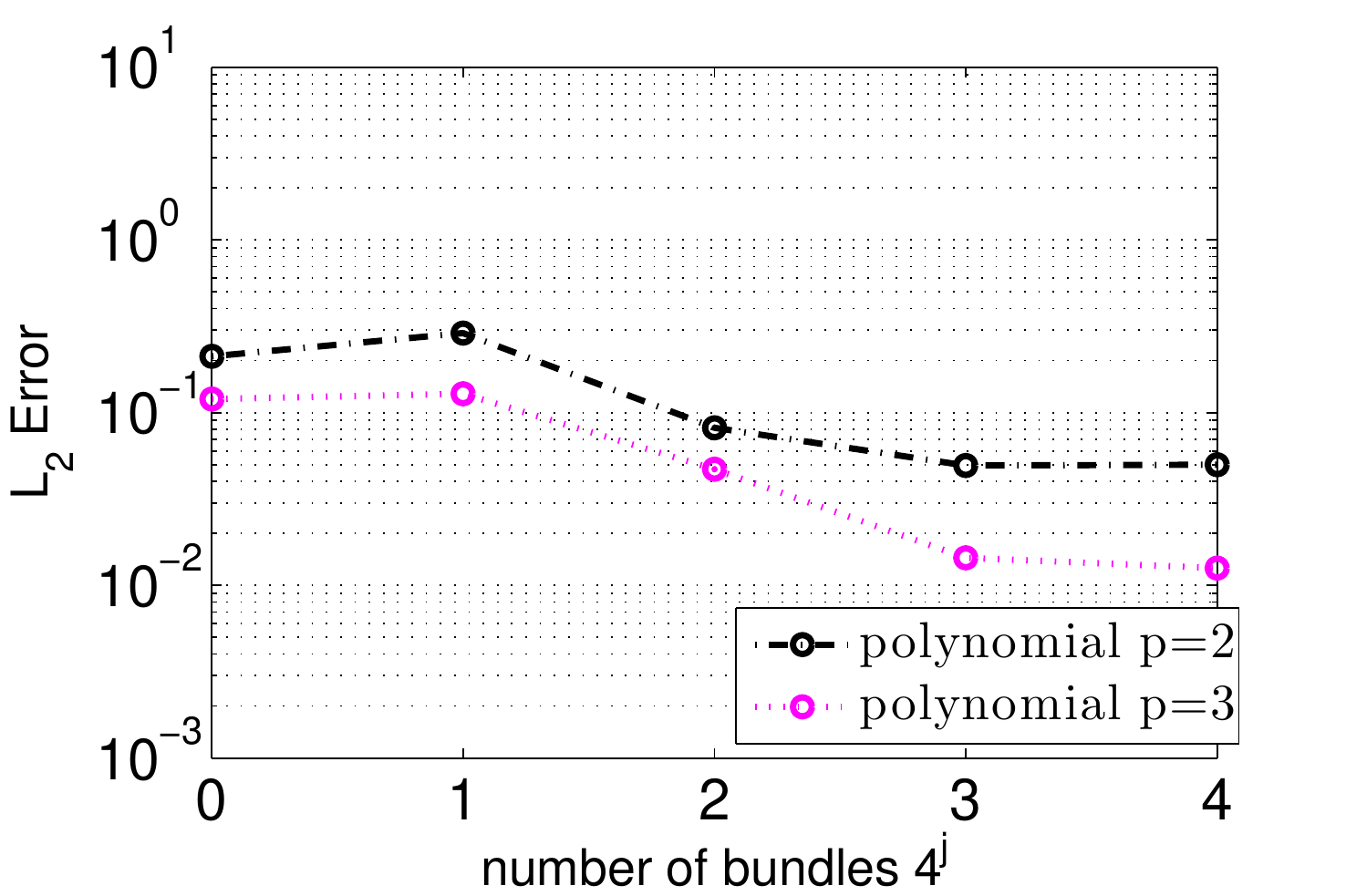}}
\caption{Relative $L_2$-error of exposure profiles and Greeks w.r.t number of bundles and the order of the basis functions. {\em Bundling method: equal-in-number bundling}.}
\label{fig:EEofSGBMbunglingmethod2}
\end{figure}

In Figures \ref{fig:EEofSGBMbunglingmethod} and \ref{fig:EEofSGBMbunglingmethod2}, we present the accuracy of SGBM for exposure quantities w.r.t the type of bundling, the number of bundles and the order of the basis functions. We check the difference of EE, PFE, $\Delta_{\text{EE}}$, and $\Gamma_{\text{EE}}$ between the SGBM and COS methods.  
When the basis function set only includes the constant $p=0$, then all derivatives w.r.t the initial asset value or variance value are equal to zero, and we cannot determine the Greeks values at all. When the basis functions are of order $1$, then we can calculate the first-derivative w.r.t  the initial asset value, but the second-derivatives values of these functions are zero.

The bundling method in Figure~\ref{fig:EEofSGBMbunglingmethod} 
is  recursive-bifurcation-with-rotation, while the results
equal-in-number bundling in Figure~\ref{fig:EEofSGBMbunglingmethod2} are very similar.

Unless stated otherwise, recursive-bifurcation-with-rotation bundling is employed in the experiments to follow. 

It is clear that increasing the number of bundles and/or the order of the basis functions can enhance the accuracy of the results; 
The impact of the polynomial order on the accuracy becomes smaller when the number of bundles increases. With $J=4^4$ bundles and $p=2$ the results are as highly satisfactory as the approximations by $p=3$.

In particular, we see that by a higher polynomial order $p$ and a larger number of bundles the accuracy of the Greek values increases. When the number of bundles is sufficiently large ($4^3$), basis functions of order $p=2$ perform as well as
$p=3$ for $\Delta_\text{EE}$, but for $\Gamma_\text{EE}$,
basis functions of order $p=3$ improve the accuracy. 

\subsection{SGBM results for the  Heston Hull-White model }
Here, we will analyze exposure results under the HHW model. For validation of the choices for the SGBM components however we first consider the pricing of options.

Our discretization of the HHW model is based on the QE Heston scheme, combined with an Euler discretization for the interest rates. We employ time step $\Delta t(\text{QE})=0.05$ and the number of paths used is  $N=5\cdot10^5$.  The following set of parameters has been used:
\\
\textbf{Test B}: $\kappa=0.3$, $\gamma=0.6$, $v_0=\bar v=0.05$, $\lambda=0.01$, $\eta=0.01$, $r_0=\theta=0.02$,
and $S_0=100$; the correlations are chosen as $\rho_{x,v}=-0.3$ and $\rho_{x,r}=\{0.2, 0.6\}$; $T=\{5,10\}$ (Feller condition not satisfied).

As explained, SGBM employs discounted moments for the H1HW model to approximate the discounted moments appearing. To determine the impact of this approximation, we will first price European options by SGBM and compare the results obtained  by the discounted cash flow plain Monte Carlo results. 

As in  \cite{Lech}, we also compare the implied volatility values for different strike values to analyze the accuracy. 

Subsequently, we present results for Bermudan options, and confirm the SGBM convergence by the comparison of the direct and the path estimator. 

\subsubsection{Pricing European options under the HHW model}
Employing SGBM for pricing European options, the problem is easier than pricing Bermudan options, as the option can not be exercised prior to maturity, i.e.,
for all $t_m<t_M$,
\begin{equation}
 V_0(\textbf{X}_0)=\mathbb{E}^{\mathbb{Q}}\left[V_M\left(\textbf{X}_M\right)D(t_0,t_M)\bigg| \textbf{X}_0\right]= \mathbb{E}^{\mathbb{Q}}\left[\mathbb{E}^{\mathbb{Q}}
 \left[V_M\left(\textbf{X}_M\right) 
 D(t_m,t_M)\bigg| \textbf{X}_m\right]D(t_0,t_m)\bigg| \textbf{X}_0\right].
\end{equation}
We can thus compute the European option estimate either directly from  the discounted averaged option values at time $t_M$; or we can use intermediate time points, $t_1,\dots,t_m$, between $t_0$ and $t_M$ to calculate the option values at the paths in a backward iteration, where the option value at time $t_0$ would be calculated based on the option values of all paths at time $t_1$.
The latter is an SGBM accuracy test which we perform here.

When the analytic formulas and the SGBM simulation are accurate, there should be no significant difference between these two approaches. However, as we use approximated HHW discounted moments derived from H1HW, 
the size of the time step will have impact on the accuracy of the results. We test this by choosing three different time steps in SGBM, with $\Delta t=\{0.05, 0.5, 10\}$ (the latter being only one time step).

\begin{figure}[ht!]
\subfigure[$\rho_{x,r}=0.2$]{\includegraphics[width=6cm]{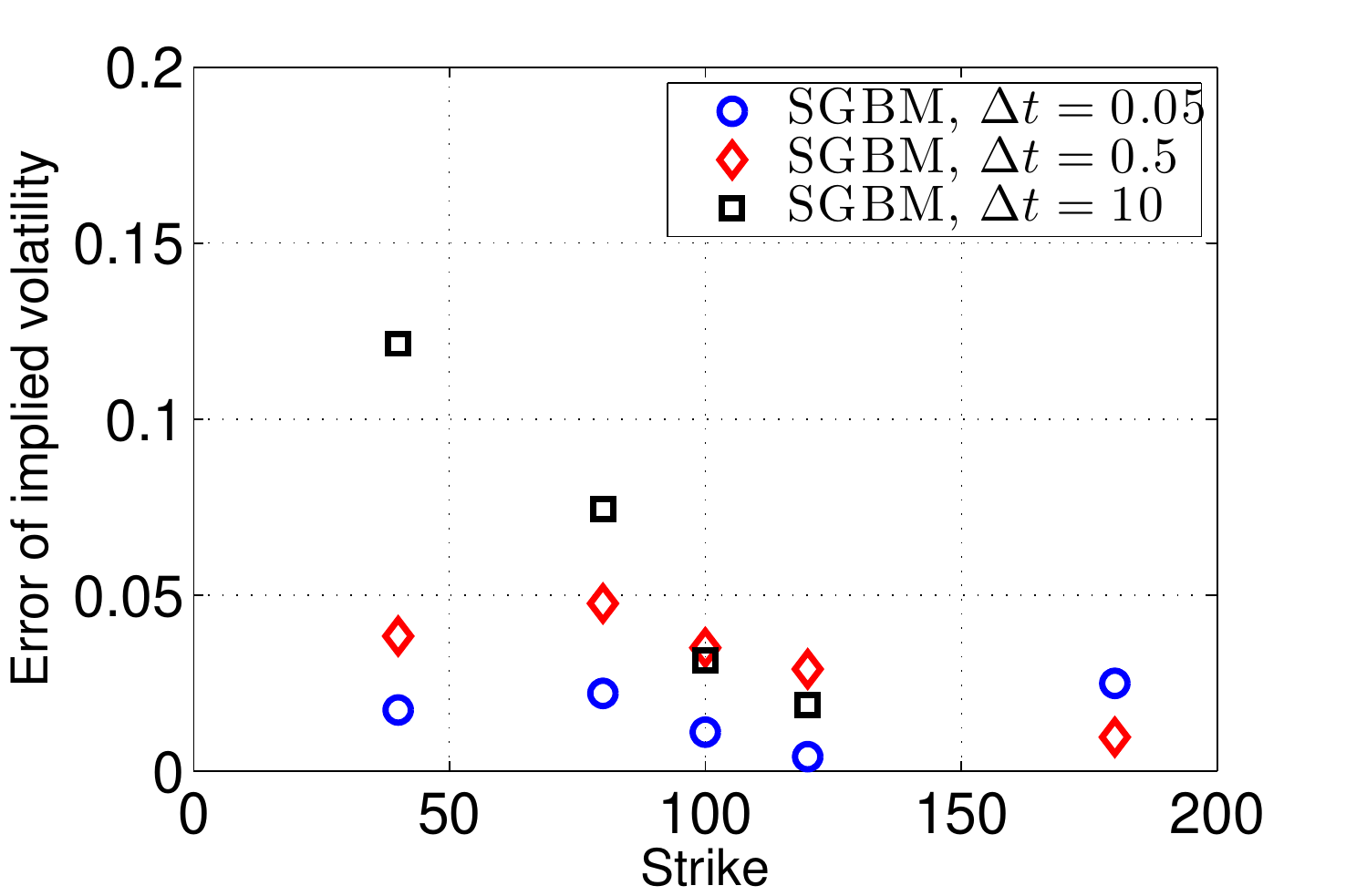}}
\hfill
\subfigure[$\rho_{x,r}=0.6$]{\includegraphics[width=6cm]{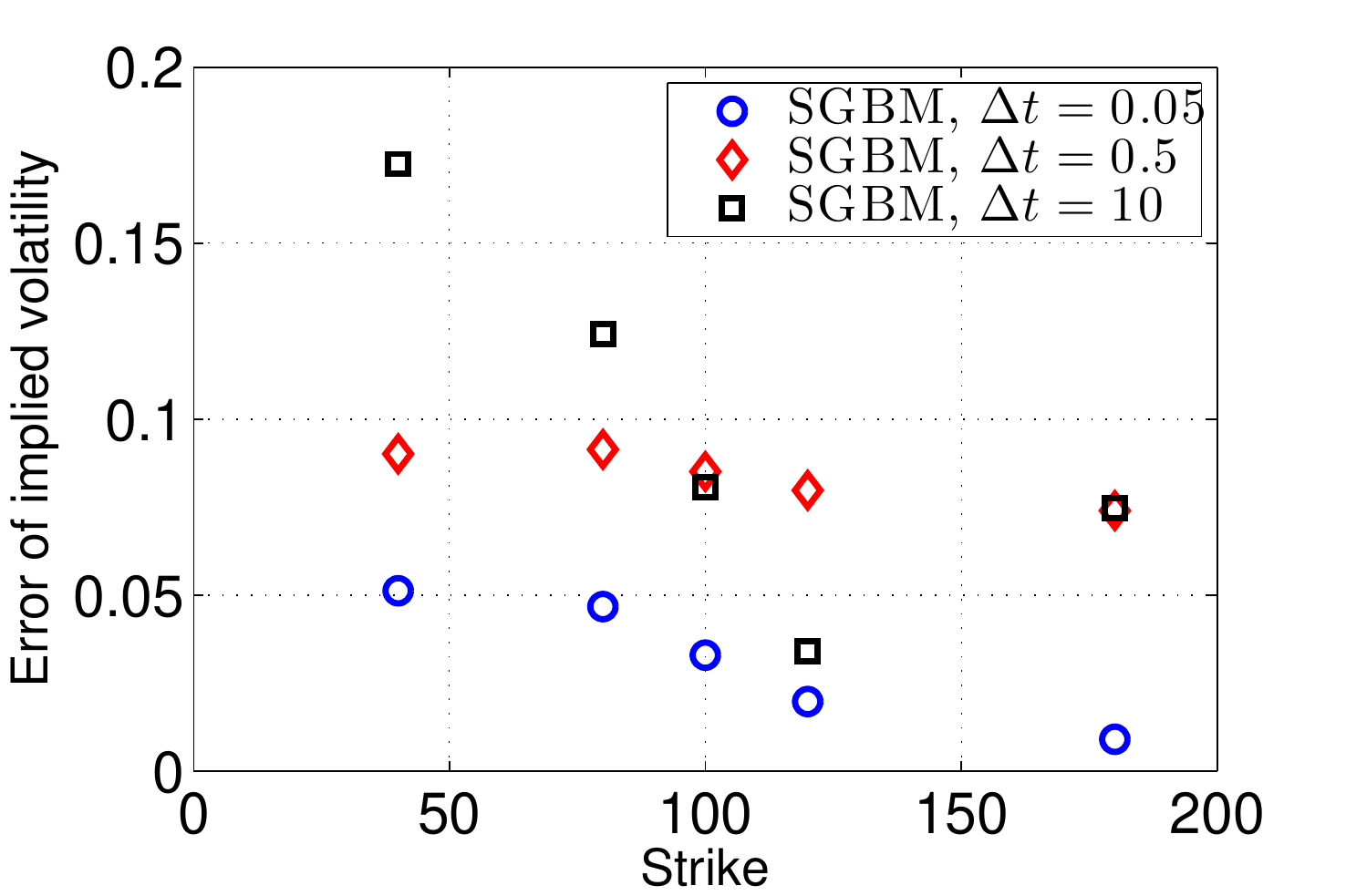}}
\caption{ Error of the implied volatility (\%) vs. strike values. obtained via the same results in Table \ref{tab:impliiedvola}. Reference values: Monte Carlo results.}
\label{fig:timestephhw}
\end{figure}

Table \ref{tab:impliiedvola}  presents the calculated implied volatility (\%) results of MC and SGBM with different time steps and 
strike values $K=\{40, 80, 100, 120, 180\}$ when $T=10$.
Figure \ref{fig:timestephhw} displays the corresponding errors in the implied volatility results. 
The reference value is the discounted cash flow results obtained via Monte Carlo.
\begin{table}[ht!]
\centering
\begin{tabular}{ r|r|r|r|r|r}
  \toprule
$\rho_{x,r}$  & Strike & Monte Carlo QE& SGBM $\Delta t=0.05$& SGBM $\Delta t=0.5$& SGBM $\Delta t=10$ \\
\midrule
\multirow{5}{*}{0.2}
 &  40 &   25.96(0.02)&  25.96(0.007)&   25.98(0.006)&   26.09(0.010)\\
 &  80 &  19.96(0.01) &  19.95(0.005)&   19.98(0.010)&   20.04(0.010)\\
 & 100 &  18.35(0.01) &  18.34(0.005)&   18.38(0.008)&   18.40(0.009)\\
 & 120 &  17.45(0.01) &  17.43(0.002)&   17.48(0.005)&   17.45(0.011)\\
 & 180 &  17.34(0.03) &  17.32(0.007)&   17.36(0.004)&   17.22(0.016)\\

 \midrule
\multirow{5}{*}{0.6}
 &  40&26.46(0.03)&   26.48(0.006)&   26.49(0.005)& 26.64(0.013)\\
 &  80&20.71(0.02)&   20.70(0.005)&   20.75(0.008)& 20.86(0.016)\\
 & 100&19.23(0.01)&   19.21(0.002)&   19.28(0.008)& 19.35(0.013)\\
 & 120&18.42(0.02)&   18.39(0.003)&   18.48(0.008)& 18.49(0.014)\\
 & 180&18.27(0.04)&   18.25(0.006)&   18.34(0.005)& 18.26(0.017)\\
   \bottomrule
\end{tabular}
\caption{Implied volatility ($\%$) results of Monte Carlo method and SGBM method. Number of bundles $J=64$ and polynomial order $p=2$. Test B with $T=10$. }
\label{tab:impliiedvola}
\end{table}

The table and figure show that when we take more time steps between $t_0$ and $t_M$, the results become more accurate. However, the results with larger time steps are also highly satisfactory. We can thus enhance the accuracy of the SGBM by using more time steps, but this will reduce the method's efficiency. 

\subsubsection{Exposure profiles of Bermudan options under HHW model}\label{sec:pricingbermhhwmodel}

We now price a Bermudan put option which can be exercised at 10 equally-space exercise date before maturity $T$. The strike is set to $K=100$. 
We use the parameters from Test B, with \{$\rho_{x,r}=0.2$, \mbox{and } $T=5$\} and \{$\rho_{x,r}=0.6$, \mbox{ and } $T=10$\} respectively. 

\begin{figure}[ht!]
\subfigure[$\rho_{x,r}=0.2$, $T=5$]{\includegraphics[width=6cm]{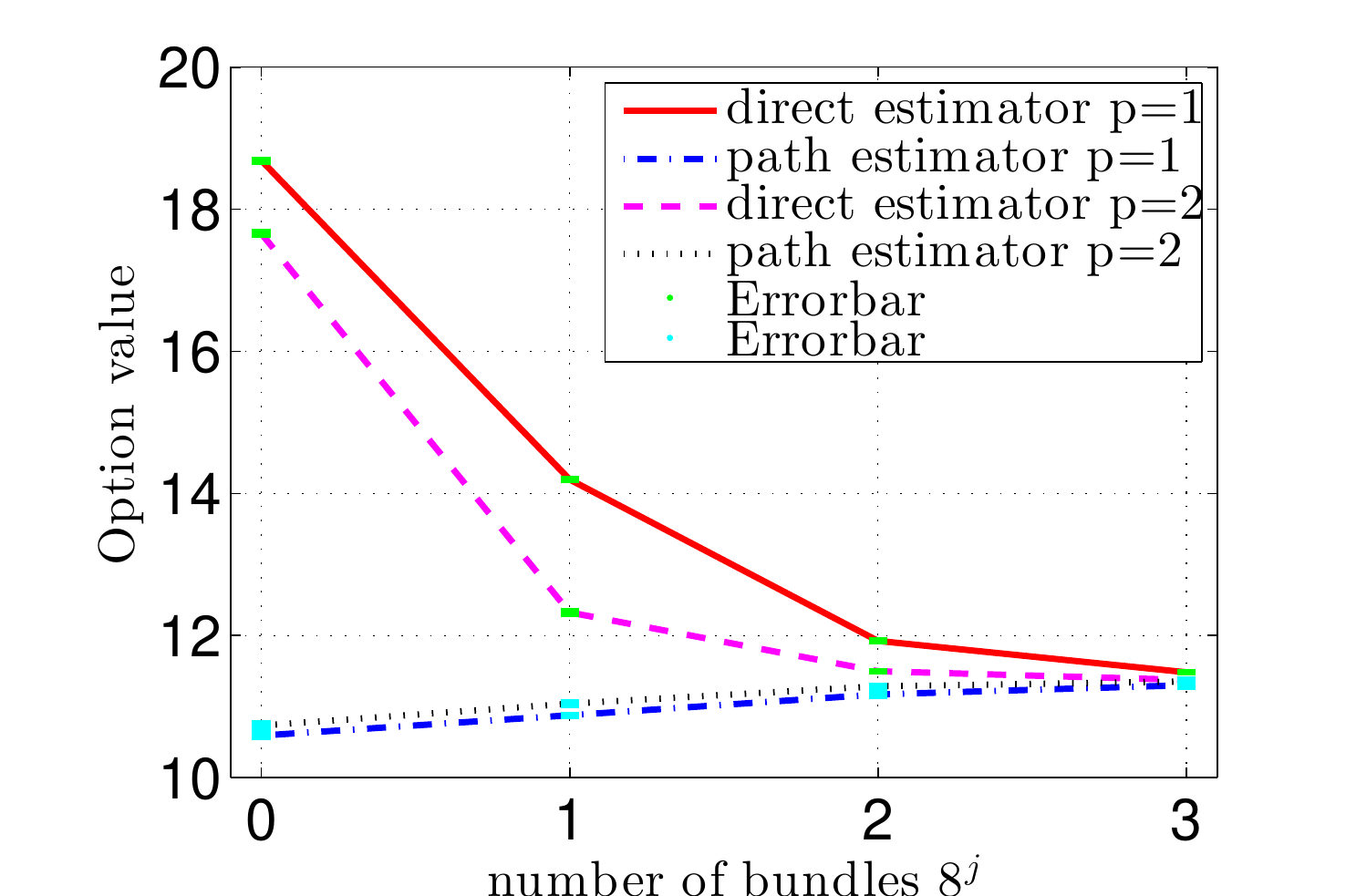}}
\hfill
\subfigure[$\rho_{x,r}=0.6$, $T=10$]{\includegraphics[width=6cm]{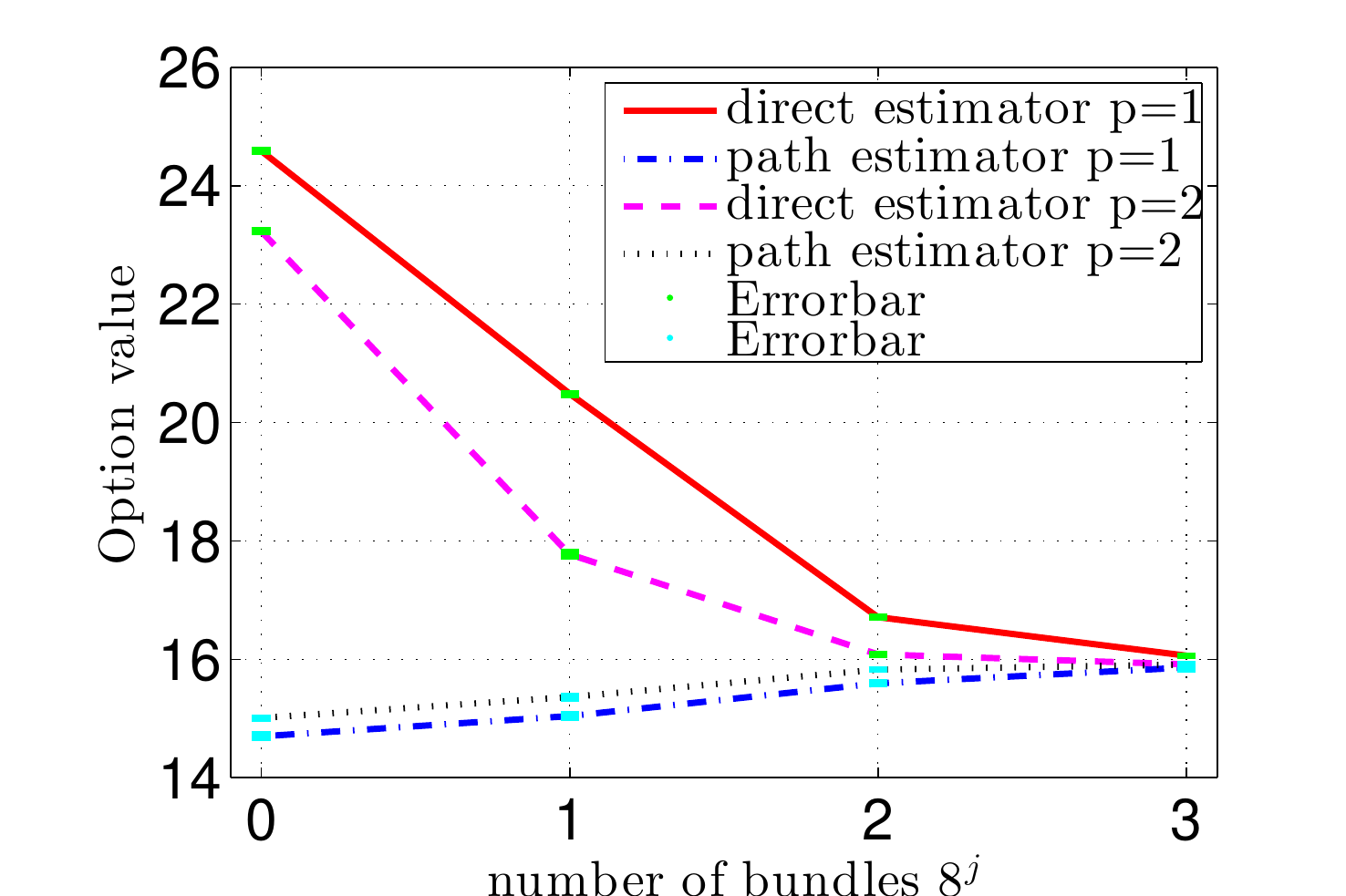}}
\caption{ Comparison of option values by the direct estimator and path estimator, when $p=1$ and $p=2$. Test B with $T=10$.}
\label{fig:H1HWoption}
\end{figure}

\begin{figure}[ht!]
\subfigure[$\rho_{x,r}=0.2$, $T=5$]{\includegraphics[width=6cm]{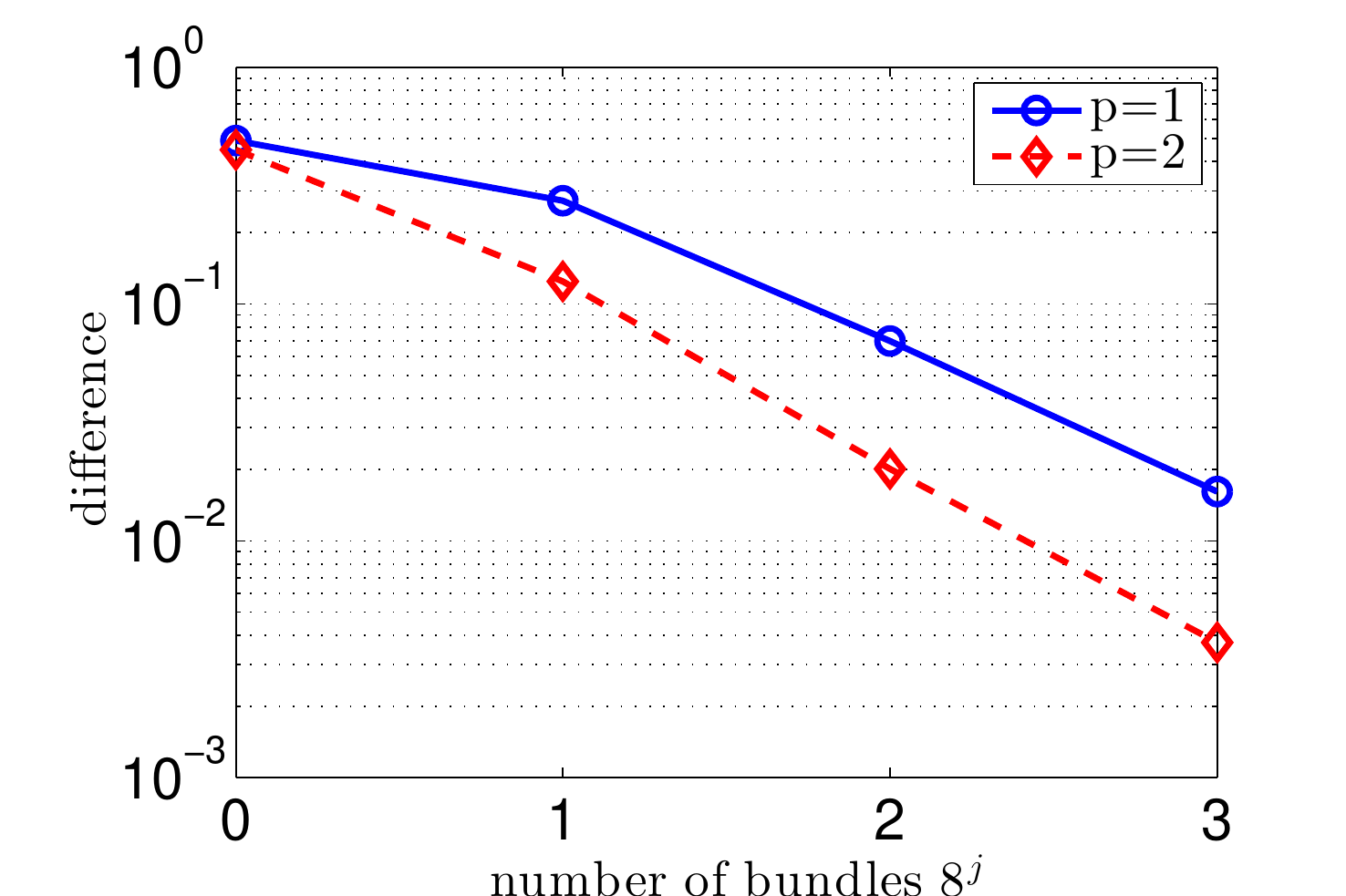}}
\hfill
\subfigure[$\rho_{x,r}=0.6$, $T=10$]{\includegraphics[width=6cm]{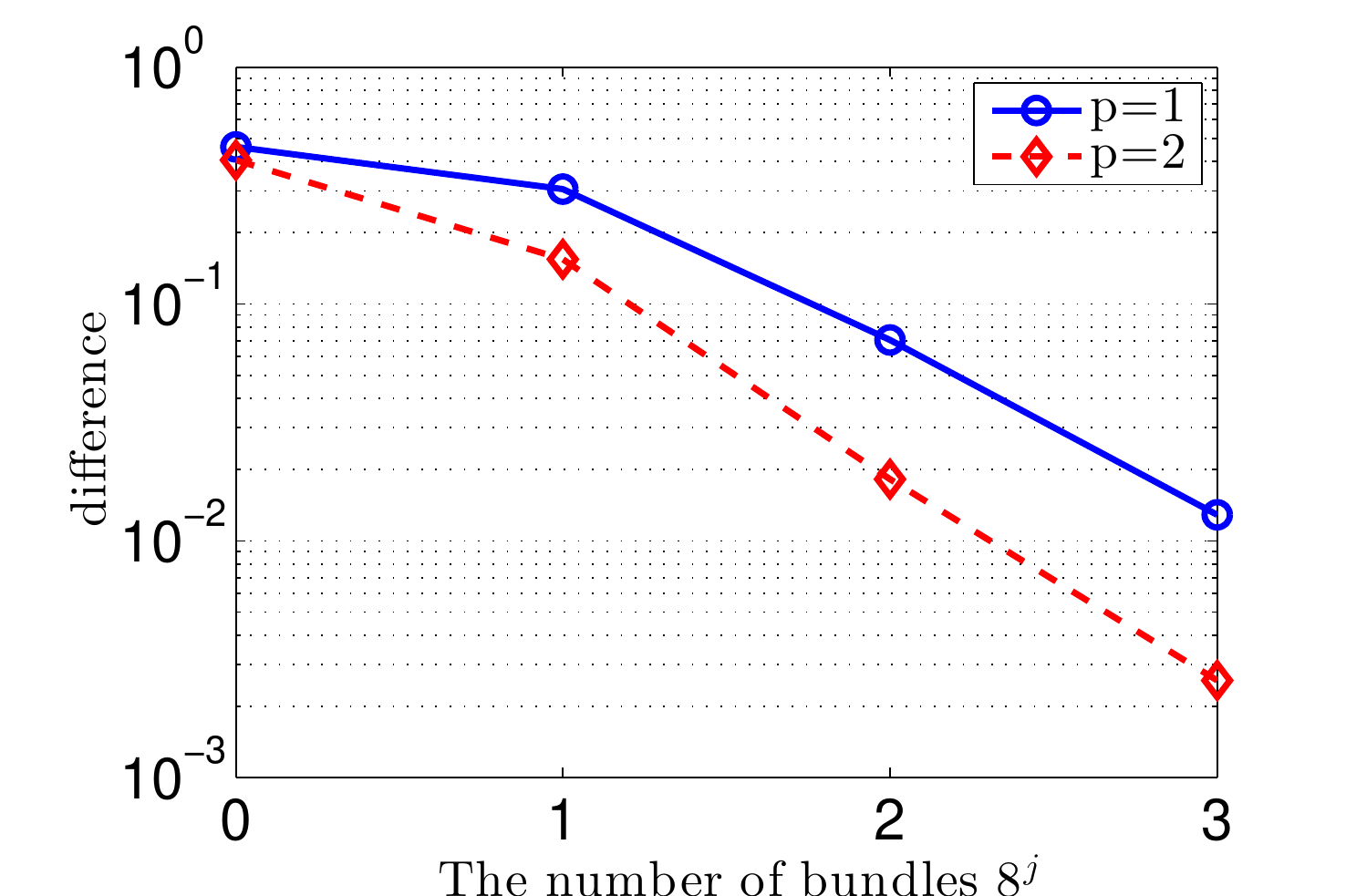}}
\caption{ Comparison of EE values obtained by the SGBM direct estimator and path estimator, when $p=1$ and $p=2$. }
\label{fig:H1HWEE}
\end{figure}

We test the convergence of SGBM by comparing the  direct estimator and the path estimator. The comparison of option value convergence with SGBM basis functions of different order  is made in Figure \ref{fig:H1HWoption}. 
Figure \ref{fig:H1HWEE} then shows the SGBM convergence of the difference of the EE values obtained by the direct and path estimators, when $p=1$ and $p=2$ w.r.t the number of bundles. The results indicate that approximation by $p=2$ is favorable, and the number of bundles is best set to $8^3=512$. The difference in the EE values of the direct and path estimator decreases with an increasing number of bundles. These HHW results support the conclusions in section \ref{sec:pricingbermheston} and thus the convergence of the SGBM.

For comparison purposes, we also present the corresponding converged results at time $t_0$  in Table \ref{tab:valueat0bermhhw}.
We also present the CVA value computed by Equation (\ref{eq:CVAdiscrete}).
\begin{table}[ht!]
\centering
\begin{tabular}{ c|r|r|r}
  \toprule
 &  & SGBM direct(std.) & SGBM path(std.) \\
 \midrule
  \multirow{4}{*}{$\rho_{x,r}=0.2$, $T=5$} 
&$V(0)$                  & 11.3747( 6.5e-04)  & 11.3507(1.5e-02)\\
&$\Delta_{\text{EE}}(0)$ & -0.2935( 3.0e-05)  &-\\
&$\Gamma_{\text{EE}}(0)$ &  0.0143(3.6e-05) &-\\
& CVA                    & 0.9829(3.1e-03)  &-\\
 \midrule
  \multirow{4}{*}{$\rho_{x,r}=0.6$, $T=10$} 
&$V(0)$                 &15.9162(1.28e-02) & 15.9310(1.9e-03) \\
&$\Delta_{\text{EE}}(0)$ & -0.2608(6.39e-04)&-\\
&$\Gamma_{\text{EE}}(0)$& 0.0085(2.08e-05) &-\\
& CVA                    &2.9678(3.42e-03)  & -\\
 \bottomrule
\end{tabular}
\caption{Values of option, Greeks and CVA; Bermudan put option; SGBM based on 5 simulations.}
\label{tab:valueat0bermhhw}
\end{table}

\subsection{Pricing barrier options and the accuracy}

In this subsection, we present results for a knock-out barrier put option under the Heston and HHW models, with barrier level $H=0.8S_0$.
Basis functions of order $p=2$ are chosen for calculation so that we obtain accurate sensitivities when applying SGBM. 
The reference values are obtained by the COS method for the Heston model. For the HHW model, we use the discounted cash flow Monte Carlo results as a reference. 
If the path does not hit the barrier, the cash flow
is equal to the payoff at the maturity; otherwise the option is knocked out at a path and the cash flow is zero. 

\begin{figure}[ht!]
\subfigure[Exposure: EE and PFE]{\includegraphics[width=6cm]{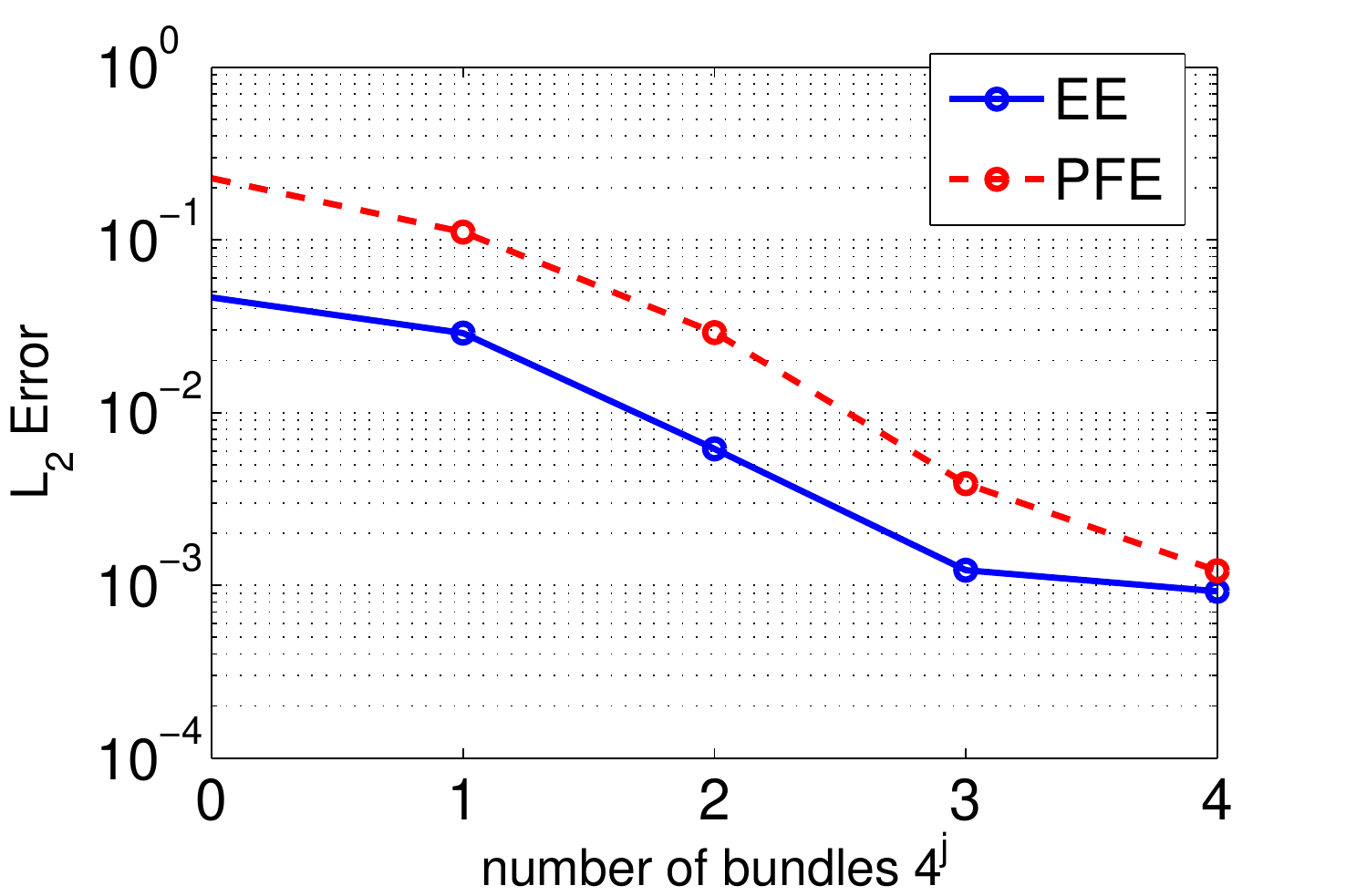}}
\hfill
\subfigure[Greeks:$\Delta_{\text{EE}}$ and $\Gamma_{\text{EE}}$ ]{\includegraphics[width=6cm]{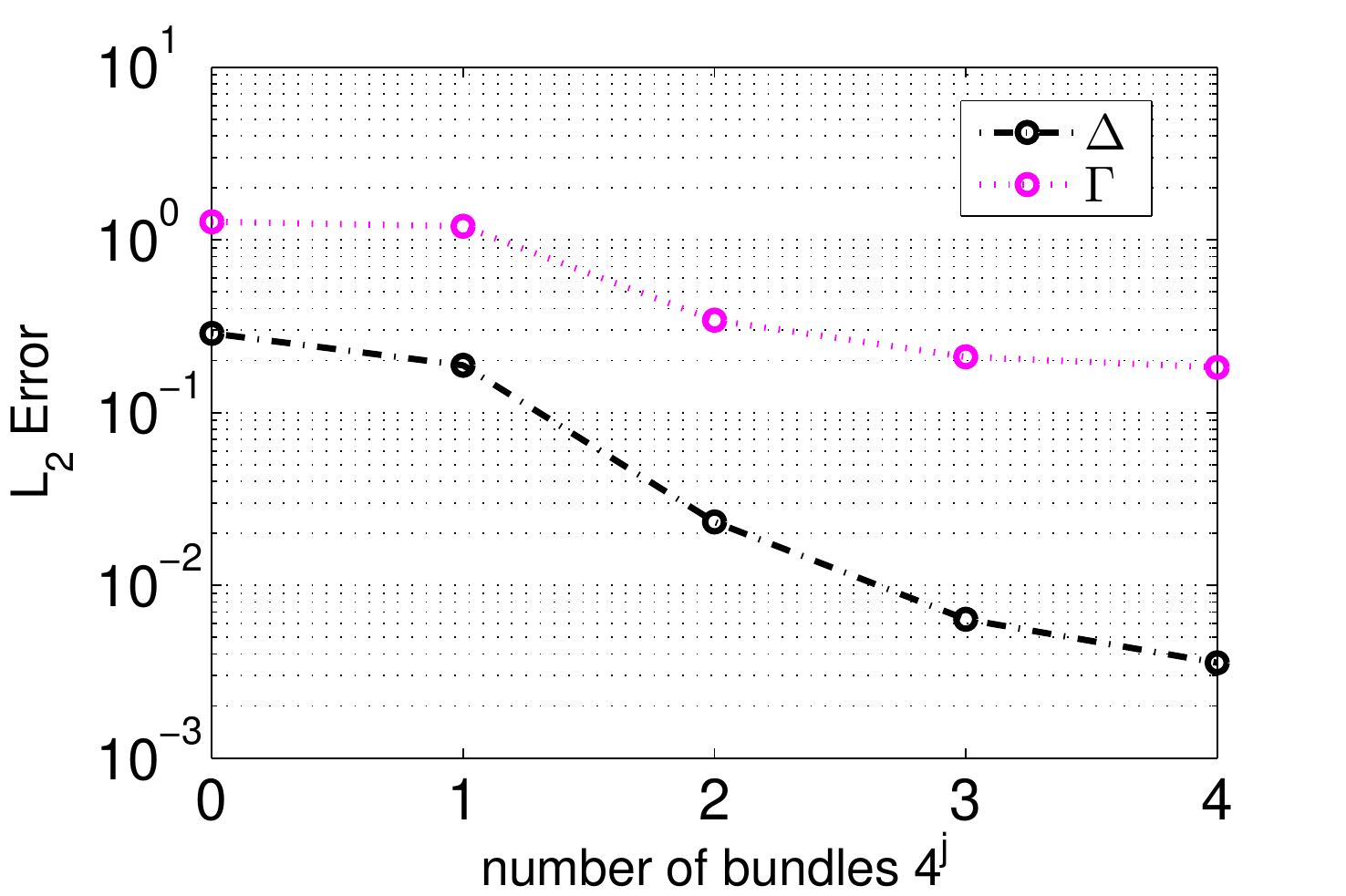}}
\caption{Relative $L_2$-error of the exposure and Greeks for a barrier option. Parameters of Test A under the Heston model.}
\label{fig:barrier}
\end{figure}

Under the Heston model, with the parameters from Test A,  Figure \ref{fig:barrier} confirms the convergence of SGBM 
by plotting the $L_2$-error of the exposure and their Greek values for barrier options w.r.t the number of bundles under the Heston model, where
COS method is available for the reference values. The corresponding values are presented in Table \ref{tab:valueat0barrier}. 

\begin{table}[ht!]
\centering
\begin{tabular}{ r|r|r|r}
  \toprule
  &COS & SGBM direct(std.) & Monte Carlo(std.)\\
 \midrule
$V(0)$                 & 1.2300 & 1.2299(1.8e-03)  &1.2283(4.9e-03)\\
$\Delta_{\text{EE}}(0)$& -0.0605   &-0.0609(1.2e-04) &-\\
$\Gamma_{\text{EE}}(0)$& 0.0031   &0.0020(3.3e-05) &-\\
CVA                   &0.0363   &0.0363(8.7e-05)  & 0.0362(1.4e-04)\\
 \bottomrule
\end{tabular}
\caption{Values of option, Greeks and CVA; knocked-out barrier put option; SGBM and MC  based on 5 simulations. Parameters of Test A under the Heston model. }
\label{tab:valueat0barrier}
\end{table}

For the HHW model computations, we use the parameters in Test B, and the resulting values are given in Table \ref{tab:valueat0barrierhhw}.
\begin{table}[ht!]
\centering
\begin{tabular}{ c|r|r|r}
  \toprule
 &  & SGBM direct(std.) & Monte Carlo(std.)\\
 \midrule
  \multirow{4}{*}{$\rho_{x,r}=0.2$, $T=5$} 
&$V(0)$                   & 0.5767( 8.6e-04) & 0.5579(2.1e-03)\\
&$\Delta_{\text{EE}}(0)$  &-0.0230(6.7e-05)&-\\
&$\Gamma_{\text{EE}}(0)$  &-2.2e-04(7.3e-06)&-\\
& CVA                     &0.9829(1.1e-04)&-\\
 \midrule
  \multirow{4}{*}{$\rho_{x,r}=0.6$, $T=10$} 
&$V(0)$                   &0.3372(1.6e-03) &0.3333(4.8e-03) \\
&$\Delta_{\text{EE}}(0)$   & -2.1e-04(2.8e-05) &-\\
&$\Gamma_{\text{EE}}(0)$  &-5.6-04(5.3e-06) &-\\
& CVA                      &0.0875(5.5e-04) &-\\
 \bottomrule
\end{tabular}
\caption{Values of option, Greeks and CVA; knocked-out put barrier option; SGBM and MC  based on 5 simulations.  Parameters of Test B under the HHW model.}
\label{tab:valueat0barrierhhw}
\end{table}

\section{Conclusion}
In this paper we have applied the Stochastic Grid Bundling Method (SGBM) for the computation of exposure profiles and Greek values for asset dynamics with stochastic volatility and stochastic interest rate. 

SGBM can be applied for the computation of expected exposure and potential full exposure functions for European, Bermudan as well as barrier options. The algorithmic structure as well as the essential method components are very similar, which makes SGBM  a suitable CVA valuation framework. 

We presented arguments for choosing the basis functions, presented two types of bundling algorithms, and showed SGBM convergence of the direct and path estimator with respect to an increasing number of bundles.
Numerical experiments demonstrate its convergence and accuracy. 

SGBM is based on the availability of analytic formulas for the discounted moments. 
When the SDE dynamics belong to the affine class, the discounted moments can be derived directly from the discounted ChF. 
Under the HHW model, these are not available and we have to resume to approximated moments. 
The impact of this on the accuracy of the results is checked.

Higher-order polynomials as the basis functions are important when accurate Greek values are needed; otherwise, polynomial order $p=1$ is sufficient for option prices and exposure quantities.

The computational efficiency is impacted  by the number of bundles used in SGBM. A parallelized algorithm will be important for reducing the calculation times. 

{\bf Acknowledgment} We thank Lech A. Grzelak, Shashi Jain, Kees de Graaf and Drona Kandhai for very helpful discussions, as well as  the CVA team at ING bank. 
Financial support by the Dutch Technology Foundation STW (project 12214) is gratefully acknowledged.

\begin{appendices}
\section{Proofs}
\subsection{Proof of proposition 1}\label{app:proof1}
Here we provide the proof of Proposition~1.
\begin{align}
 &\|c_{m}- \tilde c_{m}\|_{L_2(I_{m})}^2 \notag \\
 &= \int_{I_{m}}\left(\mathbb{E}^\mathbb{Q}\left[V_{m+1}\left(\textbf{X}_{m+1}\right)\cdot D(t_m,t_{m+1})\bigg|\textbf{X}_m\right]-
  \mathbb{E}^\mathbb{Q}\left[ \widetilde V_{m+1}\left(\textbf{X}_{m+1}\right)\cdot D(t_m,t_{m+1})\bigg|\textbf{X}_m\right]\right)^2 d \textbf{X}_m \notag \\
  &\leq \int_{I_{m}}\left(\mathbb{E}^\mathbb{Q}
  \left[\left(V_{m+1}\left(\textbf{X}_{m+1}\right)- \widetilde V_{m+1}\left(\textbf{X}_{m+1}\right)\right)^2
  \cdot \left(D(t_m,t_{m+1})\right)^2\bigg|\textbf{X}_m\right]\right) d \textbf{X}_m  \notag \\
    &\leq \int_{I_{m}}\left(\mathbb{E}^\mathbb{Q}
  \left[\left(V_{m+1}\left(\textbf{X}_{m+1}\right)- \widetilde V_{m+1}\left(\textbf{X}_{m+1}\right)\right)^2
  \bigg|\textbf{X}_m\right]\right) d \textbf{X}_m  \notag \\
      &\leq \int_{I_{m}} \int_{I_{m+1}}
  \left(V_{m+1}\left(\textbf{X}_{m+1}\right)- \widetilde V_{m+1}\left(\textbf{X}_{m+1}\right)\right)^2
  f\left(\textbf{X}_{m+1};\textbf{X}_m\right) d \textbf{X}_m d \textbf{X}_{m+1}  \notag \\
        &\leq \int_{I_{m+1}}
  \left(V_{m+1}\left(\textbf{X}_{m+1}\right)- \widetilde V_{m+1}\left(\textbf{X}_{m+1}\right)\right)^2
  \left(\int_{I_{m}} f\left(\textbf{X}_{m+1}; \textbf{X}_m\right) d \textbf{X}_m\right) d \textbf{X}_{m+1}  \notag \\
         &\leq \left(\int_{I_{m+1}}
  \left(V_{m+1}\left(\textbf{X}_{m+1}\right)- \widetilde V_{m+1}\left(\textbf{X}_{m+1}\right)\right)^2
   d \textbf{X}_{m+1}\right)\cdot\left(\int_{I_{m+1}}  \left(\int_{I_{m}} 
   f\left(\textbf{X}_{m+1}; \textbf{X}_m\right) d \textbf{X}_m\right)d \textbf{X}_{m+1} \right)\notag \\
   &=(1-\epsilon) \|V_{m+1}- \widetilde V_{m+1}\|_{L_2(I_{m+1})}^2 \cdot h(I_m),
  \end{align} 
where $h(I_m)$ is the size of the domain $I_m$,  as 
\begin{align}
&\int_{I_{m+1}} \int_{I_{m}} f\left(\textbf{X}_{m+1}; \textbf{X}_m\right) d \textbf{X}_m d \textbf{X}_{m+1} \notag \\
&=\int_{I_{m}}\int_{I_{m+1}}f\left(\textbf{X}_{m+1}; \textbf{X}_m\right)d \textbf{X}_{m+1}d \textbf{X}_m 
= \int_{I_{m}}(1-\epsilon) d \textbf{X}_m=(1-\epsilon)h(I_m).
\end{align}

\subsection{Proof of proposition 2}\label{app:proof2}

We approximate function $f(x)$ by a polynomial $\hat f(x)$ of order $p$ on interval $I$, where the residual value is defined as 

\begin{equation}
 e_p(x):=f(x)-\hat f(x), \, x \in I.
\end{equation}
Notice that $e_p^{(p+1)}(x)=f^{(p+1)}(x)$ as $\hat f^{(p+1)}(x)=0$.

The error of the 'best' estimation in $L_2$ sense can be bounded when $f(x)$ is $p+1$ times differentiable. 
Choose $p+1$ distinct points, $\{x_{0,0},x_{0,1},\dots,x_{0,p}\} \in I$, and apply interpolation, 
 $\forall x \in I$,  $\exists \; \eta(x) \in I$, such that 
 \begin{equation}
  e_p(x)=\frac{f^{(p+1)}(\eta(x))}{(n+1)!}\prod_{i=0}^p(x-x_{0,i}).
 \end{equation}
 This implies that we can find a polynomial, such that the residual function  $e_p(x_{0,0})=e_p(x_{0,1})=\cdots=e_p(x_{0,p}) $. By Rolle's theorem, 
 points $\{x_{1,0},x_{1,1},\dots, x_{1,p-1}\}$ exist, so that the first derivative 
 $e^{(1)}(x_{1,0})=e^{(1)}(x_{1,1})=\dots=e^{(1)}( x_{1,p-1})=0$. By induction, $\forall r\leq p$, there exist $p-r+1$ points $\{x_{r,0},x_{r,1},\dots,x_{r,p-r}\}$,
 From the fundamental theorem
 of calculus,$\forall r\leq p$, we have that for any point $y$ in $I$, 
 \begin{equation}
  e_p^{(r)}(y)=e_p^{(r)}(x_{r,0})+\int _{x_{r,0}}^y e_p^{(r+1)}(x)dx.
 \end{equation}
As $e_p^{(r)}(x_{r,0})=0$, we have following
 \begin{align}
 e_p^{(r)}(y)& =\int _{x_{r,0}}^y e_p^{(r+1)}(x)dx 
 \leq \int _{x_{r,0}}^y \big |e_p^{(r+1)}(x)\big |dx \notag \\
 & \leq \int _{x_{r,0}}^y \big|1\big|\cdot \big |e_p^{(r+1)}(x)\big |dx 
  \leq \left(\int _{x_{r,0}}^y \big|1\big|^2dx\right)^{\frac{1}{2}}\left(\int _{x_{r,0}}^y \big|e_p^{(r+1)}(x)\big|^2dx\right)^{\frac{1}{2}} \notag \\
 & \leq h^{\frac{1}{2}} \|e_p^{(r+1)}\|_{L^2{(I)}}.
 \end{align}
 When we square both sides, we obtain 
 \begin{equation}
  \left(e_p^{(r)}(y)\right)^2 \leq h \|e_p^{(r+1)}\|^2_{L^2{(I)}}, 
 \end{equation}
 and as a result, the $L_2$ norm of $e_p^{(r)}(y)$ in interval $I$ is given by,
 \begin{align}
 \|e_p^{(r)}\|_{L^2{(I)}}&= \left(\int_I \left(e_p^{(r)}(y)\right)^2 dy \right)^{\frac{1}{2}}\notag \\
 & \leq \left(\int_I h \|e_p^{(r+1)}\|^2_{L^2{(I)}} dy\right)^{\frac{1}{2}} \leq h \|e_p^{(r+1)}\|_{L^2{(I)}}.
 \end{align}
 By induction, it is easy to find that
 
 \begin{align}
   \|e_p\|_{L^2{(I)}}\leq h \|e_p^{(1)}\|_{L^2{(I)}}\leq \cdots \leq h^{p+1} \|e_p^{(p+1)}\|_{L^2{(I)}}=h^{p+1} \|f^{(p+1)}\|_{L^2{(I)}}
 \end{align}

\section{The joint discounted characteristic functions}
In this appendix we provide the reader with known results about characteristic function. Based on these, it is easy to determine (either by hand or by a computer program) the discounted moments, needed in SGBM.
\subsection{Heston model}\label{app:discountedchfheston}
The expression for the discounted ChF of the Heston model is given by\cite{fangfang}:
\begin{equation}
 \Phi(u_1,u_2,T|\textbf{X}_t)= \exp\left(\bar A(u_1,u_2,\tau)+\bar B(u_1,\tau)x_t+\bar C(u_1,u_2,\tau)v_t\right),
\end{equation}
%
where,
\begin{align}
&\bar A(u_1,u_2,\tau)=I_1+I_2, \; \bar B(u_1,\tau)=iu_1, \notag \\
&\bar C(u_1,u_2,\tau)=r_{+}-\frac{2D_1}{\gamma^2\left(1-ge^{-D_1 \tau}\right)}, \notag \\
\end{align}
where 
\begin{align}
 g&=\frac{iu_2-r_{-}}{iu_2-r_{+}}, \notag D_1=\sqrt{\left(\kappa-\gamma \rho_{x,v}iu_1\right)^2+\gamma^2u_1(u_1+i)}, \notag \\
 r_{\pm}&=\frac{1}{\gamma^2}\left(\kappa-\gamma\rho_{x,v}iu_1\pm D_1\right),
\end{align}
and 
\begin{align}
 &I_1=\kappa \bar v\left(r_{-}\tau-\frac{2}{\gamma^2}\log\left(\frac{1-ge^{-D_1 \tau}}{1-g}\right)\right), \notag I_2=r(iu_1-1)\tau.
\end{align}
\subsubsection{Discounted moments}\label{app:discountedmheston}
The analytic formulas for the discounted moments and their first and second derivatives are obtained with symbolic calculations in MATLAB.
We will present the discounted moments for the Heston model.

\begin{align}
 \mathbb{E}^{\mathbb{Q}}\left[x_{T}\cdot D(t,T)|\mathbf{X}_t\right]&= \left(x_t+\frac{1}{2\kappa}\left(\bar v-v_t\right)\left(1-e^{-\kappa \tau}\right)+(r-\frac{1}{2}\bar v)\tau\right)e^{-r\tau}, \\ 
 \mathbb{E}^{\mathbb{Q}}\left[x_{T}^2\cdot D(t,T)|\mathbf{X}_t\right]&=\Bigg(\left(x_t+\frac{1}{2\kappa}\left(\bar v-v_t\right)\left(1-e^{-\kappa \tau}\right)+\left(r-\frac{1}{2}\bar v\right)\tau\right)^2 \notag \\
  &\quad\quad +\frac{\bar v}{8\kappa ^3}\Omega_1+\frac{v_t}{4\kappa^3}\Omega_2 \Bigg)e^{-r\tau} ,\\
 \mathbb{E}^{\mathbb{Q}}\left[v_{T}\cdot D(t,T)|\mathbf{X}_t\right]&= \left(\bar v+\left(v_t-\bar v\right)e^{-\kappa \tau}\right)e^{-r\tau}, \\
 \mathbb{E}^{\mathbb{Q}}\left[v_{T}^2\cdot D(t,T)|\mathbf{X}_t\right]&=\Bigg(\frac{v_t\gamma^2}{\kappa}\left(e^{-\kappa \tau}-e^{-2\kappa \tau}\right)
 +\frac{\bar v \gamma^2}{2\kappa}\left(1-e^{-\kappa \tau}\right)^2 \notag \\
 &\quad \quad +\left(\bar v+\left(v_t-\bar v\right)e^{-\kappa \tau}\right)^2\Bigg)e^{-r\tau}, \\
 \mathbb{E}^{\mathbb{Q}}\left[x_{T}\cdot v_{T}\cdot D(t,T)|\mathbf{X}_t\right]&= \Bigg((\bar v+\left(v_t-\bar v\right)e^{-\kappa \tau})
 \left(x_t+\frac{1}{2\kappa}\left(\bar v-v_t\right)(1-e^{-\kappa \tau})+(r-\frac{1}{2}\bar v) \tau\right) \notag \\
 & \quad \quad +  \frac{\bar v \gamma^2}{4\kappa^2}\Omega_3+\frac{v_t\gamma^2}{2\kappa^2}\Omega_4\Bigg)e^{-r\tau},
\end{align}
with $\mathbf{X}_t=[x_t,v_t]^T$, $\tau =T-t$, where
\begin{align}
 \Omega_1=& e^{-2\kappa \tau}\gamma^2+4e^{-\kappa t}\left((1+\kappa\tau)\gamma^2-2\rho\kappa\gamma\left(2+\kappa t\right)+2\kappa^2\right) 
 +\left(2\kappa\tau-5\right)\gamma^2 \notag \\
 &\quad -8\rho\kappa\gamma\left(\kappa\tau-2\right) +8\kappa^2\left(\kappa\tau-1\right),\\
 \Omega_2=&-e^{-2\kappa\tau}\gamma^2+2 e^{-\kappa\tau}\left(-\kappa\tau\gamma^2+2\rho\kappa\gamma\left(1+\kappa\tau\right)-2\kappa^2\right)
 +\gamma^2-4\kappa\rho\gamma+4\kappa^2,\\
 \Omega_3=&e^{-2\kappa\tau}+2\kappa e^{-\kappa\tau}(\tau-\frac{2\rho}{\gamma}\left(1+\kappa\tau\right))+\frac{4\kappa\rho-\gamma}{\gamma}, \\
 \Omega_4=&e^{-\kappa\tau}\left(1-\kappa\tau+\frac{2\rho\kappa^2 \tau}{\gamma}\right)-e^{-2\kappa \tau}.
\end{align}

\subsection{Black-Scholes-Hull-White model}\label{app:discountedchfbshw}
The expression for discounted ChF of the BSHW model is given by:
\begin{equation}
 \Phi(u_1,u_2,u_3,T|\textbf{X}_t)= \exp\left(\bar A(u_1,u_2,w,\tau)+\bar B(u_1,\tau)x_t+\bar D(u_1,u_3,\tau)r_t\right),
\end{equation}
with $\bar B(u_1,\tau)$ as in \ref{app:discountedchfheston}, and
\begin{align}
&\bar A(u_1,u_2,u_3,\tau)=I_1+I_2+I_3+I_4, 
\; \bar D(u_1,u_3,\tau)=\frac{iu_1-1}{\lambda}\left(1-e^{-\lambda \tau}\right)+iu_3 e^{-\lambda \tau}, \notag \\
\end{align}
and  
\begin{align}
&I_1=\frac{1}{2}\sigma^2iu_1(iu_1-1)\tau, \notag \\
&I_2=\theta \left((iu_1-1)\tau+\frac{1}{\lambda}(e^{-\lambda \tau}-1)(iu_1-1)-iu_3\left(e^{-\lambda \tau}-1\right)\right), \notag \\
&I_3=\frac{\eta^2}{2\lambda^2}\left(\frac{2}{\lambda}(u_1+i)(e^{-\lambda \tau}-1)(\lambda u_3-u_1-i)+
\frac{1}{2\lambda}\left(e^{-2\lambda \tau}-1\right)\left(\lambda u_3-u_1-i\right)^2-(u_1+i)^2\tau\right), \notag \\
 &I_4=\frac{\eta\theta\sigma\rho_{x,r}}{\lambda}\left(-\frac{iu_1+u_1^2}{\lambda}(\lambda \tau+e^{-\lambda \tau}-1)
  +u_1u_3(e^{-\lambda \tau}-1)\right).
\end{align}
Again the discounted moments are obtained by symbolic computations in MATLAB.
\subsection{H1HW model}\label{app:discountedchfhhw}
The expression for the discounted ChF for the approximate HHW model, called the H1HW model, is given by\cite{Lech}:
\begin{equation}
 \Phi(u_1,u_2,u_3,T|\textbf{X}_t)= \exp\left(\bar A(u_1,u_2,u_3,\tau)+\bar B(u_1,\tau)x_t+\bar C(u_1,u_2,\tau)v_t+\bar D(u_1,u_3,\tau)r_t\right),
\end{equation}
where the coefficients $\bar B(u_1,\tau)$, $\bar C(u_1,u_2,\tau)$ and $\bar D(u_1,u_3,\tau)$ are the same in sections \ref{app:discountedchfheston} and \ref{app:discountedchfbshw}, and 
\begin{align}
&\bar A(u_1,u_2,u_3,\tau)=I_1+I_2+I_3+I_4, 
\end{align}
where expressions $g$, $D_1$ and $r_{\pm}$ are the same one as for the Heston model,
and 
\begin{align}
&I_1=\theta \left((iu_1-1)\tau+\frac{1}{\lambda}(e^{-\lambda \tau}-1)(iu_1-1)-iu_3\left(e^{-\lambda \tau}-1\right)\right), \\ \notag
 &I_2=\kappa \bar v\left(r_{-}\tau-\frac{2}{\gamma^2}\log\left(\frac{1-ge^{-D_1 \tau}}{1-g}\right)\right), \notag \\
&I_3=\frac{\eta^2}{2\lambda^2}\left(\frac{2}{\lambda}(u_1+i)(e^{-\lambda \tau}-1)(\lambda u_3-u_1-i)+
\frac{1}{2\lambda}\left(e^{-2\lambda \tau}-1\right)\left(\lambda u_3-u_1-i\right)^2-(u_1+i)^2\tau\right), \notag \\
 &I_4=\eta\rho_{x,r}\left(-\frac{iu_1+u_1^2}{\lambda}G_1(\tau, v_t)
  -u_1u_3G_2(\tau, v_t)\right).
\end{align}
The two  integrals appearing above are computed numerically as 
\begin{align}
 G_1(\tau, v_t)&:=\int_0^{\tau}\mathbb{E}\left[\sqrt{v_{T-s}}\big|v_t\right]\left(1-e^{-\lambda s}\right)ds 
 \approx \sum_{k=0}^{L-1} \mathbb{E}\left[\sqrt{v_{(T-k\Delta s)}}\big|v_t\right]\left(1-e^{-\lambda (k\Delta s)}\right)\Delta s, \\
 G_2(\tau, v_t)&:=\int_0^{\tau}\mathbb{E}\left[\sqrt{v_{T-s}}\big|v_t\right]e^{-\lambda s}ds
  \approx \sum_{k=0}^{L-1} \mathbb{E}\left[\sqrt{v_{(T-k\Delta s)}}\big|v_t\right]\left(e^{-\lambda (k\Delta s)}\right)\Delta s,
\end{align}
where $\tau=T-t$, $\Delta s=\frac{\tau}{L}$, $L$ is the number of integration intervals and the conditional expectation of the square root of the variance is given by 
(\ref{eq:varianceexpec}).
\end{appendices}

\end{document}